\documentclass[a4paper,traditabstract,longauth]{aa}
\usepackage[breaklinks, colorlinks, citecolor=blue]{hyperref}
\usepackage{graphicx,amsmath}
\usepackage{txfonts}
\usepackage[nonamebreak]{natbib}
\usepackage{mathrsfs}
\bibpunct{(}{)}{;}{a}{}{,} % to follow the A&A style

\def\setsymbol#1#2{\expandafter\def\csname #1\endcsname{#2}}
\def\getsymbol#1{\csname #1\endcsname}
\def\Planck{{\it Planck\/}}
\newbox\tablebox    \newdimen\tablewidth
\def\leaderfil{\leaders\hbox to 5pt{\hss.\hss}\hfil}
\def\endPlancktable{\tablewidth=\columnwidth 
    $$\hss\copy\tablebox\hss$$
    \vskip-\lastskip\vskip -2pt}
\def\endPlancktablewide{\tablewidth=\textwidth 
    $$\hss\copy\tablebox\hss$$
    \vskip-\lastskip\vskip -2pt}
\def\tablenote#1 #2\par{\begingroup \parindent=0.8em
    \abovedisplayshortskip=0pt\belowdisplayshortskip=0pt
    \noindent
    $$\hss\vbox{\hsize\tablewidth \hangindent=\parindent \hangafter=1 \noindent
    \hbox to \parindent{\sup{\rm #1}\hss}\strut#2\strut\par}\hss$$
    \endgroup}
\def\doubleline{\vskip 3pt\hrule \vskip 1.5pt \hrule \vskip 5pt}

\def\inv{\ifmmode^{-1}\else$^{-1}$\fi}
\def\mo{\ifmmode^{-1}\else$^{-1}$\fi}
\def\sup#1{\ifmmode ^{\rm #1}\else $^{\rm #1}$\fi}
\def\expo#1{\ifmmode \times 10^{#1}\else $\times 10^{#1}$\fi}
\def\,{\thinspace}
\def\lsim{\mathrel{\raise .4ex\hbox{\rlap{$<$}\lower 1.2ex\hbox{$\sim$}}}}
\def\gsim{\mathrel{\raise .4ex\hbox{\rlap{$>$}\lower 1.2ex\hbox{$\sim$}}}}

\def\simprop{\mathrel{\raise .4ex\hbox{\rlap{$\propto$}\lower 1.2ex\hbox{$\sim$}}}}
\def\deg{\ifmmode^\circ\else$^\circ$\fi}
\def\pdeg{\ifmmode $\setbox0=\hbox{$^{\circ}$}\rlap{\hskip.11\wd0 .}$^{\circ}
          \else \setbox0=\hbox{$^{\circ}$}\rlap{\hskip.11\wd0 .}$^{\circ}$\fi}
\def\arcs{\ifmmode {^{\scriptstyle\prime\prime}}
          \else $^{\scriptstyle\prime\prime}$\fi}
\def\arcm{\ifmmode {^{\scriptstyle\prime}}
          \else $^{\scriptstyle\prime}$\fi}
\newdimen\sa  \newdimen\sb
\def\parcs{\sa=.07em \sb=.03em
     \ifmmode \hbox{\rlap{.}}^{\scriptstyle\prime\kern -\sb\prime}\hbox{\kern -\sa}
     \else \rlap{.}$^{\scriptstyle\prime\kern -\sb\prime}$\kern -\sa\fi}
\def\parcm{\sa=.08em \sb=.03em
     \ifmmode \hbox{\rlap{.}\kern\sa}^{\scriptstyle\prime}\hbox{\kern-\sb}
     \else \rlap{.}\kern\sa$^{\scriptstyle\prime}$\kern-\sb\fi}
\def\ra[#1 #2 #3.#4]{#1\sup{h}#2\sup{m}#3\sup{s}\llap.#4}
\def\dec[#1 #2 #3.#4]{#1\deg#2\arcm#3\arcs\llap.#4}
\def\deco[#1 #2 #3]{#1\deg#2\arcm#3\arcs}
\def\rra[#1 #2]{#1\sup{h}#2\sup{m}}

\def\dots{\relax\ifmmode \ldots\else $\ldots$\fi}
\def\GHz{\ifmmode $\,GHz$\else \,GHz\fi}
\def\kms{\ifmmode $\,km\,s$^{-1}\else \,km\,s$^{-1}$\fi}

\newcommand{\hi}{{\sc Hi}}
\newcommand{\poker}{{\tt POKER}}
\newcommand{\healpix}{{\tt HEALPix}}

\newcommand{\msun}{\rm M_{\odot}}

\def\ben{\begin{enumerate}}
\def\een{\end{enumerate}}
\def\bi{\begin{itemize}}
\def\ei{\end{itemize}}
\def\be{\begin{equation}}
\def\ee{\end{equation}}
\def\bea{\begin{eqnarray}}
\def\eea{\end{eqnarray}}

\begin{document}

%==============================
\title{\textit{Planck} 2013 Results. XXX. Cosmic infrared background
measurements and implications for star formation}
%==============================

 \author{%This author list corresponds to \title{Author list for SVN PIP\_56\_Proj\_6\_6\_Lagache, Proj. Ref. 6\_6: The multi-frequency and multi-scale CIB anisotropy}
%Prepared by R. Leonardi (rleonardi@sciops.esa.int), ESAC/ESA
%This version is from Thu Jun 13 17:13:14 2013 CET
%\subtitle{There are 239 co-authors in this list}
\author{\small
Planck Collaboration:
P.~A.~R.~Ade\inst{89}
\and
N.~Aghanim\inst{62}
\and
C.~Armitage-Caplan\inst{94}
\and
M.~Arnaud\inst{75}
\and
M.~Ashdown\inst{72, 7}
\and
F.~Atrio-Barandela\inst{20}
\and
J.~Aumont\inst{62}
\and
C.~Baccigalupi\inst{88}
\and
A.~J.~Banday\inst{97, 11}
\and
R.~B.~Barreiro\inst{69}
% ...
}
\institute{\small
APC, AstroParticule et Cosmologie, Universit\'{e} Paris Diderot, CNRS/IN2P3, CEA/lrfu, Observatoire de Paris, Sorbonne Paris Cit\'{e}, 10, rue Alice Domon et L\'{e}onie Duquet, 75205 Paris Cedex 13, France\\
\and
Aalto University Mets\"{a}hovi Radio Observatory, Mets\"{a}hovintie 114, FIN-02540 Kylm\"{a}l\"{a}, Finland\\
\and
African Institute for Mathematical Sciences, 6-8 Melrose Road, Muizenberg, Cape Town, South Africa\\
\and
Agenzia Spaziale Italiana Science Data Center, c/o ESRIN, via Galileo Galilei, Frascati, Italy\\
\and
Agenzia Spaziale Italiana, Viale Liegi 26, Roma, Italy\\
\and
Argelander-Institut f\"{u}r Astronomie, Universit\"{a}t Bonn, Auf dem H\"{u}gel 71, D-53121 Bonn, Germany\\
\and
Astrophysics Group, Cavendish Laboratory, University of Cambridge, J J Thomson Avenue, Cambridge CB3 0HE, U.K.\\
\and
Astrophysics \& Cosmology Research Unit, School of Mathematics, Statistics \& Computer Science, University of KwaZulu-Natal, Westville Campus, Private Bag X54001, Durban 4000, South Africa\\
\and
Atacama Large Millimeter/submillimeter Array, ALMA Santiago Central Offices, Alonso de Cordova 3107, Vitacura, Casilla 763 0355, Santiago, Chile\\
\and
CITA, University of Toronto, 60 St. George St., Toronto, ON M5S 3H8, Canada\\
\and
CNRS, IRAP, 9 Av. colonel Roche, BP 44346, F-31028 Toulouse cedex 4, France\\
\and
California Institute of Technology, Pasadena, California, U.S.A.\\
\and
Centre for Theoretical Cosmology, DAMTP, University of Cambridge, Wilberforce Road, Cambridge CB3 0WA U.K.\\
\and
Centro de Estudios de F\'{i}sica del Cosmos de Arag\'{o}n (CEFCA), Plaza San Juan, 1, planta 2, E-44001, Teruel, Spain\\
\and
Computational Cosmology Center, Lawrence Berkeley National Laboratory, Berkeley, California, U.S.A.\\
\and
Consejo Superior de Investigaciones Cient\'{\i}ficas (CSIC), Madrid, Spain\\
\and
DSM/Irfu/SPP, CEA-Saclay, F-91191 Gif-sur-Yvette Cedex, France\\
\and
DTU Space, National Space Institute, Technical University of Denmark, Elektrovej 327, DK-2800 Kgs. Lyngby, Denmark\\
\and
D\'{e}partement de Physique Th\'{e}orique, Universit\'{e} de Gen\`{e}ve, 24, Quai E. Ansermet,1211 Gen\`{e}ve 4, Switzerland\\
\and
Departamento de F\'{\i}sica Fundamental, Facultad de Ciencias, Universidad de Salamanca, 37008 Salamanca, Spain\\
\and
Departamento de F\'{\i}sica, Universidad de Oviedo, Avda. Calvo Sotelo s/n, Oviedo, Spain\\
\and
Department of Astronomy and Astrophysics, University of Toronto, 50 Saint George Street, Toronto, Ontario, Canada\\
\and
Department of Astrophysics/IMAPP, Radboud University Nijmegen, P.O. Box 9010, 6500 GL Nijmegen, The Netherlands\\
\and
Department of Electrical Engineering and Computer Sciences, University of California, Berkeley, California, U.S.A.\\
\and
Department of Physics \& Astronomy, University of British Columbia, 6224 Agricultural Road, Vancouver, British Columbia, Canada\\
\and
Department of Physics and Astronomy, Dana and David Dornsife College of Letter, Arts and Sciences, University of Southern California, Los Angeles, CA 90089, U.S.A.\\
\and
Department of Physics and Astronomy, University College London, London WC1E 6BT, U.K.\\
\and
Department of Physics, Gustaf H\"{a}llstr\"{o}min katu 2a, University of Helsinki, Helsinki, Finland\\
\and
Department of Physics, Princeton University, Princeton, New Jersey, U.S.A.\\
\and
Department of Physics, University of California, Berkeley, California, U.S.A.\\
\and
Department of Physics, University of California, One Shields Avenue, Davis, California, U.S.A.\\
\and
Department of Physics, University of California, Santa Barbara, California, U.S.A.\\
\and
Department of Physics, University of Illinois at Urbana-Champaign, 1110 West Green Street, Urbana, Illinois, U.S.A.\\
\and
Dipartimento di Fisica e Astronomia G. Galilei, Universit\`{a} degli Studi di Padova, via Marzolo 8, 35131 Padova, Italy\\
\and
Dipartimento di Fisica e Scienze della Terra, Universit\`{a} di Ferrara, Via Saragat 1, 44122 Ferrara, Italy\\
\and
Dipartimento di Fisica, Universit\`{a} La Sapienza, P. le A. Moro 2, Roma, Italy\\
\and
Dipartimento di Fisica, Universit\`{a} degli Studi di Milano, Via Celoria, 16, Milano, Italy\\
\and
Dipartimento di Fisica, Universit\`{a} degli Studi di Trieste, via A. Valerio 2, Trieste, Italy\\
\and
Dipartimento di Fisica, Universit\`{a} di Roma Tor Vergata, Via della Ricerca Scientifica, 1, Roma, Italy\\
\and
Discovery Center, Niels Bohr Institute, Blegdamsvej 17, Copenhagen, Denmark\\
\and
Dpto. Astrof\'{i}sica, Universidad de La Laguna (ULL), E-38206 La Laguna, Tenerife, Spain\\
\and
European Southern Observatory, ESO Vitacura, Alonso de Cordova 3107, Vitacura, Casilla 19001, Santiago, Chile\\
\and
European Space Agency, ESAC, Planck Science Office, Camino bajo del Castillo, s/n, Urbanizaci\'{o}n Villafranca del Castillo, Villanueva de la Ca\~{n}ada, Madrid, Spain\\
\and
European Space Agency, ESTEC, Keplerlaan 1, 2201 AZ Noordwijk, The Netherlands\\
\and
Finnish Centre for Astronomy with ESO (FINCA), University of Turku, V\"{a}is\"{a}l\"{a}ntie 20, FIN-21500, Piikki\"{o}, Finland\\
\and
Haverford College Astronomy Department, 370 Lancaster Avenue, Haverford, Pennsylvania, U.S.A.\\
\and
Helsinki Institute of Physics, Gustaf H\"{a}llstr\"{o}min katu 2, University of Helsinki, Helsinki, Finland\\
\and
INAF - Osservatorio Astronomico di Padova, Vicolo dell'Osservatorio 5, Padova, Italy\\
\and
INAF - Osservatorio Astronomico di Roma, via di Frascati 33, Monte Porzio Catone, Italy\\
\and
INAF - Osservatorio Astronomico di Trieste, Via G.B. Tiepolo 11, Trieste, Italy\\
\and
INAF/IASF Bologna, Via Gobetti 101, Bologna, Italy\\
\and
INAF/IASF Milano, Via E. Bassini 15, Milano, Italy\\
\and
INFN, Sezione di Bologna, Via Irnerio 46, I-40126, Bologna, Italy\\
\and
INFN, Sezione di Roma 1, Universit\`{a} di Roma Sapienza, Piazzale Aldo Moro 2, 00185, Roma, Italy\\
\and
IPAG: Institut de Plan\'{e}tologie et d'Astrophysique de Grenoble, Universit\'{e} Joseph Fourier, Grenoble 1 / CNRS-INSU, UMR 5274, Grenoble, F-38041, France\\
\and
ISDC Data Centre for Astrophysics, University of Geneva, ch. d'Ecogia 16, Versoix, Switzerland\\
\and
IUCAA, Post Bag 4, Ganeshkhind, Pune University Campus, Pune 411 007, India\\
\and
Imperial College London, Astrophysics group, Blackett Laboratory, Prince Consort Road, London, SW7 2AZ, U.K.\\
\and
Infrared Processing and Analysis Center, California Institute of Technology, Pasadena, CA 91125, U.S.A.\\
\and
Institut N\'{e}el, CNRS, Universit\'{e} Joseph Fourier Grenoble I, 25 rue des Martyrs, Grenoble, France\\
\and
Institut Universitaire de France, 103, bd Saint-Michel, 75005, Paris, France\\
\and
Institut d'Astrophysique Spatiale, CNRS (UMR8617) Universit\'{e} Paris-Sud 11, B\^{a}timent 121, Orsay, France\\
\and
Institut d'Astrophysique de Paris, CNRS (UMR7095), 98 bis Boulevard Arago, F-75014, Paris, France\\
\and
Institute for Space Sciences, Bucharest-Magurale, Romania\\
\and
Institute of Astronomy and Astrophysics, Academia Sinica, Taipei, Taiwan\\
\and
Institute of Astronomy, University of Cambridge, Madingley Road, Cambridge CB3 0HA, U.K.\\
\and
Institute of Theoretical Astrophysics, University of Oslo, Blindern, Oslo, Norway\\
\and
Instituto de Astrof\'{\i}sica de Canarias, C/V\'{\i}a L\'{a}ctea s/n, La Laguna, Tenerife, Spain\\
\and
Instituto de F\'{\i}sica de Cantabria (CSIC-Universidad de Cantabria), Avda. de los Castros s/n, Santander, Spain\\
\and
Jet Propulsion Laboratory, California Institute of Technology, 4800 Oak Grove Drive, Pasadena, California, U.S.A.\\
\and
Jodrell Bank Centre for Astrophysics, Alan Turing Building, School of Physics and Astronomy, The University of Manchester, Oxford Road, Manchester, M13 9PL, U.K.\\
\and
Kavli Institute for Cosmology Cambridge, Madingley Road, Cambridge, CB3 0HA, U.K.\\
\and
LAL, Universit\'{e} Paris-Sud, CNRS/IN2P3, Orsay, France\\
\and
LERMA, CNRS, Observatoire de Paris, 61 Avenue de l'Observatoire, Paris, France\\
\and
Laboratoire AIM, IRFU/Service d'Astrophysique - CEA/DSM - CNRS - Universit\'{e} Paris Diderot, B\^{a}t. 709, CEA-Saclay, F-91191 Gif-sur-Yvette Cedex, France\\
\and
Laboratoire Traitement et Communication de l'Information, CNRS (UMR 5141) and T\'{e}l\'{e}com ParisTech, 46 rue Barrault F-75634 Paris Cedex 13, France\\
\and
Laboratoire de Physique Subatomique et de Cosmologie, Universit\'{e} Joseph Fourier Grenoble I, CNRS/IN2P3, Institut National Polytechnique de Grenoble, 53 rue des Martyrs, 38026 Grenoble cedex, France\\
\and
Laboratoire de Physique Th\'{e}orique, Universit\'{e} Paris-Sud 11 \& CNRS, B\^{a}timent 210, 91405 Orsay, France\\
\and
Lawrence Berkeley National Laboratory, Berkeley, California, U.S.A.\\
\and
Max-Planck-Institut f\"{u}r Astrophysik, Karl-Schwarzschild-Str. 1, 85741 Garching, Germany\\
\and
McGill Physics, Ernest Rutherford Physics Building, McGill University, 3600 rue University, Montr\'{e}al, QC, H3A 2T8, Canada\\
\and
MilliLab, VTT Technical Research Centre of Finland, Tietotie 3, Espoo, Finland\\
\and
National University of Ireland, Department of Experimental Physics, Maynooth, Co. Kildare, Ireland\\
\and
Niels Bohr Institute, Blegdamsvej 17, Copenhagen, Denmark\\
\and
Observational Cosmology, Mail Stop 367-17, California Institute of Technology, Pasadena, CA, 91125, U.S.A.\\
\and
Optical Science Laboratory, University College London, Gower Street, London, U.K.\\
\and
SB-ITP-LPPC, EPFL, CH-1015, Lausanne, Switzerland\\
\and
SISSA, Astrophysics Sector, via Bonomea 265, 34136, Trieste, Italy\\
\and
School of Physics and Astronomy, Cardiff University, Queens Buildings, The Parade, Cardiff, CF24 3AA, U.K.\\
\and
Space Research Institute (IKI), Russian Academy of Sciences, Profsoyuznaya Str, 84/32, Moscow, 117997, Russia\\
\and
Space Sciences Laboratory, University of California, Berkeley, California, U.S.A.\\
\and
Special Astrophysical Observatory, Russian Academy of Sciences, Nizhnij Arkhyz, Zelenchukskiy region, Karachai-Cherkessian Republic, 369167, Russia\\
\and
Stanford University, Dept of Physics, Varian Physics Bldg, 382 Via Pueblo Mall, Stanford, California, U.S.A.\\
\and
Sub-Department of Astrophysics, University of Oxford, Keble Road, Oxford OX1 3RH, U.K.\\
\and
Theory Division, PH-TH, CERN, CH-1211, Geneva 23, Switzerland\\
\and
UPMC Univ Paris 06, UMR7095, 98 bis Boulevard Arago, F-75014, Paris, France\\
\and
Universit\'{e} de Toulouse, UPS-OMP, IRAP, F-31028 Toulouse cedex 4, France\\
% ...
}
}

\date{Received June 17, 2013; accepted XX, 2013}

\abstract{We present new measurements of cosmic infrared background (CIB)
anisotropies using \Planck. Combining HFI data with {\it IRAS}, the angular
auto- and
cross-frequency power spectrum is measured from 143 to 3000\,\GHz, and the
auto-bispectrum from 217 to 545\,\GHz. The total areas used to compute the CIB
power spectrum and bispectrum are about 2240 and $4400\,{\rm deg}^2$,
respectively.
After careful removal of the contaminants (cosmic microwave background
anisotropies, Galactic dust and Sunyaev-Zeldovich emission), and
a complete study of systematics, the CIB power spectrum is measured with
unprecedented signal to noise ratio from angular multipoles $\ell
\sim150$ to $2500$.  The bispectrum due to the clustering of dusty,
star-forming galaxies is measured from $\ell \sim130$ to $1100$, with a total
signal to noise ratio of around 6, 19, and 29 at 217, 353, and 545\,\GHz,
respectively.  Two approaches are developed for modelling CIB power
spectrum anisotropies. The first approach takes advantage of the unique
measurements by \Planck\ at large angular scales, and models only
the linear part of the power spectrum, with a mean bias of dark matter halos
hosting dusty galaxies at a given redshift weighted by their contribution to
the emissivities. The second approach is based on a model that associates
star-forming galaxies with dark matter halos and their subhalos, using a
parametrized relation between the dust-processed infrared luminosity and
(sub-)halo mass. The two approaches simultaneously fit all auto- and
cross- power spectra very well.  We find that
the star formation history is well constrained up to redshifts around $2$,
and agrees with recent estimates of the obscured star-formation
density using {\it Spitzer\/} and {\it Herschel}.  However, at higher
redshift, the accuracy of the
star formation history measurement is strongly degraded by the uncertainty
in the spectral energy distribution of CIB galaxies.  We also find that
the mean halo mass which is most efficient at hosting star formation is
log$(M_{\rm eff}/\msun) = 12.6$ and that
CIB galaxies have warmer temperatures as redshift increases.
The CIB bispectrum is steeper than that expected from the power spectrum,
although well fitted by a power law; this gives some information about the
contribution of massive halos to the CIB bispectrum. Finally, we show that
the same halo occupation distribution can fit all power spectra
simultaneously. The precise measurements enabled by \Planck\ pose new
challenges for the modelling of CIB
anisotropies, indicating the power of using
CIB anisotropies to understand the process of galaxy formation. }
 

\authorrunning{Planck Collaboration}
\titlerunning{CIB anisotropies with \Planck}
 
 \maketitle

\section {Introduction}
%==================
This paper, one of a set associated with the 2013 release of data from the
\Planck\footnote{\Planck\ (\url{http://www.esa.int/Planck}) is a project of
the European Space Agency (ESA) with instruments provided by two scientific
consortia funded by ESA member states (in particular the lead countries
France and Italy), with contributions from NASA (USA) and telescope reflectors
provided by a collaboration between ESA and a scientific
consortium led and funded by Denmark.} mission \citep{planck2013-p01},
describes new measurements of the cosmic infrared background (CIB) anisotropy
power spectrum and bispectrum, and their use in constraining the cosmic
evolution of the star formation density and the luminous-dark matter bias.

The relic emission from galaxies formed throughout cosmic history appears as
a diffuse, cosmological background. The CIB is the far-infrared part of this
emission and it contains about half of its total energy \citep{dole2006}. 
Produced by the stellar-heated dust within galaxies, the CIB carries a wealth
of information about the process of star formation. Because dusty, star-forming
galaxies at high redshift are extremely difficult to detect individually
\citep[e.g.,][]{blain1998,lagache2003,dole2004,fernandez-conde2008,nguyen2010},
the CIB represents an exceptional tool for studying these objects and for
tracing their overall distribution \citep{knox2001}. The anisotropies
detected in this background light trace the large-scale distribution of
star-forming galaxies and, to some extent, the underlying distribution of the
dark matter halos in which galaxies reside. The CIB is thus a direct probe of
the interplay between baryons and dark matter throughout cosmic time.

The CIB has a redshift depth which complements current optical 
or near infrared measurements. This characteristic can be used to 
explore the early build-up and evolution of galaxies, one of the biggest frontiers in cosmology. 
Indeed the hope is to be able to use CIB anisotropies to improve our understanding of early gas
accretion and star formation, and to assess the impact of galaxies on
reionization. As dusty star-forming galaxies start to be found up to very high redshift (e.g., $z=6.34$, \citealt{riechers2013}), this objective may be reachable, even if quantifying the z$\lesssim 5-6$ contribution to CIB anisotropy measurements to isolate the high-redshift part will be very challenging.
As a start, CIB anisotropies can be used to measure the cosmic evolution of the star formation rate
density (SFRD) up to z$\simeq$6. Quantifying the SFRD
at high redshift ($z>2.5$) is a challenging endeavour.  Currently, most of
the measurements rely on the UV light emerging from the high-redshift galaxies
themselves \citep[e.g.,][]{bouwens2009,cucciati2012}. To estimate their
contribution to the total SFRD one needs to apply the proper conversion
between the observed UV 
rest-frame luminosity and the ongoing SFR. This conversion factor 
depends on the physical properties of the stellar population (initial mass
function, metallicities and ages) and on the amount of dust extinction, and
is thus rather uncertain. Despite the significant amount of effort aimed at
better understanding the UV-continuum slope distribution at high redshift,
this remains one of the main limitation to SFRD measurements. The uncertainty
on this conversion sometimes leads to significant revision of the SFRD
\citep[e.g.,][]{behroozi2012, bouwens2012, castellano2012}.
Remarkably, the estimates are now {\it routinely} made up to $z\sim8$
\citep[e.g.,][]{oesch2012a}, and have even been pushed up to $z\sim10$
\citep{oesch2012b}. One of the key questions, is how to quantify the
contribution of the dusty, star-forming galaxies to the SFRD at high redshift.
Since it is mostly impossible to account for this contribution on the basis
of optical/near-IR surveys, the best approach is to use the dusty galaxy
luminosity function measurements. However, such measurements at high redshift
are challenging with the current data, and this is where the CIB anisotropies,
with their unmatched redshift depth, come into play. The SFRD from dusty,
star-forming galaxies can be determined from their mean emissivity per
comoving unit volume, as derived from CIB anisotropy modelling. 

The way galaxies populate dark matter halos is another ingredient that enters
into the CIB anisotropy modelling. In particular, the galaxy {\it bias} ---
the relationship between the spatial distribution of galaxies and the
underlying dark matter density field --- is a result of the varied physics of
galaxy formation which can cause the spatial distribution of visible baryons
to differ from that of dark matter.
If galaxy formation is mainly determined by local physical processes (such as
hydrodynamics), the galaxy bias is then approximately constant on large scales
\citep{coles1993}, and the galaxy density fluctuations are thus
proportional to those of the dark matter. The proportionality
coefficient here is usually called as the linear bias factor, $b$. Its dependence on the luminosity, morphology, mass, and redshift of galaxies provides important clues to how
galaxies are formed. 
However, the linear biasing parameter is at best a crude
approximation, since the true bias is likely to be nontrivial, i.e.,
non-linear and scale dependent, especially at high redshift. At high redshift,
the biasing becomes more pronounced, as predicted by theory
\citep[e.g.,][]{kaiser1986, mo1996, wechsler1998}, and confirmed by the strong
clustering of dusty star-forming galaxies
\citep[e.g.,][]{steidel1998, blain2004, cooray2010}. CIB anisotropies can be
used to constrain the biasing scheme for dusty star-forming galaxies,
which is crucial for understanding the process and history of galaxy formation.

However, measuring the CIB anisotropies is not easy. First, the
instrument systematics, pipeline transfer function and beams have to be very
well understood and measured. One can take advantage of recent experiments such
as {\it Hershel\/} and \Planck, for which diffuse emission is measured
with better accuracy than their {\it IRAS\/} and {\it Spitzer\/}
predecessors.  Second, extracting the CIB
requires a very accurate component separation. Galactic dust, CMB anisotropies,
emission from galaxy clusters through
the thermal Sunyaev Zeldovich (tSZ) effect, and point sources all have a part
to play. In clean regions of the sky, Galactic dust dominates for multipoles
$\lesssim200$. This has a steep power spectrum (with a slope of about $-2.8$)
and exhibits spatial temperature variations, and thus spectral energy
distribution (SED) spatial
variations. Distinguishing Galactic from extragalactic dust is very difficult,
as their SEDs are quite similar, and both their spatial and spectral variations
do not exhibit any particular features. Currently, the best approach is to rely on a
Galactic template; taking another frequency is not recommended as CIB anisotropies also
contribute. Taking a gas tracer as a spatial template is the best one can do,
even if it has the drawback of not tracing the dust
in all interstellar medium phases. The CMB is very problematic at low
frequency, since its power spectrum
is about 5000 and 500 times higher at $\ell=100$ than the CIB at 143 and
217\GHz, respectively. For \Planck\, at 217\GHz\, the CMB dominates the CIB
for all $\ell$; at 353\GHz\, it dominates for $\ell<1000$; and at
545\GHz\, its power is 25 times lower at $\ell=100$ than the CIB. Any CMB
template (taken from low-frequency data or from complex component separation
algorithms) will be contaminated by residual foreground emission that will
have to be corrected for. At \Planck\ frequencies $\nu > 200$\,\GHz\, the tSZ effect can be safely ignored, being close to zero at
217\,\GHz\ and 100 times below the CIB power spectrum at 353\,\GHz. However
tSZ contamination can come from the use of a CMB template that contains residual
tSZ power. At 100\,\GHz, based on the tSZ power spectrum measured in
\cite{planck2013-p05b} and the CIB model developed in \cite{planck2011-6.6},
we estimate the tSZ power spectrum to be 10 times higher than the CIB.
The correction of
any tSZ contamination will be made difficult by the intrinsic correlation
between the CIB and tSZ signals \citep{addison2012, reichardt2012}.
Finally, as bright
point sources will put extra power at all scales in the power spectrum, point
sources need to be carefully masked up to a well-controlled flux density
level. This step is complicated by the extragalactic source confusion that
limits the depth of source detection for current far-infrared and
submillimetre space missions.

The pioneering studies in CIB anisotropy measurement with {\it Herschel}-SPIRE
\citep{amblard2011} and \Planck-HFI \citep{planck2011-6.6} are now extended in
\cite{viero2012} for the former, and in this paper for the latter.
The new \Planck\ measurements benefit from larger areas, lower instrument
systematics, and better component separation. They are not limited to
auto-power spectra but also include frequency cross-spectra, from 143 to 857\,\GHz\
(and 3000\GHz\ with {\it IRAS\/}), and extend to bispectra at 217, 353 and
857\,\GHz. These more accurate measurements pose new challenges for the
modelling of CIB anisotropies. 

Our paper is organized as follows.  We present in Sect.~\ref{data} the data we
are using and the field selection. Section~\ref{se:comp_sep} is dedicated to
the removal of the background CMB and foreground Galactic dust. We detail in
Sect.~\ref{cross_correl} how we estimate the power spectrum and bispectrum of
the residual maps, and their bias and errors. In the same section, results on CIB power spectra and bispectra are presented. In Sect.~\ref{CIB_Mod}, 
we describe our modelling and show the constraints
obtained on the SFRD, and the clustering of high-redshift, dusty galaxies.  
In Sect.~\ref{se:discuss}, we discuss the 143\,\GHz\ anisotropies, the frequency
decoherence, the comparison of our measurements with previous determinations,
the SFRD constraints, and the CIB non-Gaussianity.
We conclude in Sect.~\ref{se:cl}. The appendices give some details about the
\hi\ data used to remove Galactic dust (Appendix~\ref{hi_description}),
the CIB anisotropy modelling (Appendices~\ref{app_jnu} and
\ref{sect:effbias}), and also present the power spectra and bispectra tables
(Appendix~\ref{se:app_tables}).

Throughout the paper, we adopt the standard $\Lambda$CDM cosmological model
as our fiducial background cosmology, with parameter values derived from the
best-fit model of the CMB power spectrum measured by \Planck\
\citep{planck2013-p11}: 
$\{\Omega_{\rm m},\Omega_{\Lambda},\Omega_{\rm b} h^2,\sigma_8,h,n_{\rm s}\}=
\{0.3175, 0.6825,  0.022068, 0.8344, 0.6711, 0.9624\}$.
We also adopt a Salpeter initial mass function (IMF).

%===============
\section{Data sets and fields \label{data}}
%===============

\subsection{\Planck\ HFI data \label{planck_data} }
%---------------------------------
We used \Planck\ channel maps from the six HFI frequencies: 100, 143, 217, 353,
545, and 857\,GHz. These are $N_{\rm side}=2048$
\healpix\footnote{\url{http://healpix.sf.net}} maps \citep{gorski2005},
corresponding to a pixel size of 1.72\arcmin. We made use of the first public
release of HFI data that corresponds to temperature observations for
the nominal \Planck\ mission.  The characteristics of the maps and how they
were created are described in detail in the two HFI data processing and
calibration papers \citep{planck2013-p03, planck2013-p03b}).  At 857, 545,
and 353\GHz, we use the zodiacal light subtracted maps
\citep{planck2013-pip88}.  Some relevant numbers for the CIB analysis are
given in Table~\ref{tab_conversion}.

Maps are given in units either of ${\rm MJy}\,{\rm sr}^{-1}$ (with the
photometric convention $\nu I_{\nu}$=constant\footnote{The convention
$\nu I_{\nu}$=constant means that the
${\rm MJy}\,{\rm sr}^{-1}$ are given for a source with a spectral energy
distribution $I_{\nu}\propto \nu^{-1}$.  For a source with a different
spectral energy distribution a colour correction has to be applied
\citep[see][]{planck2013-p03d}.})
or K$_{\rm CMB}$, the conversion between the two can be exactly
computed knowing the bandpass filters. The mean coefficients used to convert
frequency maps in K$_{\rm CMB}$ units to ${\rm MJy}\,{\rm sr}^{-1}$
are computed using
noise-weighted band-average spectral transmissions. The mean conversion factors
are given in Table~\ref{tab_conversion}.  The map-making routines do not
average individual detector maps, but instead combine individual detector
data, weighted by the noise estimate, to produce single-frequency channel
maps. As portions of the sky are integrated for different times by different
detectors, the relative contribution of a given detector to a channel-average
map varies for different map pixels. The effects of this change on the
channel-average transmission spectra is very small, being of the order of
0.05\,\% for the nominal survey coverage
\citep{planck2013-p03d, planck2013-p28}.

\begin{table*}[!tbh]
\begingroup
\newdimen\tblskip \tblskip=5pt
\caption{Conversion factors, absolute calibration and inter-frequency relative
calibration errors, beam FWHM and point source flux cuts, for HFI and IRIS.
Here ``2013'' means the first public release of \Planck\ maps.}
\label{tab_conversion}
\nointerlineskip
\vskip -3mm
\footnotesize
\setbox\tablebox=\vbox{
 \newdimen\digitwidth
 \setbox0=\hbox{\rm 0}
  \digitwidth=\wd0
  \catcode`*=\active
  \def*{\kern\digitwidth}
  \newdimen\signwidth
  \setbox0=\hbox{+}
  \signwidth=\wd0
  \catcode`!=\active
  \def!{\kern\signwidth}
\halign{\tabskip=0pt\hfil#\hfil\tabskip=1.0em&
  \hfil#\hfil\tabskip=1.0em&
  \hfil#\hfil\tabskip=1.0em&
  \hfil#\hfil\tabskip=1.0em&
  \hfil#\hfil\tabskip=1.0em&
  \hfil#\hfil\tabskip=1.0em&
  \hfil#\hfil\tabskip=0pt\cr
\noalign{\doubleline}
\noalign{\vskip -2pt}
%XXX
Band& Map version& Conversion factors &
 Abs. cal. error& Rel. cal. error& FWHM& Flux cut\cr
[GHz] & & {K$_{\rm CMB} / {\rm MJy}\,{\rm sr}^{-1}$}[$\nu I_{\nu}={\rm const.}$]& [\%]&  [\%]&  [arcmin]& [mJy]\cr
\noalign{\vskip 3pt\hrule\vskip 3pt}
*100& 2013&  $244.07\pm0.22$&       *0.54& 0.3& 9.66& $400\pm50$\cr
*143& 2013&  $371.666\pm0.056$&     *0.54& 0.3& 7.27& $350\pm50$\cr
*217& 2013&  $483.4835\pm0.0074$&   *0.54& 0.3& 5.01& $225\pm50$\cr
*353& 2013&  $287.2262\pm0.0038$&   *1.24& 1.0& 4.86& $315\pm50$\cr
*545& 2013&  $*57.9766\pm0.0021$&   10.0*& 5.0& 4.84& $350\pm50$\cr
*857& 2013&  $**2.26907\pm0.00053$& 10.0*& 5.0& 4.63& $710\pm50$\cr
3000& IRIS& **$-$ & 13.5*& $-$ & 4.30& *1000\cr
\noalign{\vskip 3pt\hrule\vskip 3pt}}}

\endPlancktablewide
\endgroup
\end{table*}

As in \cite{planck2011-6.6}, the instrument noise power spectrum on the small
extragalactic fields (see Sect.~\ref{sect_field}) is estimated with the
jack-knife difference maps, which are built using the first and second halves
of each pointing period (a half-pointing period is of the order of 20 minutes).
The half-ring maps give an estimate of the
noise that is biased low (by a couple of percent) due to small correlations
induced by the way the timelines have been deglitched.
However, as discussed in \cite{planck2013-p03}, the bias is significant only
at very high multipoles. As we stop our CIB power spectra measurements at
$\ell=3000$, we can safely ignore this bias. For the larger GASS field
(see Table~\ref{tab:Fields}), we
directly compute the cross-spectra between the two half-maps to get rid of the
noise (assuming that the noise is uncorrelated between the two half-maps).

The effective beam window functions, $b_{\ell}$, are determined from planet
observations, folding in the \Planck\ scanning strategy, as described in
\cite{planck2013-p03c}. Because of the non-circular beam shape, the detector
combinations, and the \Planck\ scanning strategy,
the effective $b_{\ell}(\nu)$ of
each channel map applicable to the smallest patches considered here varies
across the sky. These variations are, however, less than 1\% at $\ell=2000$ and
average out when considering many different patches, or larger sky areas.
We will
therefore ignore them and consider a single $b_{\ell}(\nu)$ for each frequency
channel.  Because of their experimental determination, the $b_{\ell}(\nu)$ are
still affected by systematic uncertainties, which can be represented by a small
set of orthogonal eigen-modes and whose relative standard deviation can reach
0.5\% at $\ell=2000$ for $\nu=857\,$GHz.
This uncertainty is accounted for in the error budget.

\subsection{IRIS data}
%---------------------------------
Our analysis uses far-infrared data at 3000\GHz\ (100$\,\mu$m) from
{\it IRAS\/} (IRIS, \citep{mamd2005}).
During its 10-month operation period, {\it IRAS\/} made two surveys of
98\% of the sky and a third one of 75\% of
the sky. Each survey, called an HCON for Hours CONfirmation, consisted of
two coverages of the sky separated by up to 36 hours.
The first two HCONs (HCON-1 and HCON-2) were carried out concurrently, while
the third survey (HCON-3) began after the first two were completed. Due to
exhaustion of the liquid helium supply, the third HCON could not be completed.
We use the HCON difference maps to estimate the instrument noise power
spectrum (just as we use the half-pointing-period maps to estimate the
instrument noise power spectrum for HFI). As for HFI at high frequencies,
the IRIS 3000\GHz\ map is given in MJy\,sr${}^{-1}$ with the photometric
convention $\nu I_{\nu}$=constant.

The beam at  3000\GHz\ is not as well characterized as the HFI beams.
The IRIS effective FWHM is about 4.3\arcmin\ \citep{mamd2005}. We estimate
the FWHM uncertainty using different measurements of the point spread function
(PSF) coming from selected point sources used to study the PSF in the
``ISSA explanatory supplement,'' and the power spectrum analysis from
\cite{mamd2002}.  The dispersion between those estimates gives an uncertainty
of 0.5\arcmin\ on the FWHM.

\subsection{Extragalactic fields with high angular resolution \hi\ data
 \label{sect_field}}
%---------------------------------
Following the successful approach of \cite{planck2011-6.6}, we do not remove Galactic
dust by fitting for a power-law power spectrum at large angular scales, but
rather use an independent, external tracer of diffuse dust emission, the \hi\
gas. From 100$\,\mu$m to 1\,mm, at high Galactic latitude and outside molecular
clouds a tight correlation is observed between far-infrared emission from dust
and the 21-cm emission from gas\footnote{The Pearson correlation coefficient
is $>0.9$ \citep{lagache2000}.}  \citep[e.g.][]{boulanger1996, lagache1998,
planck2011-7.12}. \hi\ can thus be used as a tracer of cirrus emission in our fields,
and indeed it is the best tracer of diffuse interstellar dust emission. 

Although \Planck\ is an all-sky survey, we restricted our CIB anisotropy
measurements to a few fields at high Galactic latitude, where \hi\ data at an
angular resolution close to that of HFI are available. The 21-cm \hi\ spectra
used here were obtained with: (1) the Parkes 64-m telescope; (2) the Effelsberg
100-m radio Telescope; and (3) the 100-m Green Bank Telescope (GBT). Field
characteristics are given in Table~\ref{tab:Fields}. Further details on the
\hi\ data reduction and field selection are given in
Appendix~\ref{hi_description}. The \healpix\ HFI maps were reprojected onto
the small \hi\ GBT and EBHIS maps by binning the original \healpix\ data into
\hi\ map pixels (Sanson-Flamsteed, or ``SFL'' projection with pixel size of
3.5\arcmin\ for all
fields).  An average of slightly more than four \healpix\ pixels were
averaged for each small map pixel. For the GASS field, the HFI data were
convolved to the \hi\ angular resolution (16.2\arcmin), and then degraded to
$N_{\rm side}=512$.

We have 11 fields (one EBHIS, nine GBT and one GASS). The total area used to
compute the CIB power spectrum is about $2240\,{\rm deg}^2$, 16 times
larger than in \cite{planck2011-6.6}.

\begin{table*}[!tbh]
\begingroup
\newdimen\tblskip \tblskip=5pt
\caption{CIB field description: centre (in Galactic coordinates), size, mean
and dispersion of \hi\ column density. The given area is the size used for the
CIB power spectrum computation (i.e., removing the area lost by the masking).
The ``GASS Mask2'' is not used for CIB power spectrum analysis but for some
tests on the quality of our component separation, and for the measurement of
the bispectrum. Mask2 includes all of Mask1.}
\label{tab:Fields}
\nointerlineskip
\vskip -3mm
\footnotesize
\setbox\tablebox=\vbox{
 \newdimen\digitwidth
 \setbox0=\hbox{\rm 0}
  \digitwidth=\wd0
  \catcode`*=\active
  \def*{\kern\digitwidth}
  \newdimen\signwidth
  \setbox0=\hbox{+}
  \signwidth=\wd0
  \catcode`!=\active
  \def!{\kern\signwidth}
\halign{\tabskip=0pt#\hfil\tabskip=1.0em&
  #\hfil\tabskip=1.0em&
  \hfil#\hfil\tabskip=1.0em&
  \hfil#\hfil\tabskip=1.0em&
  \hfil#\hfil\tabskip=1.0em&
  \hfil#\hfil\tabskip=1.0em&
  \hfil#\hfil\tabskip=0pt\cr
\noalign{\doubleline}
\noalign{\vskip -2pt}
Radio Telescope& Field name& $l$& $b$& Area& Mean $N$(\hi)& $\sigma$ $N$(\hi)\cr
& & [deg]& [deg]& [deg$^2$]& [$10^{20}\,{\rm cm}^{-2}$]&
 [$10^{20}\,{\rm cm}^{-2}$]\cr
\noalign{\vskip 3pt\hrule\vskip 3pt}
Effelsberg& EBHIS& 225& !63& 91.6& 1.6& 0.3\cr
\noalign{\vskip 3pt\hrule\vskip 3pt}
       GBT&     N1& *85&   !44& 26.4& 1.2& 0.3\cr
          &     AG& 165&   !66& 26.4& 1.8& 0.6\cr
          &     SP& 132&   !48& 26.4& 1.2& 0.3\cr
          &    LH2& 152&   !53& 16.2& 0.7& 0.2\cr
          & Bootes& *58&   !69& 54.6& 1.1& 0.2\cr
          &   NEP4& *92&   !34& 15.7& 2.4& 0.4\cr
          &   SPC5& 132&   !31& 24.6& 2.3& 0.6\cr
          &   SPC4& 133&   !33& 15.7& 1.7& 0.3\cr
          &     MC& *57& $-$82& 31.2& 1.4& 0.2\cr
\noalign{\vskip 3pt\hrule\vskip 3pt}
Parkes& GASS Mask1& 225& $-$64& 1914& 1.4& 0.3\cr
      & GASS Mask2& 202& $-$59& 4397& 2.0& 0.8\cr
\noalign{\vskip 3pt\hrule\vskip 3pt}}}
\endPlancktablewide
\endgroup
\end{table*}

\subsection{Point sources: flux cut and masks \label{sec:flux_cut}}
%---------------------------------
We use the \Planck\ Catalogue of Compact Sources \citep[PCCS,][]{planck2013-p05}
to identify point sources with signal-to-noise ratio greater or equal to 5
in the maps, and we create point-source masks at each frequency.  We mask out a
circular area of 3$\,\sigma$ radius around each source (where $\sigma$=FWHM/2.35).
The point sources to be removed have flux densities above a chosen threshold. 
The threshold is determined using the number counts $dN/dS$ on the cleanest 30\%
of the sky and by measuring the flux density at which we observe a departure
from a Euclidean power law; this departure is a proxy for measuring the flux
density regime where the incompleteness starts to be measurable. In practice,
this departure often corresponds to about 80\% of completeness
\citep[e.g.,][]{planck2012-VII}.
The uncertainty on flux densities comes from the form of the $dN/dS$
slope. Flux density cuts and all-sky effective beam widths
are given in Table~\ref{tab_conversion}. For IRIS at 3000\GHz, we use the
{\it IRAS\/} faint source catalogue \citep{moshir92} and
we mask all sources with $S_{3000}>1\,$Jy.

%===============
\section{Extracting CIB from \textit{Planck}-HFI and IRIS maps
 \label{se:comp_sep}}
%===============

One of the most difficult steps in extracting the CIB is the removal of the
Galactic dust and the CMB.
CMB anisotropies contribute significantly to
the total HFI map variance in all channels at frequencies up to and including
353\,\GHz. Galactic dust contributes at all frequencies but is dominant at high
frequencies. 
 One approach is to keep all the components and
search for the best-fit model of the CIB
in a likelihood approach, accounting for CMB and dust (e.g., as a power
law).  This is the philosophy of the method developed for the
low-frequency data analysis to extract the cosmological parameters
\citep{planck2013-p08, planck2013-p11}. However, with such an approach, the
complexity of the likelihood and the number of parameters and their degeneracies
prevent the use of advanced models for the clustered CIB
(models beyond a simple power law).  
We thus decided to use another approach, based on template removal.
To remove the CMB in the fields retained for our analysis, we
used a simple subtraction technique (as in \citealt{planck2011-6.6}). This method
enables us to reliably evaluate CMB and foreground component residuals, noise contamination, and to
easily propagate errors (for example cross-calibration errors). It also
guarantees that high-frequency CIB anisotropy signals will not leak into lower
frequency, CMB-free maps. For the Galactic dust, the present work focuses on very clean regions of the sky, for
which Galactic foregrounds can be monitored using ancillary
\hi\ observations. 

We describe in this section the removal of the two components. Some corrections are made at the map level, others can only be done at the power spectrum level.

\subsection{CMB removal}
%---------------------------------
\label{se:cmb-removal}

\subsubsection{A low-frequency map as a CMB template \label{cmb}}
The extraction of CIB anisotropies at low frequency is strongly limited by our
ability to separate the CIB from the CMB. As a matter of fact, at multipole
$\ell \sim100$, the CIB anisotropy power spectrum represents about 0.2\%
and 0.04\% of the CMB power spectrum at 217 and 143\,\GHz, respectively. 
We decided in this paper to use the HFI lowest frequency channel (100\,\GHz) as a
CMB template. It has the advantage of being an ``internal'' template,
meaning its noise, data reduction processing, photometric calibration,
and beam are all well
known. It also has an angular resolution close to the higher HFI frequency
channels. Following \cite{planck2011-6.6}, we removed the CMB contamination in the maps at $\nu \le 353\,$GHz. At 545\,\GHz\, as the CMB power represents less than 5\% of the CIB power, we removed the CMB in harmonic space. 

Using a lowest possible frequency channel as a CMB template
ensures the lowest CIB
contamination, since the CIB SED is decreasing as $\nu^{3.4}$
\citep{gispert2000}. 
Contrary to what was done in \cite{planck2011-6.6}, where they
used the 143\,\GHz\ channel as a CMB template, we used the HFI 100\,\GHz\ map. This is a good compromise
between being a low frequency template, and having an angular resolution close
to the higher frequency HFI channels.
Note that using the 100\,\GHz\ channel was not feasible in \cite{planck2011-6.6} as the
data were too noisy, and the field area too small ($140$ versus
$2240\,{\rm deg}^2$).

As detailed in \cite{planck2011-6.6}, we applied a Wiener filter to the 100\,\GHz\ map,
designed to minimize the contamination of the CMB template by instrument noise.
Errors in relative photometric calibration (between channels) are accounted for
in the processing, as detailed in Sect.~\ref{cross_correl1} and
Sect.~\ref{cross_correl2}.

Following \cite{planck2011-6.6}, two corrections have to be applied to
the measured CIB power spectra when using such a CMB template.  First we 
need to remove the extra instrument noise that has been introduced by the CMB
removal. This is done through:
\begin{equation}
N^{\text{CMBres}}_{\ell}(\nu)= N_{\ell}(\nu_{100}) \times w_\ell^2 \times
 \left( \frac{b_{\ell}(\nu)}{b_\ell(\nu_{100})} \right )^2,
\label{eq:xtranoise}
\end{equation}
with $\nu$ equal to 143, 217 or 353\,GHz.  $b_{\ell}(\nu)$ is the beam window function,
$w_\ell$ is the Wiener filter, and $N_{\ell}(\nu_{100})$ is the
noise power spectrum of the 100\,\GHz\ map. It is computed in the same way as
in the other frequency channels, using the half-pointing period maps.
Second, owing to the lower angular resolution of the 100\,\GHz\ channel compared
to 143, and 217 and 353\,\GHz, we also have to remove the CMB
contribution that is left close to the angular resolution of the 143, 217
and 353\,GHz channels:
\begin{equation}
C^{\text{CMBres}}_{\ell}(\nu) = C^{\text{CMB}}_{\ell}(\nu) \times
 F_{\rm \ell}^2 \times b^2_\ell(\nu)\times \left( 1 - w_\ell \right)^2,
\label{eq:cmb_left}
\end{equation}
with $F_{\rm \ell}$ being the pixel and reprojection transfer function
(detailed in Sect.~\ref{cross_correl1}).

Note that Eqs.~\ref{eq:xtranoise} and \ref{eq:cmb_left} are the corrections
for the auto-power spectra. They are easily transposable to cross-power spectra.
 
\subsubsection{CMB template contamination to the CIB \label{cmb_res}}
The low-frequency channel CMB template has the disadvantage of being contaminated by the
CIB and the tSZ effect (the Galactic dust is removed using our dust model detailed in the following section,
and IR and radio point sources are masked). Indeed the cross-spectra between the estimated
CIB maps at $\nu$ and $\nu^\prime$ involve the $a_{\ell m}$ product:
\begin{eqnarray}
a_{\ell m}^\nu \times  a_{\ell m}^{\nu^\prime\ast} &=&
 \left\{ a^{{\rm CIB},\nu}_{\ell m} +
  a^{{\rm SZ},\nu}_{\ell m} - w_\nu
 \left( a^{{\rm CIB},100}_{\ell m} + a^{{\rm SZ},100}_{\ell m}
  \right) \right\} \nonumber\\
&\times &
\left\{ a^{{\rm CIB},\nu^\prime}_{\ell m} + a^{{\rm SZ},\nu^\prime}_{\ell m}
 - w_{\nu^\prime} \left( a^{{\rm CIB},100}_{\ell m}
  + a^{{\rm SZ},100}_{\ell m} \right) \right\}^\ast,
\label{eq:cross_alm}
\end{eqnarray}
where $w_\nu$ is either zero if $\nu \geq 545$, or the Wiener filter applied to
the 100\GHz\ map ($w_\ell$) if $\nu \leq 353$. 

Besides the signal $C_{\rm CIB}^{\nu \times \nu^\prime}$ that we want to measure,
Eq.~\ref{eq:cross_alm} involves
three additional contributions that we have to correct for: tSZ$\times$tSZ;
spurious CIB$\times$CIB; and CIB$\times$tSZ correlations. We discuss
each of them in this section.

\begin{table*}[!tbh]
\begingroup
\newdimen\tblskip \tblskip=5pt
\caption{SZ correction (Eq.~\ref{eq:SZcorr}) 
to be applied to the CIB measurements at 217 and
353\GHz\,, in ${\rm Jy}^2\,{\rm sr}^{-1} [\nu I_\nu={\rm constant}]$. 
The uncertainty on $C_{\rm SZcorr}$ is of order 10\%,
while the uncertainty on $C_{\rm CIB-SZcorr}$ is about a factor of two.
\label{tab:sz-corr}}
\nointerlineskip
\vskip -3mm
\footnotesize
\setbox\tablebox=\vbox{
 \newdimen\digitwidth
 \setbox0=\hbox{\rm 0}
  \digitwidth=\wd0
  \catcode`*=\active
  \def*{\kern\digitwidth}
  \newdimen\signwidth
  \setbox0=\hbox{+}
  \signwidth=\wd0
  \catcode`!=\active
  \def!{\kern\signwidth}
  \newdimen\pointwidth
  \setbox0=\hbox{\rm .}
  \pointwidth=\wd0
  \catcode`?=\active
  \def?{\kern\pointwidth}
\halign{\tabskip=0pt\hfil#\hfil\tabskip=1.0em&
  \hfil#\hfil\tabskip=1.0em&
  \hfil#\hfil\tabskip=1.0em&
  \hfil#\hfil\tabskip=1.0em&
  \hfil#\hfil\tabskip=1.0em&
  \hfil#\hfil\tabskip=1.0em&
  \hfil#\hfil\tabskip=0pt\cr
\noalign{\doubleline}
\noalign{\vskip -2pt}
$\ell$& $C_{\rm SZcorr}^{353\times353}$& $C_{\rm CIB-SZcorr}^{353\times353}$&
 $C_{\rm SZcorr}^{217\times353}$& $C_{\rm CIB-SZcorr}^{217\times353}$&
 $C_{\rm SZcorr}^{217\times217}$& $C_{\rm CIB-SZcorr}^{217\times217}$\cr
\noalign{\vskip 3pt\hrule\vskip 3pt}
**53& 643?*& $-$395?*& 446?*& $-$170?*& 309?*& $-$46?*\cr
*114& 301?*& $-$182?*& 209?*& *$-$78?*& 145?*& $-$21?*\cr
*187& 184?*& $-$110?*& 128?*& *$-$47?*& *89?*& $-$13?*\cr
*320& 107?*& *$-$64?*& *74?*& *$-$27?*& *51?*& *$-$7.4\cr
*501& *66?*& *$-$40?*& *46?*& *$-$17?*& *31?*& *$-$4.7\cr
*683& *46?*& *$-$29?*& *31?*& *$-$12?*& *21?*& *$-$3.5\cr
*890& *32?*& *$-$22?*& *21?*& **$-$9.2& *14?*& *$-$2.6\cr
1157& *18?*& *$-$15?*& **9.2& **$-$5.6& **4.7& *$-$1.7\cr
1504& **7.5& **$-$9.4& **1.3& **$-$2.2& **0.2& *$-$0.5\cr
1956& **4.4& **$-$6.7& **0.2& **$-$1.2& **0.0& *$-$0.1\cr
\noalign{\vskip 3pt\hrule\vskip 3pt}}}
\endPlancktablewide
\endgroup
\end{table*}

\paragraph{CIB$\times$CIB spurious correlations --}

From Eq.~\ref{eq:cross_alm}, the CIB$\times$CIB spurious contribution reads
\begin{equation}
C_{\rm CIBcorr}^{\nu\times\nu^\prime} = -w_\nu C_{\rm CIB}^{100\times\nu^\prime} -
w_{\nu^\prime}C_{\rm CIB}^{100\times\nu} + w_\nu w_{\nu^\prime}C_{\rm CIB}^{100\times100}.
\label{eq:corr_CMB}
\end{equation}

Using a model of CIB anisotropies, we can compute
$C_{\rm CIBcorr}^{\nu\times\nu^\prime}$. This correction includes both the shot
noise and
the clustered CIB anisotropies. The correction will be taken into account when
searching for the best-fit CIB model: for each realization of our CIB
anisotropy models (detailed in Sects.~\ref{mod_lin} and \ref{mod_hod}) we will
compute $C_{\rm CIBcorr}^{\nu \times \nu^\prime}$.

\begin{figure}
\hspace{-0.5cm}
\includegraphics[width=9.5cm]{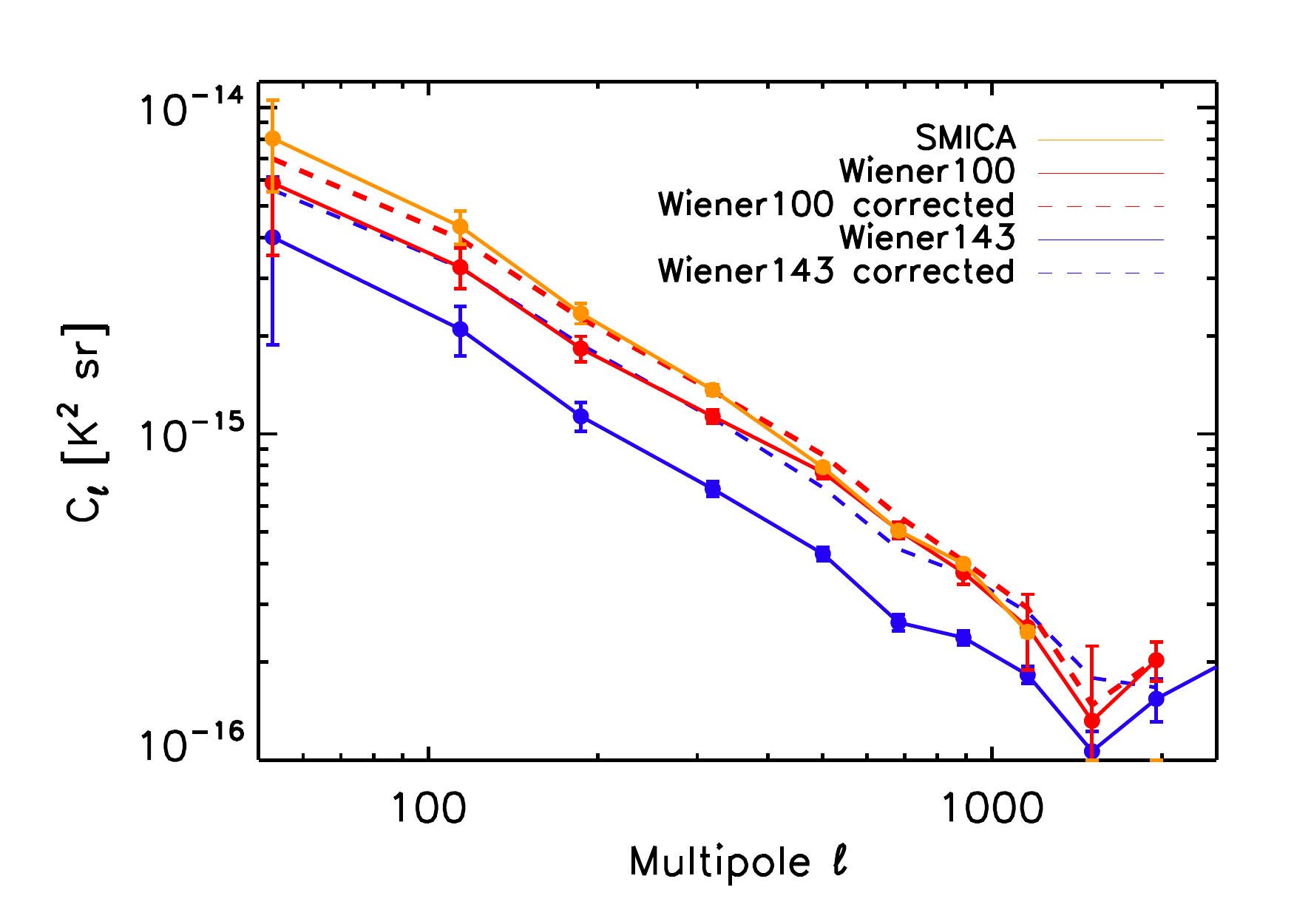}
\caption{Power spectrum of the residual map (map $-$ dust $-$ CMB) at
  217\GHz\ obtained in the GASS field with different CMB templates: the
  100\GHz\ Wiener-filtered map (red points); the 143\,\GHz\ Wiener-filtered
  map (blue points); and the SMICA map (orange points). These CMB maps are
  contaminated by both CIB anisotropies and shot noise. For the CIB measured
  using Wiener-filtered CMB maps, we can easily compute the correction to
  apply to recover the {\it true} CIB, using a model of the CIB anisotropies.
  Such corrected
  power spectra are shown with the dashed lines (using the CIB model from
  \citealt{planck2011-6.6}). They are only indicative, since the correction strongly
  depends on the CIB model.}
\label{fig:CIB_217_CMB}
\end{figure}

To illustrate the order of magnitude of the correction, we can use the model
constructed in \cite{planck2011-6.6}
to fit the early \Planck\ CIB measurements, and the shot-noise levels
given in
Sect.~\ref{sec:sn}, and compute $C_{\rm CIBcorr}^{\nu \times \nu^\prime}$
when subtracting
the 143\,GHz map instead of the 100\,GHz map as a CMB template. Note that this
comparison is only indicative, since this CIB model at 143 and 100\,\GHz\ is an
extrapolation of the $217\times217$ power spectrum using the clustering
parameters of the 217\,\GHz\ best-fit model, and the emissivities computed using
the same empirical model of galaxy evolution. We show this comparison for the
$217\times217$ power spectrum in Fig~\ref{fig:CIB_217_CMB}. We see from this
figure that the CIB obtained using the 143\,\GHz\ map as a CMB template is
about $1.6$ times lower than the CIB obtained using the 100\,\GHz\ map. Applying
the correction (Eq.~\ref{eq:corr_CMB}) largely decreases this
discrepancy (compare the two dashed lines\footnote{Note that the two curves have not exactly the same level as the corrections linked to the tSZ contributions are not the same when using the two CMB templates, and they are not applied for this plot.}). Note that
the correction is also non negligible when using the 100\,\GHz\ map: it is a
factor $1.15$ for $50<\ell<700$. This justifies applying the correction
systematically when fitting for our best CIB models.

We can extend the check on
the impact of the choice of CMB map using the cross-correlation between the
CIB and the distribution of dark matter, via the lensing effect on the CMB
\citep{planck2013-p13}. We specifically
compared the cross-correlation for the CIB at 217\,\GHz\ obtained using our two
Wiener-filtered 100 and 143\,\GHz\ maps.
We reached the same conclusion as before, and
retrieved the same underestimate of the CIB when using the Wiener-filtered
143\,\GHz\ rather than 100\,\GHz\ CMB map.

Finally, we note that an alternative method of removing the CMB contamination
was extensively tested; this was
based on an internal linear combination of frequency maps, combined (or not)
with a needlet analysis.  However, such CMB maps are often not
suited to our purposes because they are, among other problems, contaminated
by the CIB that has leaked from the high-frequency channels that are used
in the component
separation process.  We compare the CIB obtained using the 100\,\GHz\ and SMICA
CMB maps \citep{planck2013-p06} in Fig.~\ref{fig:CIB_217_CMB}.  We see that they
are compatible within $1\,\sigma$ (red-dashed line and orange points).
Again this comparison is only indicative,
since the SMICA CMB map main contaminants are SZ,
shot noise from point sources, and CIB\footnote{The contamination of the SMICA
CMB map by foregrounds has been computed by running SMICA on the FFP6
simulations.}; those contaminants have not been corrected for here, because the
correction would need fully realistic simulations.

\paragraph{tSZ$\times$tSZ correlation -- }
From Eq.~\ref{eq:cross_alm}, the tSZ$\times$tSZ spurious contribution
reads
\begin{equation}
C_{\rm SZcorr}^{\nu \times \nu^\prime} = C_{\rm SZ}^{\nu \times \nu^\prime}  + w_\nu
w_{\nu^\prime} C_{\rm SZ}^{100\times100} -  w_{\nu^\prime} C_{\rm SZ}^{\nu \times100}
 - w_{\nu} C_{\rm SZ}^{\nu^\prime \times100}.
\label{eq:SZ1}
\end{equation}

We can make explicit the frequency dependence of the tSZ and write
\begin{equation}
\Delta T_{\rm SZ}^\nu(\theta,\phi) = g_\nu \sum_{\ell m} a_{\ell m}^{\rm SZ},
\end{equation}
where $g_\nu$ is the conversion factor from the tSZ Compton parameter
$y$ to CMB temperature units.  Hence
\begin{equation}
C_{\rm SZ}^{\nu \times \nu}  = g_\nu^2 C_{\rm SZ}.
\label{eq:SZ_SED}
\end{equation}
Eq.~\ref{eq:SZ1} then becomes
\begin{equation}
C_{\rm SZcorr}^{\nu \times \nu^\prime} = C_{\rm SZ}
 \left\{ g_\nu g_{\nu^\prime} + w_\nu w_{\nu^\prime} g_{100}^2
  -g_{100}(w_{\nu^\prime}g_{\nu}+w_{\nu}g_{\nu^\prime})\right\}.
\label{eq:SZ2}
\end{equation}

We compute $C_{\rm SZcorr}$ using the tSZ power spectrum, $C_{\rm SZ}$, and
the conversion factors, $g_\nu$, given in \cite{planck2013-p05b}. The
uncertainty on $C_{\rm SZ}$ is about 10\%.

\paragraph{tSZ$\times$CIB correlation -- }
From Eq.~\ref{eq:cross_alm}, the tSZ$\times$CIB spurious contribution
reads
\begin{eqnarray}
C_{\rm CIB\times SZcorr}^{\nu \times \nu^\prime} &=& C_{\rm CIB \times SZ}^{\nu
  \times \nu^\prime } + C_{\rm CIB \times SZ}^{\nu^\prime \times \nu}
- w_\nu   C_{\rm CIB \times SZ}^{100\times\nu^\prime}
- w_{\nu^\prime}C_{\rm CIB \times SZ}^{100\times\nu}  \nonumber\\
&-& w_\nu   C_{\rm CIB \times SZ}^{\nu^\prime\times100}
 - w_{\nu^\prime}C_{\rm CIB \times SZ}^{\nu\times100}
 + 2w_\nu w_{\nu^\prime}C_{\rm CIB \times SZ}^{100 \times 100},
\label{eq:SZ3}
\end{eqnarray}
where $C_{\rm CIB \times SZ}^{\nu \times \nu^\prime}$ is the notation for the
cross-spectrum between ${\rm CIB}(\nu)$ and ${\rm SZ}(\nu^\prime)$.
This correction is highly dependent on the model used to compute the
cross-correlation between tSZ and CIB; we use the model from
\cite{addison2012}. We made the assumption that
\begin{equation}
C_{\rm CIB \times SZ}^{\nu \times \nu^\prime} =
 \langle a_{\ell m}^{\rm CIB}(\nu)
 a_{\ell m}^{\rm SZ\ast}(\nu^\prime) \rangle
 = g_{\nu^\prime} \phi(\nu) C_{\rm CIB \times SZ} \, ,
\end{equation}
where $\phi(\nu)$ is the amplitude of the power spectrum of the CIB correlated
with the tSZ, $C_{\rm CIB \times SZ}$, taken from \cite{addison2012}.
This paper also provides examples of cross-spectra, with a reference frequency
at 150\GHz. We use this reference frequency and power spectrum ratios to
compute $C_{\rm SZ \times CIB}^{\nu \times \nu^\prime}$ following
\begin{equation}
C_{\rm SZ \times CIB}^{\nu \times \nu^\prime} = \frac{\phi(\nu^\prime)}{\phi(150)}
 \frac{g{_{\nu}}}{g_{150}} C_{\rm SZ \times CIB}^{150 \times 150}.
\end{equation}

\begin{figure}
\hspace{-0.3cm}
\includegraphics[width=9.5cm]{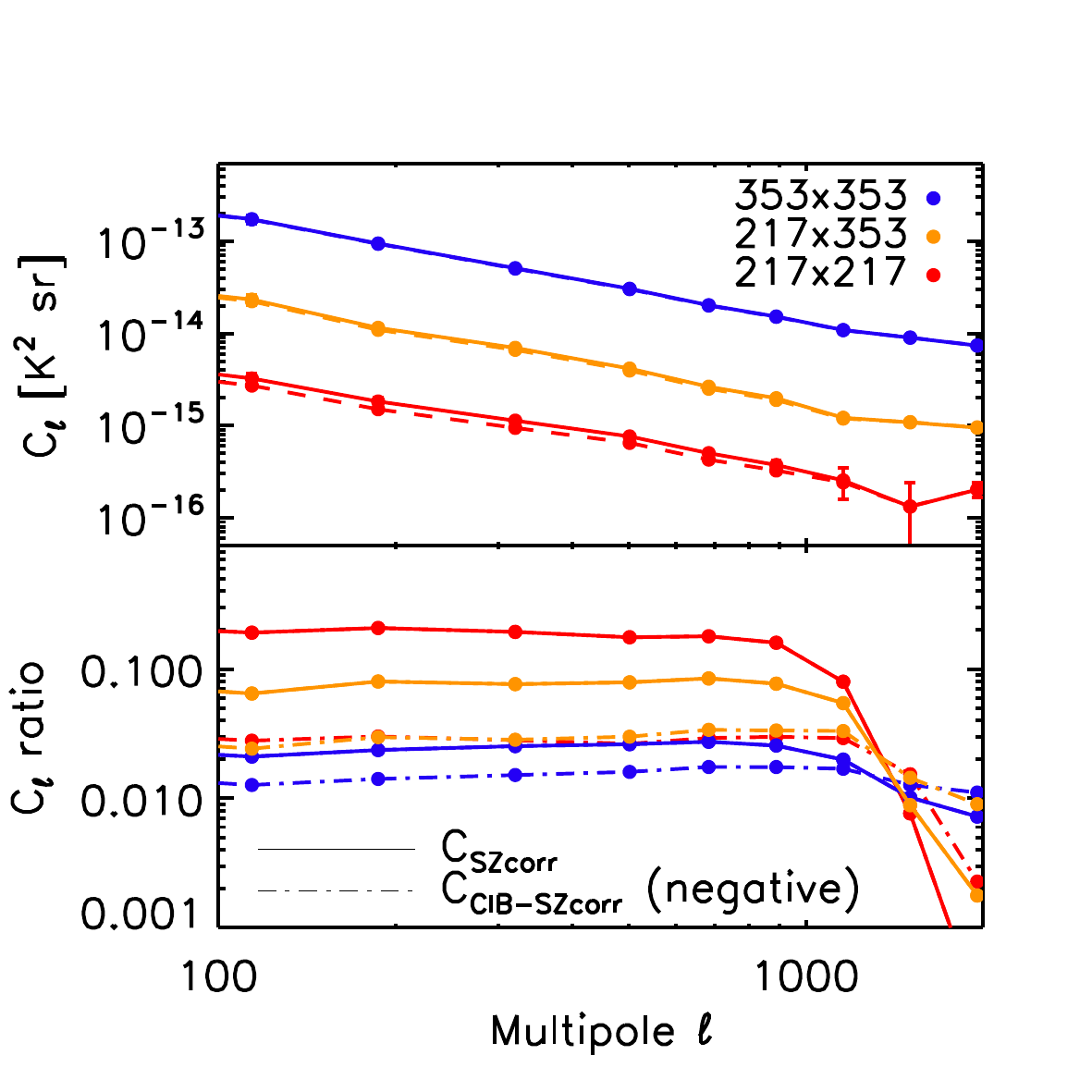}
\caption{{\it Top:} Residual map (a.k.a CIB) auto- and cross-spectra measured
at 217 and 353\,\GHz\ (circles). The dashed lines represent the measured power
spectra corrected for the tSZ contamination (both tSZ and tSZ$\times$CIB,
see Eq.~\ref{eq:SZ2} and \ref{eq:SZ3}, respectively). {\it Bottom:} Absolute
value of the ratio
${C_{\rm SZcorr}^{\nu \times \nu^\prime}}/{ C_{\rm CIB}^{\nu \times \nu^\prime}
[\rm measured]}$ (continuous line) and
${C_{\rm CIB-SZcorr}^{\nu \times \nu^\prime}}/{ C_{\rm CIB}^{\nu \times \nu^\prime}
[\rm measured]}$ (dot-dashed line, negative). The highest contamination is for
the $217\times217$ power spectrum, where the total correction represents about
15\% of the measured CIB power spectrum.
For $C_{\rm SZcorr}^{\nu \times \nu^\prime}$, the uncertainty is about 10\%.
On the contrary, $C_{\rm CIB-SZcorr}^{\nu \times \nu^\prime}$ is poorly constrained,
as the model used to compute it is uncertain by a factor of two.}
\label{fig:cib_sz}
\end{figure}

We show in Fig.~\ref{fig:cib_sz} the measured CIB,
$C_{\rm CIB}^{\nu \times \nu^\prime} (\rm measured)$,
together with the corrected one:
\begin{equation}
 C_{\rm CIB}^{\nu \times \nu^\prime}= C_{\rm CIB}^{\nu \times \nu^\prime}
 (\rm measured)
 - \left\{C_{\rm SZcorr}^{\nu \times \nu^\prime}
 + C_{\rm CIB\times SZcorr}^{\nu \times \nu^\prime} \right\}.
\label{eq:SZcorr}
\end{equation}
We also show the ratio of the corrections to the measured CIB, and give the
values of the corrections at 217 and 353\GHz\ in Table~\ref{tab:sz-corr}.
We see that tSZ contamination is the highest for the $217\times217$
combination, with a contamination of the
order of 15\%. It is less than 5\% and 1\% for $217\times353$ and
$353\times353$, respectively.  The tSZ power spectrum measured by
\cite{planck2013-p05b} is uncertain by 10\%, while
tSZ$\times$CIB is uncertain by a factor of two \citep{addison2012}.  Hopefully,
where the contamination is important (i.e., $217\times217$), the dominant
contribution to the CIB comes from $C_{\rm SZcorr}^{\nu \times \nu^\prime}$.
Hereafter, we therefore apply the correction coming from the tSZ
contamination to the measured CIB, and add the uncertainties of the correction
quadratically to the CIB uncertainties.

When cross-correlating maps at 353\,\GHz\ and above
545\,\GHz, the correction linked to the tSZ contamination is dominated by the
term $C_{\rm CIB\times SZcorr}^{\nu \times \nu^\prime}$, which is highly uncertain.
The correction is about 3\%, $<1$\%, 4\% and $<2$\% for the $217\times545$,
$353\times545$, $217\times857$ and $353\times857$ cross-power spectra,
respectively. Although small, for consistency with the case $\nu \leqslant
353$\GHz\ and $\nu^\prime \leqslant 353$\GHz, we also apply the tSZ-related
corrections to the measured CIB when $\nu \leqslant 353$\GHz\ and $\nu^\prime
\geqslant 545$\GHz.

\subsection{Dust model}
%---------------------------------
\label{sec:dust_model}
Many studies, using mostly {\it IRAS\/} and {\it COBE\/} data, have revealed
the strong
correlation between the far-infrared dust emission and 21-cm integrated emission
at high Galactic latitudes. In particular, \cite{boulanger1996} studied this
relation over the whole high Galactic latitude sky and reported a tight
dust-\hi\ correlation for $N_{\rm HI} < 4.6\times10^{20}\,{\rm cm}^{-2}$. For
higher column densities the dust emission systematically exceeds that expected
by extrapolating the correlation. Examining specific high Galactic latitude
regions, \cite{arendt1998}, \cite{reach1998}, \cite{lagache1998} and
\cite{lagache1999} found infrared excesses
with respect to $N_{\rm HI}$, with a threshold varying from 1.5 to
$5.0\times10^{20}\,{\rm cm}^{-2}$. \cite{planck2011-7.12} presented the results from the
comparison of \Planck\ dust maps with GBT \hi\ observations in 14 fields
covering more than $800\,{\rm deg}^2$. They showed that the brighter fields in
their sample, with an average \hi\ column density greater than
about $2.5\times10^{20}\,{\rm cm}^{-2}$, show significant excess dust emission
compared to the \hi\ column density. Regions of excess lie in organized
structures that suggest the presence of hydrogen in molecular form.  Because of
this, we restrict our CIB analysis to the cleanest part of the sky, with mean
$N_{\rm HI} <2.5\times10^{20}\,{\rm cm}^{-2}$.

\subsubsection{Constructing dust maps}
\label{sect:const_dust_maps}
As detailed in \cite{planck2011-7.12}, \cite{planck2011-6.6} and in Appendix~\ref{hi_description}, we
constructed integrated \hi\ emission maps of the different \hi\ velocity
components observed in each individual field: the local component, typical of
high-latitude \hi\ emission, intermediate-velocity clouds (IVCs), and
high-velocity clouds (HVCs), if present.
To remove the cirrus contamination from HFI maps, we determined the far-IR to
millimetre emission of the different \hi\ components. We assumed that the
HFI maps, $I_\nu(x,y)$, at frequency $\nu$ can be represented by the following
model
\begin{equation}
\label{eq_regress}
  I_\nu(x,y) = \sum_i \alpha^{i}_\nu N_{\rm HI}^{i}(x,y) + C_\nu(x,y),
\end{equation}
where $N_{\rm HI}^i(x,y)$ is the column density of the $i$th \hi\ component,
$\alpha^i_\nu$ is the far-IR to mm versus \hi\ correlation coefficient of
component $i$ at frequency $\nu$ and $C_\nu(x,y)$ is an offset. 
The correlation coefficients $\alpha_\nu^i$ (often called emissivities) were
estimated using $\chi^2$ minimization
given the \hi, HFI and IRIS data, as well as the model (Eq.~\ref{eq_regress}).
We removed from the HFI and IRIS maps the \hi\ velocity maps multiplied by the
correlation coefficients. For EBHIS and GBT fields, we considered only one correlation coefficient per field and per frequency.
The removal was done at the HFI and IRIS angular
resolutions, even though the \hi\ map is of lower resolution ($\sim$10\arcmin).
This is not a problem because cirrus, with a roughly $k^{-2.8}$ power-law power
spectrum \citep{mamd07}, has negligible power between the \hi\ (GBT and EBHIS) and HFI and IRIS angular resolutions, in comparison to the power in the CIB. The correlation of the dust emission 
with the different \hi\ velocity components and its variation from field to field is illustrated in Fig. 5 of \cite{planck2011-6.6}.
\\

For the GASS field, due to its large size the dust model  needs 
to take into account variations of the dust  emissivity across the field.
We make use of an analysis of the dust-to-gas correlation over
the southern Galactic cap ($b < -30^\circ$) \citep{planck2013-p31} using \hi\ data from the GASS
southern sky survey \citep{2010A&A...521A..17K}.  The \Planck-HFI maps are
linearly correlated with \hi\ column density over an area of $7500\,{\rm deg}^2$
covering all of the southern sky
($\delta < 0^\circ$) at $b < -30^\circ$ (17\% of the sky). 
We use HFI maps corrected for the mean value of the CIB \citep{planck2013-p03b}.
The \Planck\ maps and \hi\ emission at Galactic velocities are correlated over circular patches with $15^\circ $
diameters, centered on a \healpix\ grid with Nside=32. The linear regression is iterated to identify and mask 
sky pixels that depart from the  correlation.
At microwave frequencies the correlation coefficients ($\alpha_\nu$) and offsets
($C_\nu$) derived from this linear correlation analysis include a significant CMB
contribution that comes from
the chance correlation of the cosmic background with the \hi\ emission.
This contribution is estimated by assuming that the SED of
dust emission follows a modified blackbody spectrum for
$100 \le \nu \le 353\,$GHz.
The fit is performed on the differences
$\alpha_\nu -\alpha_{\rm 100GHz}$ that
are CMB-free for each sky area when expressed in units of K$_{\rm CMB}$.
This yields values of the correlations coefficients corrected for CMB, $\alpha_\nu^{\rm c}$. The
detailed procedure is described in Planck Collaboration et al (2013).
In this section we explain how the results of this
study are used to build a model of the dust contribution to the sky emission.

To make the dust model in the GASS field, we start by building a map of the
dust emission to \hi\ column density ratio, interpolating the values of
$\alpha_\nu^{\rm c}$, corrected for the CMB, using \healpix\ sky pixels with a
Gaussian kernel. The $1\sigma$ width of this convolution kernel is
equal to the pixel size $1.8^\circ$ of the \healpix\ grid for Nside=32. To reduce the data
noise at $\nu < 353\,$GHz we use the modified blackbody fits to
$\alpha_\nu-\alpha_{\rm 100GHz}$ and not the measured values of $\alpha_\nu$.  This yields a set
of six maps of the dust emission per unit \hi\ column density
for all HFI frequencies from 100 to 857\GHz.  We also build a set of Galactic
offset maps from the offsets $C_\nu$ of the \Planck-\hi\ correlation. These
offsets comprise contributions from Galactic dust and the CMB.
We subtract the CMB contribution assuming that the SED of the dust
contribution to $C_\nu$ is the same as that of $\alpha_\nu^{\rm c}$ for each
sky area. 
For each frequency, the dust model is the product of the dust emission per unit
\hi\ column density times the \hi\ map, plus the Galactic offset map.
The angular resolution of the model is that of the \hi\ map ($16.2\arcmin$). 

As for the smaller EBHIS and GBT fields, the model in the GASS field only
accounts for the emission of dust in \hi\ gas. Clouds with a significant
fraction of molecular gas produce localized regions with positive residual
emission. The histogram of residual emission at 857\GHz\ also shows a
non-Gaussian extension towards negative values (as observed in the EBHIS
field, see Appendix~\ref{EBHIS_HI}). In the maps these pixels correspond to
localized \hi\ clouds with no (or a
weak) counterpart in the \Planck\ map.  In the GASS survey, these clouds
are likely to be part of the Magellanic Stream, with radial velocities within
the range used to build the
Galactic~\hi\ map.  For the analysis of the CIB we mask pixels with positive
and negative residuals larger than $3\,\sigma$.  To be conservative this first
mask is slightly enlarged, and apodized. It covers about $4400\,{\rm deg}^2$
(Mask2 in Table~\ref{tab:Fields}). This mask still contains some regions
with $N_{\rm HI} \ge 2.5\times10^{20}\,{\rm cm}^{-2}$ and thus potential IR
emission from molecular gas clouds. To measure the CIB power spectrum, all
regions with $N_{\rm HI} \ge 2.1 \times10^{20}\,{\rm cm}^{-2}$ are further
masked (Mask1). The final area is about $1900\,{\rm deg}^2$.

\begin{figure}
\hspace{-0.45cm}
\includegraphics[width=9.5cm]{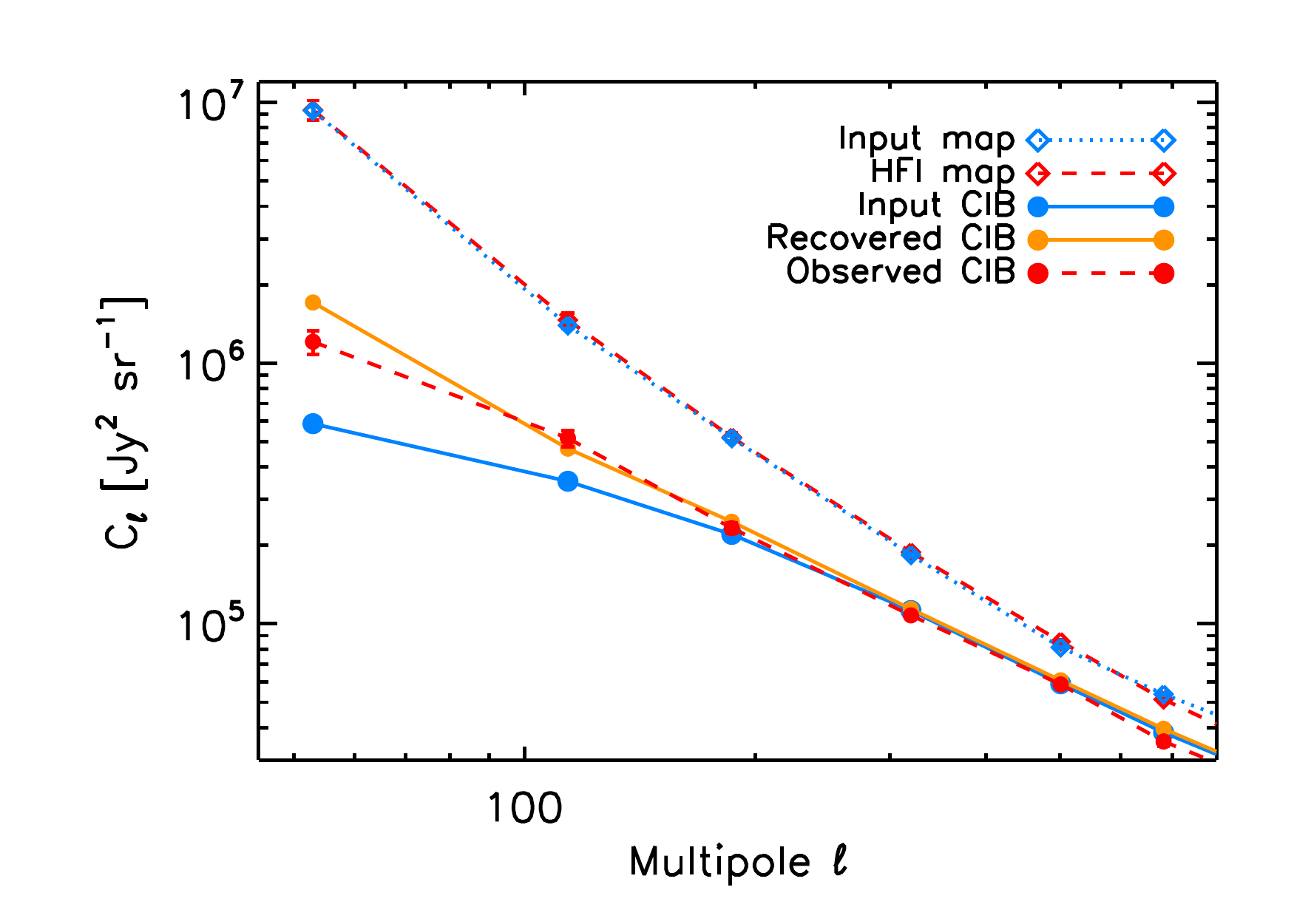}
\caption{\label{resid_dust} Power spectrum analysis of the simulations made at
857\,GHz in the GASS field to characterize cirrus residuals. The blue and red
diamonds compare the power spectrum of our simulated and HFI maps,
respectively, while the blue and orange dots are the simulated and recovered CIB,
respectively. The recovered CIB is biased by cirrus residuals at
low multipoles. The measured CIB (obtained on the GASS Mask1 field, displayed
with all but cirrus error bars) shows the same behaviour at low multipoles.
Thus the measurements in the first two $\ell$ bins have to be considered as
upper limits. As discussed in Sect.~\ref{dust_error}, the simulations are
used to compute the error bars linked to the cirrus removal.}
\end{figure}

\subsubsection{Dust map uncertainties \label{dust_error}}

The uncertainties estimated for the emissivities by the least-squares fit method are substantially underestimated, as they do not take into account systematic effects associated with the CIB and the CMB, nor the spatial variation of the emissivity 
inside a patch (or a field).  We use the GASS field to estimate the error we have on the dust model. For this field, we have the large-scale variations (at $\sim$15\deg) of the dust emissivities. We make the hypothesis that the measured variations extend to smaller angular scales, with a distribution on the sky given by a power spectrum with a slope $-2.8$, similar to that of the dust emission. Within this assumption, we simulate multiple maps of the dust emissivity that all match the dispersion of the dust emissivity  measured at 857\,\GHz. For each realization, we obtain a dust map at 857\,\GHz\ by multiplying the dust emissivity map by the GASS map of Galactic \hi.
Dust maps at other frequencies are obtained by scaling the 857\,\GHz\ flux using the mean dust SED given in \cite{planck2013-p31}.   
These simulations provide a good match to the Galactic residuals of the dust-\hi\ correlation characterized in \cite{planck2013-p31}.
We obtain simulated maps of the sky emission adding realizations of the HFI instrument noise, CIB and CMB to the dust maps.
We perform on the simulated sky maps the same correlation analysis with \hi\ as that done on the \Planck\ maps. 
For each simulation, we obtain values of the dust emissivity that we compare with the  input emissivity map averaged over each sky patch.
We find that there is no systematic difference between the values derived from the correlation analysis and the input values of the dust emissivities.  
The fractional error, i.e., the standard deviation of the difference between measured and the input values  divided by the  mean dust emissivity,  is 13\% of the mean dust emissivity  at 857\,\GHz\ \citep{planck2013-p31}.
This error increases  slightly towards lower frequencies up to 16 and 21\% for the 143 and 100\,\GHz\ channels.

We show in Fig.~\ref{resid_dust} the power spectrum of the recovered CIB, compared to the input CIB. We see that it is strongly biased by Galactic dust residuals in the first two multipole bins. These points have thus to be considered as upper limits. For the other points, we use the simulations to set the error bars linked to the Galactic dust removal. The observed linear correlation between the sigmas of dust residuals and $N_{\rm HI}$ is used to compute the power spectrum of Galactic dust residuals for each field, following
\begin{equation}
 C_{\ell}^{\rm Field} = C_{\ell}^{\rm GASS} \times \left( \frac{<N_{\rm HI}^{\rm Field}>}{<N_{\rm HI}^{\rm GASS}>} \right)^2 .
\end{equation}

\subsubsection{Galactic dust residuals \label{dust_res}}
One of the main issues when using \hi\ column density as a dust tracer is the
presence of ``dark gas'' \citep{planck2011-7.0,planck2011-7.12}, ionised gas \citep{lagache2000}, and emissivity variation at scales smaller that those probe by the correlation analysis.
The dust contribution of the dark gas becomes rapidly visible
(not only at high frequencies), when
${\rm N}_{\rm HI}\gtrsim2.5\times10^{20}\,{\rm cm}^{-2}$.
In addition to the simulations discussed in the previous section, we investigated the contribution of dust residuals by computing
the CIB power spectra on the GASS field, using Mask1 and Mask2. Mask1 is very
conservative and is our nominal mask for CIB analysis. Mask2 contains
higher-column density regions, so it is not suitable for CIB-only analysis,
but is useful, for example, when cross-correlating the CIB with other
large-scale structure tracers (when dust contamination is less of
a problem, e.g., \citealt{planck2013-p13}) and for the bispectrum measurement at low frequencies. As expected, the power spectra
of residual maps computed using Mask2 are in excess compared to the CIB.
Whatever the frequency, this excess represents
about 5-10\% of the cirrus power spectrum at low $\ell$. This leads to a CIB from
Mask2 that is overestimated by a factor of 1.5 at 857\,\GHz\ and 1.2 at 217\,\GHz,
for $\ell \le 200$, compared to the CIB from Mask1. 
Taking such a residual at the 5-10\% level of the dust
power spectrum (as observed between Mask1 and Mask2) would only strongly affect the
first two bins, at $\ell=53$ and $114$. This is consistent with the simulation analysis (see Fig.~\ref{resid_dust}). 
We will thus not use those bins when searching for the best CIB model, and
we consider those points as upper limits.

\begin{figure*}
\begin{center}
\includegraphics[width=17cm]{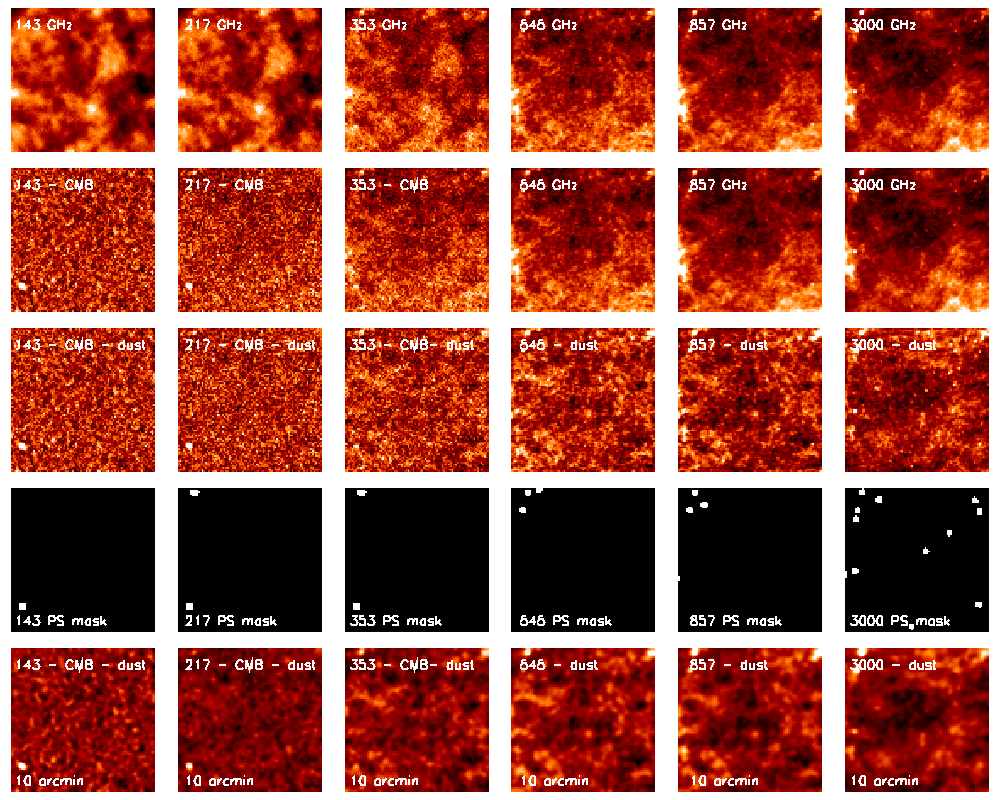}
\end{center}
\caption{\label{fig_CIB_flat_field} Maps of the roughly $25\,{\rm deg}^2$ of
the SPC5 field, from {\it left} to {\it right}: 143, 217, 353, 545, 857 and
3000\,\GHz. From {\it top} to {\it bottom}: raw HFI and IRIS maps; CMB-cleaned
maps; residual maps (CMB- and cirrus-cleaned); point source masks; and residual
maps smoothed at 10\arcm\ to highlight the CIB anisotropies. The joint
structures clearly visible (bottom row) correspond to the anisotropies of the
CIB.}
\end{figure*}

\begin{figure*}
\includegraphics[width=6cm]{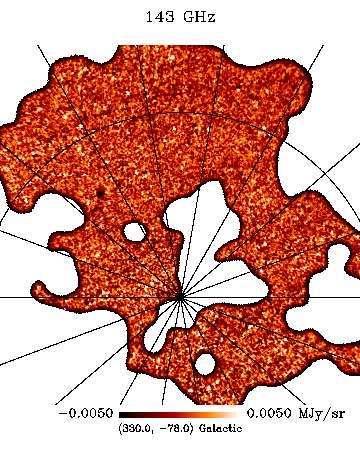}
\includegraphics[width=6cm]{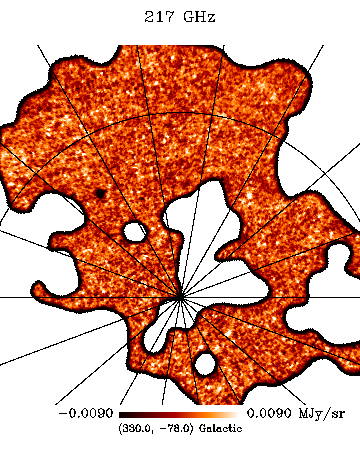}
\includegraphics[width=6cm]{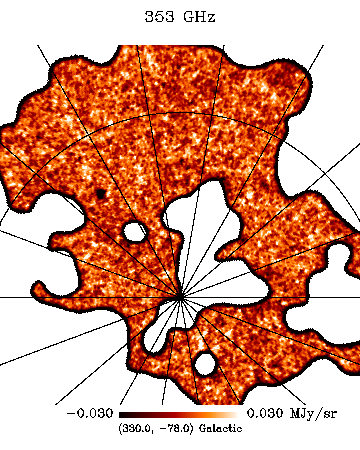}
\begin{center}
\includegraphics[width=6cm]{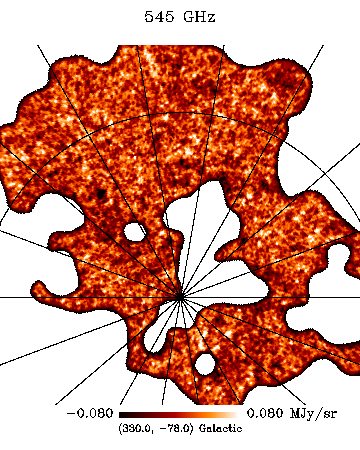}
\includegraphics[width=6cm]{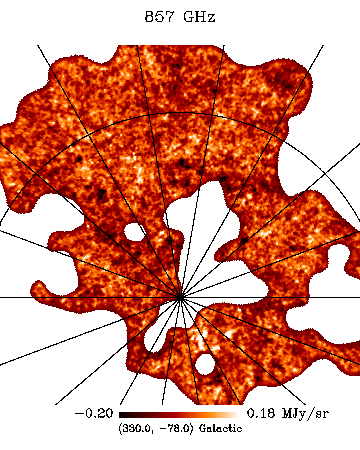}
\end{center}
\caption{\label{GASS_CIB_map} Residual maps (Galactic dust and CMB removed)
at 16.2\arcmin\ angular resolution, extracted from the area covered by the
GASS \hi\ data (on Mask1). The patch covers $60\deg \times60\deg$.}
\end{figure*}

\subsection{CMB and Galactic dust cleaned maps \label{sec:CIB_maps}}
%---------------------------------
We show in Fig.~\ref{fig_CIB_flat_field}, one example of the cleaning process,
from the frequency to the CIB maps, for
the SPC5 field. The bottom row shows the residual CIB maps, smoothed
to 10\arcmin. In
Fig.~\ref{GASS_CIB_map} are shown the CIB maps in a part of the GASS field, at
16.2\arcmin\ resolution. We see from both figures that common structures,
corresponding to CIB anisotropies, are clearly visible. As previously noticed \citep{planck2011-6.6},
the three intermediate frequencies (353, 545 and 857\,\GHz) show highly correlated
structures. On the contrary, the 3000\,\GHz\ data on one hand, and the 217\,\GHz\
on the other, reveal a decoherence, which can be attributed to the redshift
distribution of the CIB anisotropies. We will come back to this decoherence by
measuring the correlation coefficients in Sect.~\ref{sec:decohence}.

There are two frequencies where extracting the CIB from the frequency maps
is particularly challenging.
\begin{itemize}
\item
At 3000\,\GHz, the cirrus contamination is the highest and we observe more
spatial variations of the dust--\hi\ emissivity which are difficult to manage.
This is due
to the contamination at 3000\,\GHz\ data by a hotter dust component, which may be
linked to the so-called ``very small grains'' in some interstellar
filaments. Consequently, we have not tried to extract the CIB in the GASS field
at 3000\,\GHz, and the Bootes field has been discarded for the CIB analysis
due to interstellar dust residuals (due to one IVC component, which has both a
high emissivity and a hot spectrum; P. Martin, private communication). Moreover,
the EBHIS field is right in the middle of the missing {\it IRAS\/} observation.
It is thus also not used for the CIB analysis. In the end, the total area used
to compute the CIB power spectrum at 3000\,\GHz\ is $183\,{\rm deg}^2$.
\item
At 143\,\GHz, we expect the correlated CIB anisotropies in brightness to be
about 1--2\% of the CMB anisotropies for multipoles $\ell\,{\sim}\,100$--1000
(while it is about 3--15\% at 217\GHz). The removal of the CMB has thus to be
extremely accurate. Moreover, the expected CIB is lower than the instrument
noise. Constraints can only be obtained on the large GASS field that is
more immune to noise. We see from
Fig.~\ref{GASS_CIB_map} that the 143\GHz\ CIB map shows some structures that are
correlated with the higher CIB frequency maps.  We discuss in
Sect.~\ref{sec:143} our attempt to obtain some constraints at this frequency.
\end{itemize}

We used two different approaches to measure the CIB power spectra according to
the size of the fields: (i) for the EBHIS and GBT flat-sky fields, we used
an updated version of \poker~\citep{ponthieu2011}, and we computed the error
bars using Monte Carlo simulations (Sect.~\ref{cross_correl1});
(ii) for the GASS field, we used {\tt Xspect} \citep{tristram2005}, a method
that was first developed for the Archeops experiment to obtain estimates of
the angular power spectrum of the CMB temperature anisotropies, including
analytical error bars (Sect.~\ref{cross_correl2}). Having two completely
different pipelines and a many fields with various dust and CMB contaminations
(and noise contributions), is extremely valuable, as it allows us to test the
robustness of our approach.

%================================
\section {Angular power spectrum and bispectrum \label{cross_correl}}
%================================

\subsection {Cross-correlation pipeline for the EBHIS and GBT fields \label{cross_correl1}}
%---------------------------------
To determine the cross-correlation between the CIB observed at two frequencies,
$\nu_1$ and $\nu_2$, we used a modified version of
\poker~\citep{ponthieu2011}. \poker~is an algorithm that determines the power
spectrum of a map, corrects for mask aliasing and computes the covariance
between each power spectrum bin via a Monte Carlo approach.
It has been used to measure the
power spectrum of the CIB as observed by \Planck-HFI \citep{planck2011-6.6} and is
described in detail therein. In the following, we will call \emph{auto-power
spectrum} $C^{\nu_1}_\ell$ the usual angular power spectrum
and \emph{cross-power spectrum} $C^{\nu_1\nu_2}_\ell$ its generalization to two
different frequencies defined as
\begin{equation}
\frac{1}{2}\langle a^{\nu_1}_{\ell m} a^{\nu_2\ast}_{\ell^\prime m^\prime}
 + a^{\nu_1\ast}_{\ell m}a^{\nu_2}_{\ell^\prime m^\prime}\rangle = (2\pi)^2
C^{\nu_1\nu_2}_\ell \delta_{\ell \ell^\prime}\delta_{mm^\prime},
\end{equation}
where $a^\nu_{\ell m}$ are the Fourier coefficients of the CIB anisotropy at
the observation frequency $\nu$. We take a common mask for both frequencies and
it is straightforward to generalize the algebra presented in \cite{ponthieu2011}
and \cite{planck2011-6.6} to obtain an unbiased estimate of the cross-power spectrum:
\begin{equation}
C^{\nu_1\nu_2}_b \simeq \sum_{b'}M^{-1}_{bb'}\hat{P}^{\nu_1\nu_2}_{b'}
-N^{\nu_1\nu_2,\,{\rm instr}}_b - N^{\nu_1\nu_2,\,{\rm res}}_{b}.
\label{eq:pb_res}
\end{equation}
Here $\hat{P}^{\nu_1\nu_2}_b$ denotes the binned pseudo-cross-power spectrum of
the maps at frequencies $\nu_1$ and $\nu_2$, and $M_{bb'}$ is
the so-called mode-mixing matrix that is described in detail in Eq.~24
of \cite{planck2011-6.6} and its appendix. We recall that it includes the
mode-coupling effects induced by the mask, the smoothing by the instrumental
beam, and the map pixelization (see below). The two noise terms,
$N^{\nu_1\nu_2,\,{\rm instr}}_b$ and $N^{\nu_1\nu_2,\,{\rm res}}_{b}$,
refer, respectively, to
the instrument noise and to the contribution of what can be considered as
random components at the map level, such as the CMB residuals of our component
separation and the noise of the 100\GHz\ map that propagates to our maps (see
Sect.~\ref{se:cmb-removal}). Of course, Eq.~\ref{eq:pb_res} applied to
$\nu_1=\nu_2$ gives the auto-power spectrum of the anisotropy, and we computed
the two auto-spectra at $\nu_1$ and $\nu_2$ at the same time as the cross-power
spectrum.

We built very similar simulation and analysis pipelines to those
of \cite{planck2011-6.6} to obtain our measures and their associated error bars. We here
briefly recall the main steps of these pipelines and highlight the modifications
introduced by the generalization from auto-power spectra estimation to that of
cross-power spectra.

\paragraph{Analysis pipeline.}

\begin{enumerate}
\item In order to combine our measurements from different fields, we define a
  common multipole binning. Above $\ell \sim 200$, we choose a logarithmic binning
  $\Delta \ell/ \ell=0.3$, while below $\ell \sim 200$,
  we respect the generic prescription
  that bins of multipoles should be larger than twice the multipole corresponding to the largest angular scale contained in
  the field.
\item We define a common weight map $W^{\nu_1\nu_2}$ that masks out bright point
  sources found at both frequencies. Together with the mask, three different
  transfer functions must be accounted for in the computation of $M_{bb'}$
  \citep[Eq.~24 of][]{planck2011-6.6}:
  (i) the instrument beam transfer function that
  depends on the exact beam shape and the scan pattern on the observed
  field, although since the variation of beam transfer function is less than
  1\% between our fields, we take the same one for all of them
  (see Sect.~\ref{planck_data}); (ii) the \healpix~pixel window function;
  and (iii) the reprojection from \healpix~maps to the flat-sky maps used
  by \poker. The first transfer function comes from a dedicated
  analysis \citep{planck2013-p03c}. The second one is provided by the
  \healpix~library. The third one is computed via Monte Carlo simulations: we
  simulate full sky \healpix~maps of diffuse emission with a typical CIB power
  spectrum, reproject them on our observed patches and compute the angular power
  spectra; the ratio between the measured and the input power spectrum gives the
  transfer function. This ratio is the same for all fields (all in SFL
  projection with 3.5\arcmin\ pixels) within statistical error
  bars, so we compute the average and apply it to all fields.
\item An estimate of the noise auto-power spectrum at $\nu_1$ and $\nu_2$ is
  obtained from jack-knife maps. Indeed, at each frequency, two maps can
  be built using only the first (respectively, the second) half of each
  \Planck\ observation
  ring. The difference between these maps is dominated by instrumental
  noise. Applying \poker~to these difference maps gives an estimate of the
  instrument noise auto-power spectrum at each frequency
  $N^{\nu_i,\,{\rm instr}}_b$.  The noise contribution to the
  auto-power spectrum are negligible at high frequencies.
\item $N^{\nu_1\nu_2,\,{\rm res}}_{b}$ is the contribution of
  CMB residuals both from the component separation and from the propagation
  of noise in the 100\GHz\ CMB
  map. We have estimates of each component of this residual power spectrum in
  $\ell$ space on the sphere (see. Sect.~\ref{se:cmb-removal}) and combine
  them into our measurement bins.
\item We now apply Eq.~\ref{eq:pb_res} to our data and determine their auto-
  and cross-power spectra.
\end{enumerate}

\paragraph{Simulation pipeline.}

The simulation pipeline is essential to provide the error bars on our estimates
coming from the analysis pipeline. We created 100 simulations of our data maps
and computed their auto- and cross-power spectra. The dispersion of these
spectra gives the complete covariance matrices
of our binned auto- and cross-power spectra and their associated error bars.
For each realization we follow these steps.

\begin{enumerate}
\item The measured auto- and cross-power spectra are used as inputs to simulate
  CIB anisotropy maps at each frequency, with the appropriate correlated
  component. To
  do so, we generate random Gaussian amplitudes $x_{\vec{\ell}}$ and
  $y_{\vec{\ell}}$ in Fourier space, such that:
\begin{eqnarray}
a^{\nu_1}_{\vec{\ell}} &=& x_{\vec{\ell}}(C^{\nu_1}_\ell)^{1/2}; \nonumber \\
a^{\nu_2}_{\vec{\ell}} &=&
 x_{\vec{\ell}}C^{\nu_1\nu_2}_\ell/(C^{\nu_1}_\ell)^{1/2}
 + y_{\vec{\ell}} \left\{C^{\nu_2\nu_2}_\ell -
  \left(C^{\nu_1\nu_2}_\ell\right)^2/C^{\nu_1}_\ell\right\}^{1/2}. \nonumber
\end{eqnarray}
\item For each frequency $\nu_i$, we simulate a noise map with the appropriate
  power spectrum $\hat{N}^{\nu_i,\,{\rm instr}}_\ell$ and add it to the
  simulated signal map.
\item We add the simulations of CMB and 100\,\GHz\ noise residuals (Eqs.\ref{eq:xtranoise} and \ref{eq:cmb_left}), and of CMB residuals induced by relative calibration errors, so
  that we have a final pair of maps at $\nu_1$ and $\nu_2$ that are a faithful
  representation of our data.
\end{enumerate}

\paragraph{Error estimation.}
Statistical uncertainties due to instrument noise and component separation
residuals are derived using the simulation pipeline. Systematic uncertainties
include mask aliasing effects, imperfect subtraction of foreground templates,
and beam and projection transfer function errors. We studied some fields in common with
\cite{planck2011-6.6}, with very similar masks, and showed that the correction of mask
effects by \poker\ leads to no more than 2\% uncertainty on the result. 

The transfer function due to the projection of spherical \healpix\ maps to our
square patch is determined via Monte Carlo (see step 2 of the description of the
analysis pipeline). This process provides estimates $F_b$ of the transfer
function in our measurement bins.  However, we need an estimate of the transfer
function $F_\ell$ for all $\ell$ modes, to include it in the derivation of
$M_{bb'}$. We therefore construct a smooth interpolation of our measures.
The projection transfer function plays the same role as an extra instrumental
beam and enters the derivation of $M_{bb'}$. 
Roughly speaking, it damps the signal at high $\ell$ and correcting for
this effect corresponds to dividing the
measured power spectrum by the damping function. 
The statistical uncertainty on the determination of $F_\ell$ is
smaller than 1\%, and hence leads to the same uncertainty on the $C_\ell$.

Between the first analysis of CIB anisotropies with \Planck-HFI and this work,
much progress has been achieved on the beam measurements
\citep{planck2013-p03c}.  Based on this progress, we have better estimates of
the beam shapes
and better assessments of the uncertainties on these beam transfer
functions, both for auto-spectra and for cross-spectra between different
frequency bands. We used the eigenmodes to compute the beam transfer
function uncertainties, which are all smaller than 0.55\% in our angular range.
On the contrary, auto- and cross-spectra involving IRIS suffer from a
much larger beam uncertainty (0.5\arcmin\ for a FWHM of 4.3\arcmin).

The systematic uncertainties on the contribution of Galactic dust residuals
(see Sect.~\ref{dust_error}) are added linearly to the statistical error bars.
An example of the error budget is given in Table~\ref{tab:cross-spec}.

\begin{table*}[!tbh]
\begingroup
\newdimen\tblskip \tblskip=5pt
\caption{Example of a power spectrum averaged for the flat fields.
This illustrates the order of magnitude of the different errors.
The total error contains the cosmic variance and instrumental noise, and also
the cirrus residuals (CMB is negligible at this frequency). The contribution
of cirrus errors only is given in the sixth column for comparison. The
errors linked to the projection of the fields on the tangential plane and
those due to the beam, although systematic and not stochastic, are given
for convenience
in the same units as the total power $C_\ell$, in columns 7
and 8, respectively.}
\label{tab:cross-spec}
\nointerlineskip
\vskip -3mm
\footnotesize
\setbox\tablebox=\vbox{
 \newdimen\digitwidth
 \setbox0=\hbox{\rm 0}
  \digitwidth=\wd0
  \catcode`*=\active
  \def*{\kern\digitwidth}
  \newdimen\signwidth
  \setbox0=\hbox{+}
  \signwidth=\wd0
  \catcode`!=\active
  \def!{\kern\signwidth}
\halign{\tabskip=0pt\hfil#\hfil\tabskip=1.0em&
  \hfil#\hfil\tabskip=1.0em&
  \hfil#\hfil\tabskip=1.0em&
  \hfil#\hfil\tabskip=1.0em&
  \hfil#\hfil\tabskip=1.0em&
  \hfil#\hfil\tabskip=1.0em&
  \hfil#\hfil\tabskip=1.0em&
  \hfil#\hfil\tabskip=0pt\cr
\noalign{\doubleline}
\noalign{\vskip -2pt}
\multispan8\hfil{$857\GHz\, \times\, 857\GHz$}\hfil\cr
\noalign{\vskip 3pt\hrule\vskip 3pt}
$\ell$& $\ell_{\rm min}$& $\ell_{\rm max}$& $C_\ell$&
 Total error& Cirrus error&
 \multispan2\hfil{Systematic errors [${\rm Jy}^2\,{\rm sr}^{-1}$]}\hfil\cr
\noalign{\vskip -5pt}
& & & & & & \multispan2\hrulefill\cr
& & & [${\rm Jy}^2\,{\rm sr}^{-1}$]& [${\rm Jy}^2\,{\rm sr}^{-1}$]&
 [${\rm Jy}^2\,{\rm sr}^{-1}$]& **Projection& Beam\cr
\noalign{\vskip 3pt\hrule\vskip 3pt}
53 & 23 & 84 & $1.49\times10^6$ &$1.27\times10^6$ &$1.12\times10^6$ &$1.99\times10^3$ &$2.31\times10^1$ \cr
114 & 84 & 145 & $6.37\times10^5$ &$1.62\times10^5$ &$1.16\times10^5$ &$1.84\times10^3$ &$4.48\times10^1$ \cr
187 & 145 & 229 & $2.87\times10^5$ &$3.70\times10^4$ &$2.73\times10^4$ &$8.00\times10^2$ &$5.29\times10^1$ \cr
320 & 229 & 411 & $1.34\times10^5$ &$8.04\times10^3$ &$5.63\times10^3$ &$4.55\times10^2$ &$6.70\times10^1$ \cr
502 & 411 & 592 & $7.20\times10^4$ &$2.58\times10^3$ &$1.49\times10^3$ &$1.62\times10^2$ &$7.59\times10^1$ \cr
684 & 592 & 774 & $4.38\times10^4$ &$1.81\times10^3$ &$6.14\times10^2$ &$8.93\times10^1$ &$6.82\times10^1$ \cr
890 & 774 & 1006 & $3.23\times10^4$ &$9.48\times10^2$ &$2.85\times10^2$ &$9.41\times10^1$ &$6.07\times10^1$ \cr
1158 & 1006 & 1308 & $2.40\times10^4$ &$4.88\times10^2$ &$1.29\times10^2$ &$5.23\times10^1$ &$4.52\times10^1$ \cr
1505 & 1308 & 1701 & $1.83\times10^4$ &$2.58\times10^2$ &$5.97\times10^1$ &$2.49\times10^1$ &$3.41\times10^1$ \cr
1956 & 1701 & 2211 & $1.46\times10^4$ &$1.63\times10^2$ &$2.77\times10^1$ &$2.09\times10^1$ &$4.16\times10^1$ \cr
2649 & 2211 & 3085 & $1.16\times10^4$ &$1.38\times10^2$ &$1.12\times10^1$ &$7.89\times10^1$ &$6.52\times10^1$ \cr
\noalign{\vskip 3pt\hrule\vskip 3pt}}}
\endPlancktablewide
\endgroup
\end{table*}

\subsection {Cross-correlation pipeline for the GASS field \label{cross_correl2}}
%---------------------------------
For the large GASS field, we used another strategy:
\begin{itemize}
\item Due to the size of the field, which violates the flat-sky
  approximation used for the EBHIS and GBT fields, we compute the angular
  auto- and cross-power spectra on the \healpix\ maps,
  using the {\tt Xspect} algorithm \citep{tristram2005}. We apply the
  Galactic dust mask discussed in Sect.~\ref{sect:const_dust_maps}. The power spectra
  are computed using maps at the HFI angular resolution ($N_{\rm side}=2048$).
\item As the angular resolution of the \hi\ map is 16\arcmin, we use a hybrid method to remove the Galactic dust. On large angular scales ($\ell \le 590$), we remove the cirrus from the maps using \hi. On small angular scales ($\ell > 590$),
  we remove an estimate of the dust power spectrum. This estimate comes from
  the dust model that is fit on large angular scales ($120 \le \ell \le590$) by a power law, following
  \cite{mamd07}, and then extrapolated to small angular scales. The power-law 
  fit has been shown to be valid for the whole range of angular scales covered by our measurements \cite[e.g.,][]{mamd2010}.
\item As for the flat-sky fields, the CMB is removed as described in
Sect.~\ref{se:cmb-removal}. 
\item Statistical error bars are not computed using simulations but using analytical formulations (see below).
\end{itemize}

\paragraph{Auto- and cross-power spectra.}
We used the maps built from the first and second halves of each pointing
period.
As described in \cite{tristram2005}, the spherical harmonic coefficients
from the cut-sky maps are corrected from the mode-coupling introduced by the
mask, as well as the beam smoothing effect in the harmonic domain.
The cross-power spectra are unbiased estimates of the
angular power spectrum, avoiding any correction for the instrument noise,
contrary to our measurements done in Sect.~\ref{cross_correl1}. We computed the
spectra using the same multipole binning as that of the flat fields. We checked
that the bin to bin correlation was smaller than 1\%.

\begin{figure}
\hspace{-0.5cm}
\includegraphics[width=9.5cm]{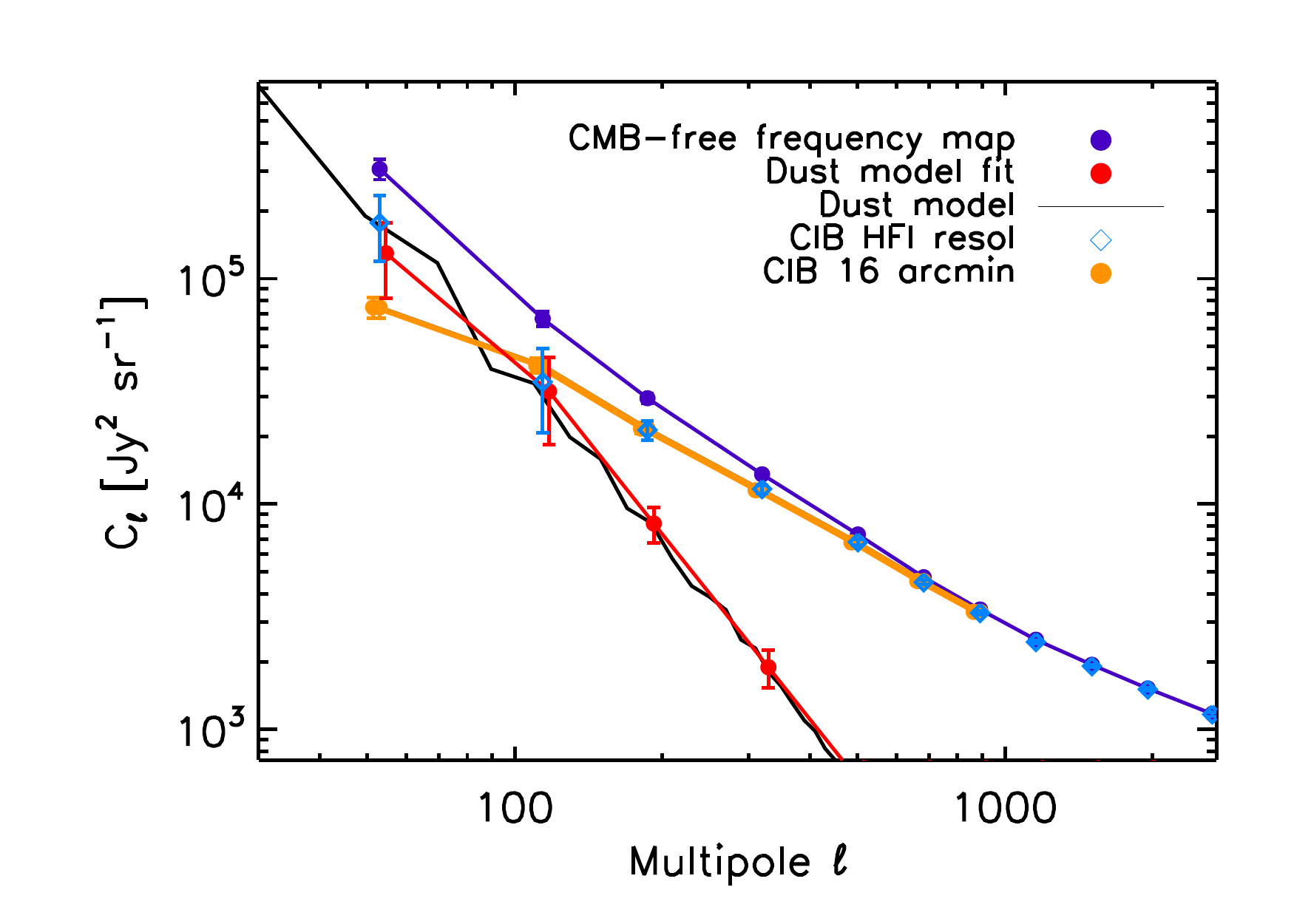}
\caption{\label{fig:gass_decomp} $353\times545$ cross-power spectrum in the GASS
  field (using Mask1). The black line is the cross-power spectrum of the dust
  model, with error bars not shown for clarity. The red points are the result of
  a power-law fit to the dust model, using the bins of the CIB power
  spectrum. The CIB obtained by the spatial removal of the dust is shown in
  orange (note that it stops at $\ell\simeq1000$ due to the angular resolution
  of the \hi\ data), while the CIB obtained from the spectral removal of the
  dust (i.e., on the power spectrum) is shown in light blue (diamonds).
  We see that the two methods give identical CIB for $\ell\gtrsim300$.
  In dark blue is the cross-power spectrum of the CMB-free frequency
  maps; the dust removal is negligible for $\ell\ge700$.  For clarity the
  points have been shifted by $\ell\pm3$\% and the error linked to the cirrus bias (see Sect~\ref{dust_error}) have not been added.}
\end{figure}

We show in Fig.~\ref{fig:gass_decomp} an example of a cross-power spectrum
($545\times353$) computed on maps for which the CMB has been removed.
We also show
the power spectrum of the dust model, fitted by a broken power law.  The
slope of the dust power spectrum appears to flatten at $\ell\sim110$,
with a slope in the range $-2.1$ to $-1.8$
for $\ell<110$, and $-2.8$ to $-2.7$ for $\ell>110$, depending on the
frequency. We compare the CIB power spectrum obtained by removing the dust at the map
level to that obtained by subtracting the fit of the dust model power spectrum. For $\ell\ge200$
we have excellent agreement. We also see that for $\ell\ge700$ the
dust removal has a negligible impact. At low $\ell$, where the dust correction
is important, we chose to use the dust removal on the map, since it leads to a
lower variance on the residual power spectrum, as shown in
\cite{penin2012b}. Indeed, the spatial subtraction removes each moment of
the statistics, whereas the subtraction of the power spectrum only removes the
first two moments.  At low $\ell$, errors on the fit are also quite
large.  We arbitrary decide to take the transition between the dust removal on
the map and on the power spectrum at $\ell=510$. The exact choice of the
multipole for the transition has no consequence on the resulting
CIB power spectrum.
 
\paragraph{Error bars.}
We quadratically combine the following error terms to obtain the final uncertainty
on the GASS CIB power spectra:
\begin{itemize}
\item The error bars on the spectrum computed analytically as described in
  \cite{tristram2005}, from the auto-power and cross-power spectra of the two
  maps. They include both the sampling variance (which dominates at large
  scales) and the instrumental noise (which dominates at small scales).
\item The errors linked to the CMB removal. Errors on the extra instrumental
  noise that have been introduced by the CMB removal
  (see Eq.~\ref{eq:xtranoise})
  are computed using the errors on the noise measurements at 100\GHz. For the
  error on the CMB contribution that is left close to the angular resolution of
  the 143 and, 217 and 353\GHz\ channels (see Eq.~\ref{eq:cmb_left}), we use
  the theoretical cosmic variance estimate on the determination of the CMB power
  spectrum. Finally, we add the errors that come from the relative
  photometric calibration in the CMB removal. 
\end{itemize}
As detailed in Sect.~\ref{dust_error}, the uncertainty on the dust model is derived from simulations. 
It is added linearly to the statistical error bars. For $\ell \ge$510, we also add the error on the dust model fit.
Beam errors are computed using the eigenmodes. They are the same as those
detailed in Sect.~\ref{planck_data} and \ref{cross_correl1}.

\subsection{Bispectrum pipeline for the GASS field}\label{Sect:bisp_pipeline}
The bispectrum $b_{\ell_1 \ell_2 \ell_3}$ is the 3-point correlation function
in harmonic space:
\begin{equation}
\langle a_{\ell_1 m_1} \, a_{\ell_2 m_2} \, a_{\ell_3 m_3}\rangle
 = b_{\ell_1 \ell_2 \ell_3} \times G_{\ell_1 \ell_2 \ell_3}^{m_1 m_2 m_3},
\end{equation}
with $G_{\ell_1 \ell_2 \ell_3}^{m_1 m_2 m_3}$ the Gaunt integral \citep{Spergel1999}. 
It is a lowest-order indicator of the non-Gaussianity of the field.

The maps used are the same as for the power spectrum analysis, but degraded to
$N_{\rm side}=512$. As the signal-to-noise ratio for bispectra is quite low compared to that 
of power spectra, bispectra have to be measured on the largest possible clean area of the sky. Here,
we measure the bispectrum on GASS Mask2.
We apply the binned bispectrum estimator described
in \cite{Lacasa2012} and used for the \Planck\ tSZ map
analysis \citep{planck2013-p05b} and non-Gaussianity constraints
\citep{planck2013-p09a}.  A large bin size $\Delta \ell=128$ has been adopted
to minimize multipole correlations due to the mask. We used a multipole range
$\ell_{\rm min}=129$ to $\ell_{\rm max}=896$, leaving six multipole bins
and 43 bispectrum configurations $(\ell_1,\ell_2,\ell_3)$ (accounting for
permutation invariance and the triangular condition). We only considered
auto-bispectra for simplicity, i.e., at a single frequency (see
Table~\ref{Table:217n353n545bispvaluesjy}).

For each frequency, we computed the auto-bispectra of the two maps built using
the two half-pointing period rings, and average these bispectra for a raw
estimate. The raw estimate has been then debiased from mask and beam effects
using simulations.  Specifically we generated simulations with a high level of
non-Gaussianity and a bispectrum corresponding to the CIB prescription from
\cite{Lacasa2012}, computed the ratio of the bispectrum of the masked map to
the full-sky bispectrum, and finally averaged this ratio over simulations,
finding a quick convergence especially at high multipoles. We found this ratio
to be very close to
$f_\mathrm{sky} \, b_{\ell_1}(\nu)\,b_{\ell_2}(\nu)\,b_{\ell_3}(\nu)$
(with $b_{\ell}(\nu)$ the beam window function), showing that the multipole
correlation is indeed negligible for this bin size and bispectrum. We checked
on the two half maps that the \Planck\ noise is close to Gaussian, hence it
does not bias our bispectrum estimates; it however increases their variance.

The error estimates are the sum in quadrature of:
\begin{itemize}
\item cosmic variance computed with analytical formulae and including the
noise;
\item dust residuals from (the absolute value of) the bispectrum of the dust
model, scaled to the residual dust amplitude found in Sect.~\ref{dust_error}.
\end{itemize}
Note that beam errors are completely negligible in the range of $\ell$
considered.
The full-sky cosmic variance of the bispectrum is composed of four terms:
\begin{equation}
\mathrm{Cov}(b_{\ell_1\ell_2\ell_3}\, ,\,
 b_{\ell_1^\prime\ell_2^\prime\ell_3^\prime}) = \tens{C}_{2\times2\times2}
 + \tens{C}_{3\times3} + \tens{C}_{2\times4} + \tens{C}_{6},
\end{equation}
with \citep[see][]{Lacasa2012}
\begin{equation}
\tens{C}_{2\times2\times2} =
 \frac{C_{\ell_1}\,C_{\ell_2}\,C_{\ell_3}}{\frac{(2\ell+1)_{123}}{4\pi}
\begin{pmatrix}
\ell_1& \ell_2& \ell_3 \\ 0& 0& 0
\end{pmatrix}^2 } \; \delta_{\ell_1 \ell^\prime_1} \,
 \delta_{\ell_2 \ell^\prime_2}
 \, \delta_{\ell_3 \ell^\prime_3} \times \Delta_{\ell_1 \ell_2 \ell_3}.
\end{equation}
Here
\begin{equation}
\Delta_{\ell_1 \ell_2 \ell_3} = \left\{ \begin{array}{ll} 6&
 \mathrm{equilateral\ triangles}\\ 2&
 \mathrm{isosceles\ triangles}\\ 1&
 \mathrm{general\ triangles} \end{array}\right.
\end{equation}
and when the bispectrum is factorizable
\begin{eqnarray}
\nonumber\tens{C}_{3\times3}=
b_{\ell_1\ell_2\ell_3}\,
 b_{\ell_1^\prime\ell_2^\prime\ell_3^\prime}
 &\times& \Bigg( \frac{\delta_{\ell_1 \ell^\prime_1}}{2\ell_1+1}
 +\frac{\delta_{\ell_1 \ell^\prime_2}}{2\ell_1+1}
 +\frac{\delta_{\ell_1 \ell^\prime_3}}{2\ell_1+1}\\
\nonumber &&+\frac{\delta_{\ell_2 \ell^\prime_1}}{2\ell_2+1}
 +\frac{\delta_{\ell_2 \ell^\prime_2}}{2\ell_2+1}
 +\frac{\delta_{\ell_2 \ell^\prime_3}}{2\ell_2+1}\\
 &&+ \frac{\delta_{\ell_3 \ell^\prime_1}}{2\ell_3+1}
 +\frac{\delta_{\ell_3 \ell^\prime_2}}{2\ell_3+1}
 +\frac{\delta_{\ell_3 \ell^\prime_3}}{2\ell_3+1} \Bigg)
\end{eqnarray}
$\tens{C}_{2\times4}$ and $\tens{C}_{6}$ involve, respectively, the map
trispectrum and 6-point function. $\tens{C}_{2\times2\times2}$ is the cosmic
variance in the weak non-Gaussianity limit, as considered, e.g., by
\cite{Crawford2013} in their tSZ and CIB non-Gaussianity study; it produces a
diagonal covariance matrix, since it does not couple the different
configurations.
$\tens{C}_{3\times3}$ is the bispectrum correction to the covariance matrix
and couples bispectrum estimates in different configurations. In the cosmic
variance limited case, we found that including only
$\tens{C}_{2\times2\times2}$ would noticeably overestimate the detection
significance. However when considering all error sources,
$\tens{C}_{3\times3}$ has a small impact on the SNR; hence we include it in
the following analysis but neglect higher order corrections
$\tens{C}_{2\times4}$ and $\tens{C}_{6}$.

We use the $f_\mathrm{sky}$ approximation, that is
$\mathrm{Cov} = \mathrm{Cov}_\mathrm{full-sky}/f_\mathrm{sky}$, as the first
multipoles were discarded and we saw no mode coupling with our large bin
size. The power spectrum used for $\tens{C}_{2\times2\times2}$ is the map
auto-power spectrum, including the noise as necessary, debiased from the mask
and beam effects.

\begin{figure*}
\begin{center}
\includegraphics[width=18cm]{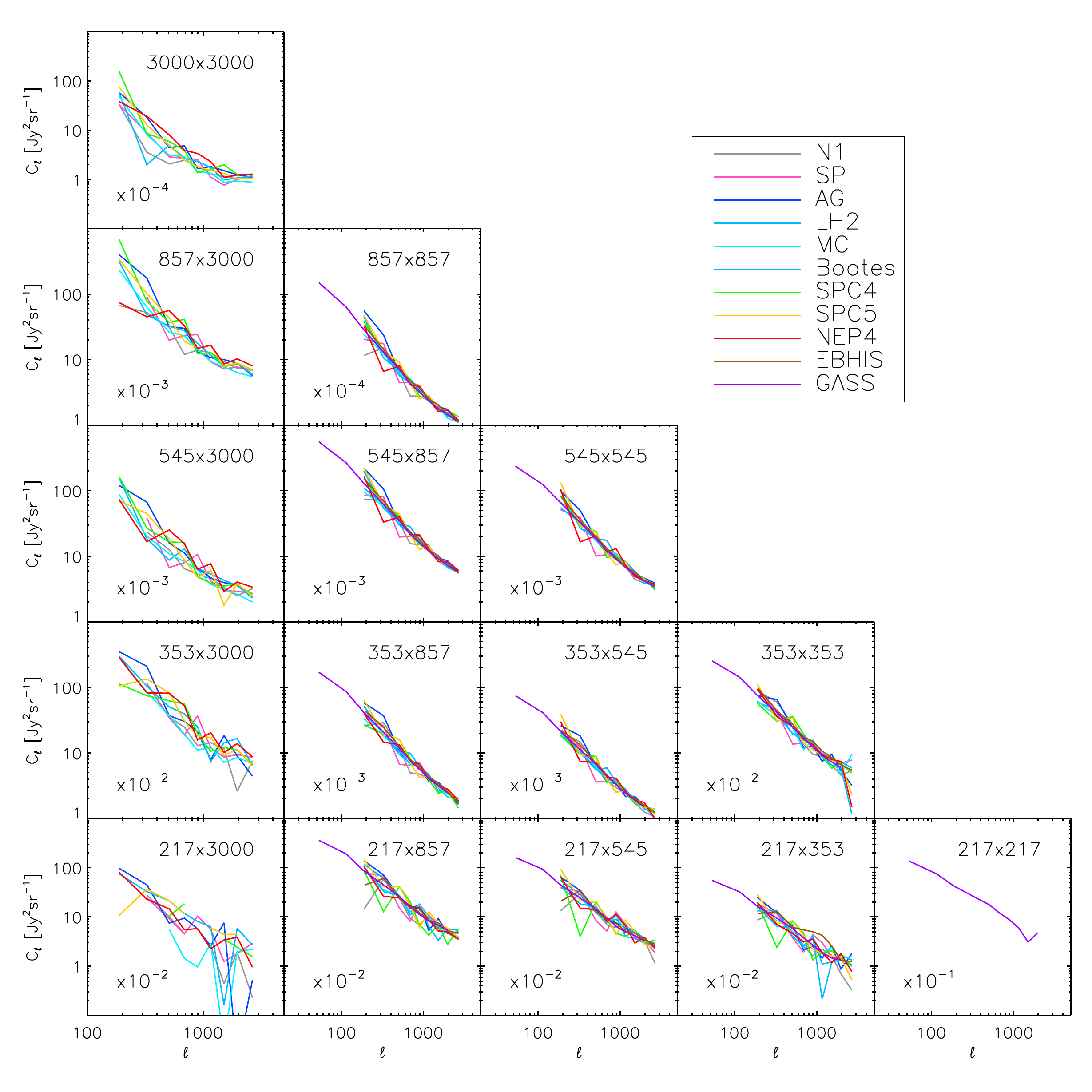}
\end{center}
\caption{Auto- and cross-power spectra of the CIB for each field (but EBHIS,
  Bootes and GASS at 3000\GHz, see Sect.~\ref{sec:CIB_maps}). For readability,
  error bars on individual measurements are not plotted. For the $217\times
  217$ case, measurements on the flat-sky fields are noise dominated, and we
  thus use only the results from the GASS field. For display purpose, power spectra have been multiplied by the number given at the bottom-left side of each panel.}
\label{fig:all_alapep}
\end{figure*}

As described in Sect.~\ref{cmb_res}, the residual maps are contaminated by
tSZ contribution at 100\,\GHz\ leaking through the CMB cleaning process. While
this contamination is negligible compared to the CIB signal at 353\,\GHz\ and
above, it is important at 217\,\GHz, where the contamination can be as large as 40\%, depending 
on the configuration and multipole. We derived the SZ bispectrum by measuring
the bispectrum of the map of the \Planck\ SZ cluster catalogue (both clusters
and candidate clusters from \cite{planck2013-p05a} are considered).
A 80\% error on the amplitude of the tSZ correction is
added to the covariance measurement described previously. This error
is based on the relative difference between the bispectrum of the
\Planck\ SZ catalog of confirmed clusters \citep{planck2013-p05a}, the
bispectrum of the \Planck\ estimated tSZ map, and the bispectrum of the Planck FFP6 SZ simulations \citep[see the
bispectrum analysis in][]{planck2013-p05b}. The error is conservative as it takes for 
all multipoles and configurations the maximum observed difference.

\subsection{CIB power spectrum}
%---------------------------------
Figure~\ref{fig:all_alapep} presents a summary of all the measured auto- and
cross-power spectra on residual maps.  Following the same covariance studies as in \cite{planck2011-6.6},
we combine our cross-power spectrum estimates on individual fields
$f$ for each bin $b$ into an
average cross-power spectrum, using inverse variance weights,
\begin{equation}
C_b^{\nu \times \nu^\prime} =
 \frac{\sum_f W_b^f P_b^{f,\nu \times \nu^\prime}}{\sum_f W_b^f},
\end{equation}
with $W_b^f = 1/\sigma^2(P_b^{f,\nu \times \nu^\prime})$.  
These weights are estimated in the following way:
\begin{itemize}
\item[-] Step 1: Only statistical errors are used to compute a weighted average of the power spectra and their associated error bars on small fields.
\item[-] Step 2: The projection error is added linearly to the error bar of step 1.
\item[-] Step 3: For frequencies where observations on GASS are available, we average the GASS power spectrum and the small fields power spectrum. We here use inverse statistical variance weights for GASS, and the inverse variance derived from step 2 for the small fields. This gives the final power spectrum estimate and a pseudo-statistical error bar.
\item[-] Step 4: We compute the error due to beam uncertainties using the average power spectrum. We add this error linearly to the error derived on Step 3.
\item[-] Step 5 : The bias induced by Galactic dust residuals is linearly added to the error derived on Step 4 to obtain the final total error bar.
\end{itemize}
The resulting power spectra and their errors are given in Table~\ref{tab:res_map_pow_spec}. For the $217\times217$ auto-power
spectrum, only the measurement on the GASS field is considered, as the
measurement in the flat-sky fields is noise dominated. Note also that the last bin for
all measurements involving the subtraction of the 100\,\GHz\ CMB template is
$\ell=1956$, due to the angular resolution of \Planck-HFI at 100\,\GHz.
To obtain the CIB power spectra, the estimates obtained from the CMB- and dust-free maps (Table~\ref{tab:res_map_pow_spec}) have to be corrected for SZ
contaminations (Eqs.~\ref{eq:SZ1} and \ref{eq:SZ3}), and
for the spurious CIB contamination (Eq.~\ref{eq:corr_CMB}) induced by our CMB template. This last contamination is computed using our best-fit
 model described in Sect.~\ref{mod_hod}. Moreover, following  Sects.~\ref{dust_error} and \ref{dust_res}, the first two bins at multipoles $\ell=53$ and $114$ have to be considered as upper limits.
CIB power spectrum values are given in Table~\ref{tab:cib_final}. The errors contain all the terms: statistical uncertainty; beam and projection uncertainty; cirrus bias; and errors from the SZ correction. CIB power spectra are shown on the figure comparing the measurements with the model (Fig.~\ref{fig:modmes_halomod}). Comparison with previous recent measurements are shown and discussed in Sect.~\ref{sect_compar_previous}.\\ 

\begin{figure}
\hspace{-0.5cm}
\includegraphics[width=9.5cm]{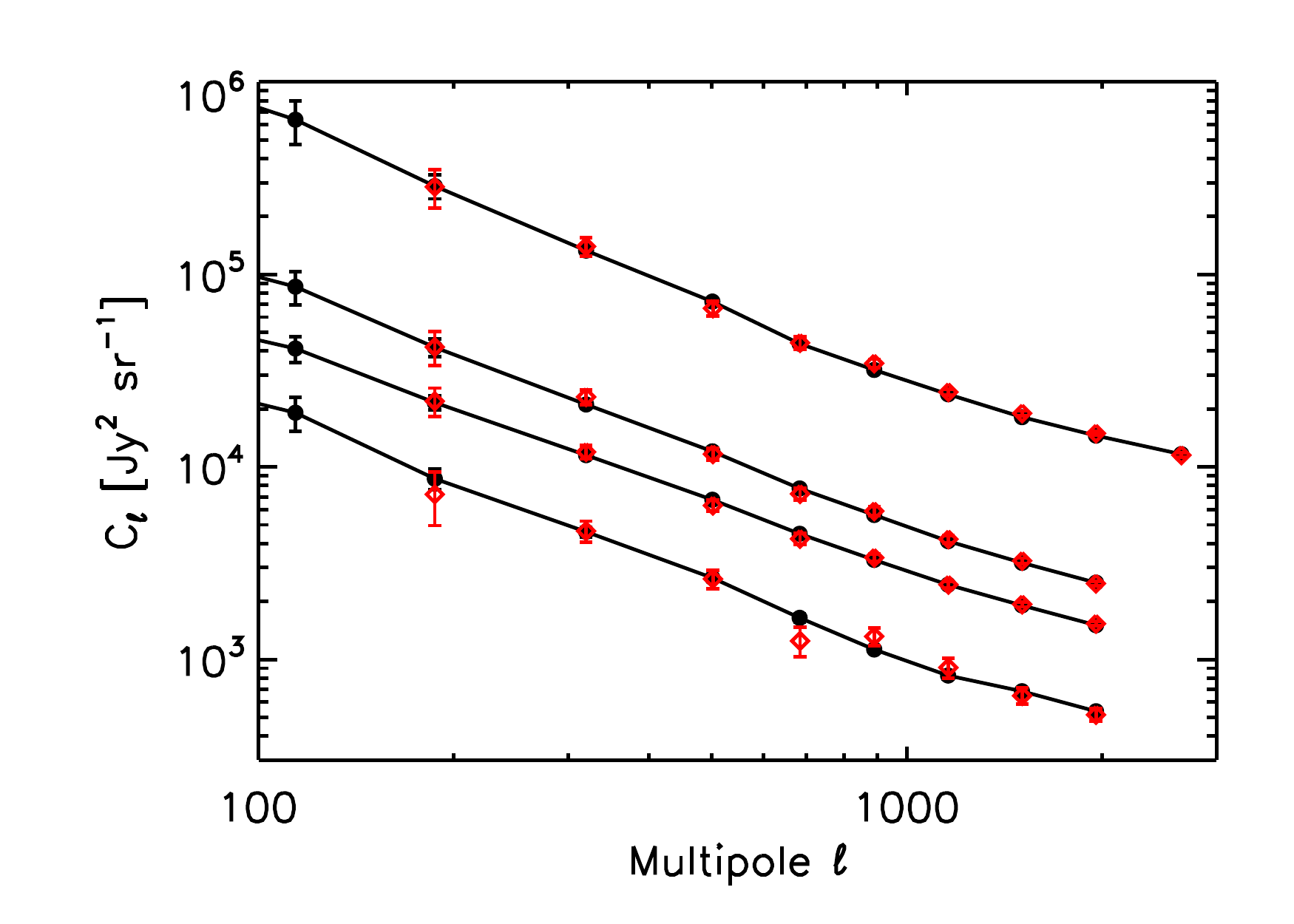}
\caption{Comparison between the CMB- and dust-free map power spectra obtained from the analysis of the flat
  fields ($328\,{\rm deg}^2$, red diamonds) and GASS ($1914\,{\rm deg}^2$,
  black dots). From top to bottom: $857\times857$; $857\times353$;
  $545\times353$; and $857\times217$.}
\label{fig:powspec_gass_flat}
\end{figure}

The power spectra from the flat-sky and GASS fields have been obtained using
two independent pipelines. The fields have different Galactic dust and point
sources contamination, as well as instrument noise levels, and between GASS
and the flat-sky fields, different pixelization and projection. 
Comparing combined power spectra obtained on all flat-sky fields to that
obtained on GASS is thus a powerful consistency check on our determination
of power spectra and error
bars. We show on Fig.~\ref{fig:powspec_gass_flat} this comparison for an
arbitrary set of
frequencies. The power spectra are always compatible within 1$\,\sigma$. Of
course, due to the much larger area of the GASS field, the CIB measured in
GASS has much smaller cosmic variance errors.

\subsection{CIB bispectrum}
\begin{figure}
\centering
\includegraphics[width=0.99\linewidth]{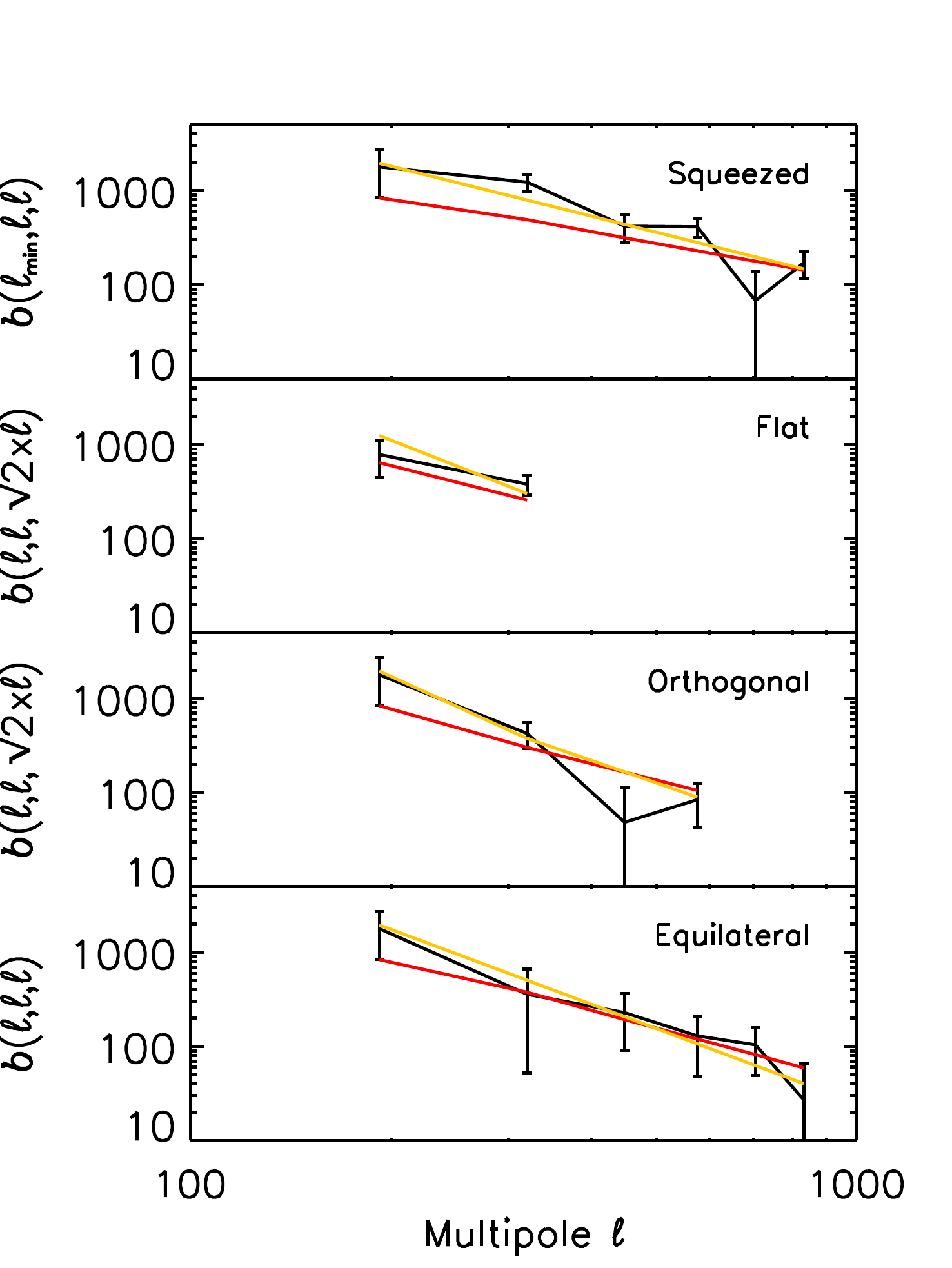}
\caption{CIB bispectrum at 353\,\GHz\ in some particular configurations (black
points, in Jy$^3$ sr$^{-1}$). The red curve is the CIB bispectrum predicted from the power spectrum
(best-fit model from Sect.~\ref{mod_hod}) following \cite{Lacasa2012}. The
yellow curve is a power-law fit (as given in
Table~\ref{Table:fitpowerlawbisp}).
See Sect.~\ref{discussbisp} for more details.}
\label{Fig:bisp353} 
\end{figure}

We measure the bispectrum only at 217, 353 and 545\,\GHz. The bispectrum at
143\,\GHz\ is noise-dominated and is moreover highly contaminated by tSZ and
extragalactic radio point sources. At 857\,\GHz, as we are using Mask2,
Galactic dust residuals contaminate the bispectrum in most configurations. In particular, 
the residuals produce a rising bispectrum at high multipoles. In the following, we are thus not considering 
the 857\,\GHz\ frequency for the analysis. 
At 217, 353 and 545\,\GHz, the bispectrum is measured in 38, 40 and 36 configurations, respectively.
We show in Fig.~\ref{Fig:bisp353} the measured CIB bispectrum at 353\,\GHz\
for some particular configurations, namely equilateral $(\ell,\ell,\ell)$,
orthogonal isosceles $(\ell,\ell,\sqrt{2}\ell)$, flat isosceles
$(\ell,\ell,2\ell)$, and squeezed $(\ell_{\rm min},\ell,\ell)$.
The bispectrum decreases with scale and exhibits a peak in the squeezed
configurations, as predicted by \cite{Lacasa2012}.

In Table~\ref{Table:bispSNR}, we give the significance of the detection for
the three frequencies used, with either the average significance per
configuration or the significance of the total bispectrum, when accounting for
the whole covariance matrix. The bispectrum is significantly detected at each
frequency individually. Moreover, these measurements represent the first detection
of the CIB bispectrum per configuration, permitting us to
probe the scale and configuration dependence of the bispectrum, as well as its
frequency behaviour. The bispectrum values are given in
Table~\ref{Table:217n353n545bispvaluesjy}.

\begin{table}[!tbh]
\begingroup
\newdimen\tblskip \tblskip=5pt
\caption{Detection significance of the bispectra at each frequency. Note that the mean SNR 
per configuration and the total SNR are not directly linked by the square root of the configuration 
numbers as the covariance matrix is not diagonal.}
\label{Table:bispSNR}
\nointerlineskip
\vskip -3mm
\footnotesize
\setbox\tablebox=\vbox{
 \newdimen\digitwidth
 \setbox0=\hbox{\rm 0}
  \digitwidth=\wd0
  \catcode`*=\active
  \def*{\kern\digitwidth}
  \newdimen\signwidth
  \setbox0=\hbox{+}
  \signwidth=\wd0
  \catcode`!=\active
  \def!{\kern\signwidth}
\halign{\tabskip=0pt\hfil#\hfil\tabskip=1.0em&
  \hfil#\hfil\tabskip=1.0em&
  \hfil#\hfil\tabskip=0pt\cr
\noalign{\doubleline}
\noalign{\vskip -2pt}
Band& Mean SNR per configuration& Total SNR\cr
\noalign{\vskip 3pt\hrule\vskip 3pt}
217\GHz& 1.24& *5.83\cr
353\GHz& 2.85& 19.27\cr
545\GHz& 4.59& 28.72\cr
\noalign{\vskip 3pt\hrule\vskip 3pt}}}
\endPlancktable
\endgroup
\end{table}

%================================
\section{Interpreting CIB power spectrum measurements \label{CIB_Mod}}
%================================
Once the CMB and Galactic dust have been removed, there are three astrophysical
contributors to the power spectrum at the HFI frequencies: two from dusty
star-forming galaxies (with both shot noise, $C_{{\rm d, shot}}^{\nu \times
  \nu^\prime}$, and clustering, $ C_{{\rm d, clust}}^{\nu \times \nu^\prime}(\ell)$,
components); and one from radio galaxies (with only a shot-noise component,
$C_{{\rm r, shot}}^{\nu \times \nu^\prime} $, the clustering of radio sources being
negligible, e.g., \citealt{hall2010}).  The measured CIB power spectrum
$C_{\rm measured}^{\nu \times \nu^\prime}(\ell) $ is thus
\begin{equation}
\label{eq:pow_spec_all_compo}
C_{\rm measured}^{\nu \times \nu^\prime}(\ell)
 = C_{{\rm d, clust}}^{\nu \times \nu^\prime}(\ell)
 + C_{{\rm d, shot}}^{\nu \times \nu^\prime}
 + C_{{\rm r, shot}}^{\nu \times \nu^\prime}.
\end{equation}

In this section, we first discuss the shot-noise contributions. We then present how we can model
$C_{{\rm d, clust}}^{\nu \times \nu^\prime}(\ell)$. Two approaches are considered. The first one is the simplest, and use only the large scale CIB measurements to fit for a linear model. The second one is based on the halo-model formalism. Our main goal is to use CIB anisotropies to measure the SFRD and effective bias redshift evolution.

\subsection{Shot noise from dusty star forming and radio galaxies
 \label{sec:sn}}
The shot noise arises from sampling of a background composed of a finite
number of sources, and as such is decoupled from the correlated term.
The angular resolution of the HFI instrument is not high
enough to
measure the shot-noise levels. As demonstrated in \cite{planck2011-6.6}, the non-linear
contribution to the power spectrum is degenerate with the shot-noise
level (on the scales probed by \Planck).
Since our data by themselves are not sufficient to explore this
degeneracy, we need to rely on a model to compute the shot noise.

\subsubsection{Auto-power spectrum}
The shot-noise level at frequency $\nu$ can be easily computed using
monochromatic galaxy number counts.  Let us consider a flux interval
$[S_k,S_k+\Delta S_k]$. The number of sources per unit solid angle,
$n_{k}$ in this flux interval is
\begin{equation}
n_{k} = \frac{dN}{dS} \Delta S_k 
\end{equation}
and the variance is,
\begin{equation}
\sigma_{B_k}^2 = n_k S_k^2.
\end{equation}
Summing all flux intervals gives the variance on the total contribution
to the CIB:
\begin{equation}
\sigma_{B}^2 =  \sum_{k} n_k S_k^2 =  \frac{dN}{dS} S_k^2 \Delta S_k.
\end{equation}
When we take the limit of $\Delta S_k$ tending to zero, this sum becomes
the integral
\begin{equation}
\label{rms_fluc}
\sigma_{B}^2 =  \int_{0}^{S^{\rm c}}  \frac{dN}{dS} S^2 dS.
\end{equation}
Here $S^{\rm c}$ is the flux cut above which bright sources are detected and
can be removed. This cut is mandatory, since the integral does not converge
in the Euclidian regime, $\frac{dN}{dS} \propto S^{-2.5}$, which is the case
at bright fluxes for star-forming galaxies. 
 
\subsubsection{Cross-power spectrum}
We now consider two frequencies $\nu$ and $\nu^\prime$. 
The number of sources $n_{kl}$ in the flux density and redshift intervals
$[S_k,S_k+\Delta S_k]$ and $[z_l,z_l+\Delta z_l]$, is
\begin{equation}
n_{kl} = \frac{dN}{dS dz} \Delta S_k \Delta z_l
\end{equation}
Considering a small redshift interval, the covariance between the two
frequencies can be approximated as:
\begin{equation}
\sigma_{\nu_1 \nu_2,kl} = n_{kl} S_{\nu,kl} S_{\nu^\prime,kl} = n_{kl} S_{\nu,kl}^2
 R_{\nu \nu^\prime,kl},
\end{equation}
where $R_{\nu \nu^\prime,kl}$ is the mean colour for the considered galaxy
population in the considered flux density and redshift interval.
Using a mean colour per flux density and redshift interval is not a strong
assumption as long as $\Delta S_k$ and $\Delta z_l$ are small.
Summing over flux densities, redshifts and the galaxy population gives
\begin{equation}
\sigma_{\nu \nu^\prime} = \sum_{\rm pop} \sum_{k} \sum_{l} n_{{\rm pop},kl}
 S_{\nu,kl}^2 R_{\nu \nu^\prime,{\rm pop},kl}\, ,
\end{equation}
with the integral limit
\begin{align}
\sigma_{\nu \nu^\prime} & = \sum_{\rm pop} \int_{S_{\nu}=0}^{\infty}
 \int_{z=0}^{\infty} H(S_{\nu}\,{<}\,S_{\nu}^{\rm c},
 R_{\nu^\prime \nu ,{\rm pop}}
 S_{\nu }\,{<}\,S_{\nu^\prime}^{\rm c})  \nonumber \\
& \frac{dN_{\rm pop}}{dS_{\nu} dz}
 S_{\nu}^2 R_{\nu^\prime\nu,{\rm pop}}\, dS_{\nu} dz.
\label{eq:sn}
\end{align}
Here $H(P_1,P_2)$ is equal to 1 when $P_1$ and $P_2$ are both true, and 0
otherwise, and
$S_{\nu}^{\rm c}$ and $S_{\nu^\prime}^{\rm c}$ are the flux cuts in the frequency
bands $\nu$ and $\nu^\prime$. They are given in Table~\ref{tab_conversion}.

We use this formalism to compute the radio galaxy shot noise. For the
star-forming dusty galaxy shot noise, we rely on the formalism detailed in \citealt{bethermin2013} (their appendix B).

\subsubsection{Shot-noise values}
\label{se:sn}
We use the \cite{bethermin2012model} model to compute the star-forming dusty
galaxy shot noise, $C_{{\rm d, shot}}^{\nu \times \nu^\prime}$
(Eq.~\ref{eq:pow_spec_all_compo}). The model is in rather good agreement
with the number counts measured by {\it Spitzer\/} and {\it Herschel}
\citep[e.g.,][]{glenn2010}. It also gives a reasonable CIB
redshift-distribution, which is important for the cross-spectra. Since this
model is based on observations that have typical calibration uncertainties of
$<$10\%, the estimations of the shot-noise levels (being proportional to the
square of calibration factor) cannot be accurate to more than $\sim$20\%.
As uncertainties in the flux cuts induce uncertainties in the shot noise that are negligible
(less than 3\% at all frequencies), we take 20\% as the shot-noise uncertainty.

For extragalactic radio sources, we use the \cite{tucci11} model (more
specifically, the one referred as ``C2Ex'' in the paper) to compute
$C_{{\rm r, shot}}^{\nu \times \nu^\prime}$ (Eq.~\ref{eq:pow_spec_all_compo}).
The predictions for high--frequency number counts are based on a
statistical extrapolation of flux densities of radio sources from
low--frequency data (1--5\,GHz). In particular, this model considers
physically based recipes to describe the complex spectral behaviour
of blazars, which dominate the mm-wave counts at bright flux
densities. It is able to give a good fit to all bright
extragalactic radio source data available so far: number counts up to
600\,GHz; and spectral index distributions up to at least 200--300\,\GHz\
\cite[see][]{tucci11,planck2012-VII,lopez2012}. As for the dusty galaxies, we consider an error 
of 20\% on the shot-noise computation from the model. But contrary to dusty galaxies,
the shot noise for the radio population depends strongly on the flux cut
\citep{planck2011-6.6}.  Accordingly, we add to the 20\% mentioned above, a shot-noise error taken to be the
shot-noise variations as we change the flux cut, considering the flux cut
errors given in Table~\ref{tab_conversion}.

The shot-noise levels for the HFI flux cuts given in Sect.~\ref{sec:flux_cut},
are listed in Tables~\ref{SN_IR_jy} and ~\ref{SN_rad_jy}.

\begin{table*}[!tbh]
\begingroup
\newdimen\tblskip \tblskip=5pt
\caption{Shot-noise levels $C_{{\rm d, shot}}^{\nu \times \nu^\prime}$
(flat power-spectra) for star-forming galaxies (in ${\rm Jy}^2\,{\rm sr}^{-1}$)
computed using the \cite{bethermin2012model} model. To obtain the shot noise in
${\rm Jy}^2\,{\rm sr}^{-1}$ for our photometric convention
$\nu I_\nu={\rm constant}$, a colour correction, given in
Sect.~\ref{se:CIB_model}, has to be applied. \label{SN_IR_jy}}
\nointerlineskip
\vskip -3mm
\footnotesize
\setbox\tablebox=\vbox{
 \newdimen\digitwidth
 \setbox0=\hbox{\rm 0}
  \digitwidth=\wd0
  \catcode`*=\active
  \def*{\kern\digitwidth}
  \newdimen\signwidth
  \setbox0=\hbox{+}
  \signwidth=\wd0
  \catcode`!=\active
  \def!{\kern\signwidth}
  \newdimen\pointwidth
  \setbox0=\hbox{\rm .}
  \pointwidth=\wd0
  \catcode`?=\active
  \def?{\kern\pointwidth}
\halign{\tabskip=0pt\hfil#\hfil\tabskip=1.0em&
  \hfil#\hfil\tabskip=1.0em&
  \hfil#\hfil\tabskip=1.0em&
  \hfil#\hfil\tabskip=1.0em&
  \hfil#\hfil\tabskip=1.0em&
  \hfil#\hfil\tabskip=1.0em&
  \hfil#\hfil\tabskip=1.0em&
  \hfil#\hfil\tabskip=0pt\cr
\noalign{\doubleline}
\noalign{\vskip -2pt}
& 3000& 857&  545& 353& 217& 143& 100 \cr
\noalign{\vskip 3pt\hrule\vskip 3pt}
3000& 10064$\pm$2013&         \dots&        \dots&          \dots&
        \dots&         \dots&          \dots\cr
*857& *4427$\pm$*885& 5628$\pm$1126&        \dots&          \dots&
        \dots&         \dots&          \dots\cr
*545& *1525$\pm$*305& 2655$\pm$*531& 1454$\pm$291&          \dots&
        \dots&         \dots&          \dots\cr
*353& **434$\pm$**87& *913$\pm$*183& *543$\pm$109& 225?*$\pm$45?*&
        \dots&         \dots&          \dots\cr
*217& ***96$\pm$**19& *216$\pm$**43& *135$\pm$*27& *59?*$\pm$12?*&
 16?*$\pm$3?*&         \dots&          \dots\cr
*143& ***26$\pm$***5& **56$\pm$**11& **35$\pm$**7& *15?*$\pm$*3?*&
 *4.3$\pm$0.9& 1.2*$\pm$0.2*&          \dots\cr
*100& ***11$\pm$***2& **20$\pm$***4& **12$\pm$**2& **5.4$\pm$*1.1&
 *1.5$\pm$0.3& 0.42$\pm$0.08&  0.15$\pm$0.03\cr
\noalign{\vskip 3pt\hrule\vskip 3pt}}}
\endPlancktablewide
\endgroup
\end{table*}

\begin{table*}[!tbh]
\begingroup
\newdimen\tblskip \tblskip=5pt
\caption{Shot-noise levels $C_{{\rm r, shot}}^{\nu \times \nu^\prime}$
(flat power-spectra)  for radio galaxies (in ${\rm Jy}^2\,{\rm sr}^{-1}$)
computed using the \cite{tucci11} model. To obtain the shot noise in
${\rm Jy}^2\,{\rm sr}^{-1}$ for our photometric convention
$\nu I_\nu={\rm constant}$, a colour correction has to be applied.
It is however lower than 1\% (see Sect.~\ref{se:CIB_model}).
 \label{SN_rad_jy}}
\nointerlineskip
\vskip -3mm
\footnotesize
\setbox\tablebox=\vbox{
 \newdimen\digitwidth
 \setbox0=\hbox{\rm 0}
  \digitwidth=\wd0
  \catcode`*=\active
  \def*{\kern\digitwidth}
  \newdimen\signwidth
  \setbox0=\hbox{+}
  \signwidth=\wd0
  \catcode`!=\active
  \def!{\kern\signwidth}
  \newdimen\pointwidth
  \setbox0=\hbox{\rm .}
  \pointwidth=\wd0
  \catcode`?=\active
  \def?{\kern\pointwidth}
\halign{\tabskip=0pt\hfil#\hfil\tabskip=1.0em&
  \hfil#\hfil\tabskip=1.0em&
  \hfil#\hfil\tabskip=1.0em&
  \hfil#\hfil\tabskip=1.0em&
  \hfil#\hfil\tabskip=1.0em&
  \hfil#\hfil\tabskip=1.0em&
  \hfil#\hfil\tabskip=0pt\cr
\noalign{\doubleline}
\noalign{\vskip -2pt}
& 857&  545& 353& 217& 143& 100\cr
\noalign{\vskip 3pt\hrule\vskip 3pt}
857& $4.28\pm0.90$&         \dots&         \dots&         \dots&         \dots&
        \dots\cr
545& $2.28\pm0.56$& $2.86\pm0.70$&         \dots&         \dots&         \dots&
        \dots\cr
353& $2.10\pm0.53$& $2.59\pm0.63$& $3.28\pm0.82$&         \dots&         \dots&
        \dots\cr
217& $1.53\pm0.46$& $1.92\pm0.56$& $2.40\pm0.70$& $3.12\pm0.79$&         \dots&
        \dots\cr
143& $2.38\pm0.58$& $2.86\pm0.68$& $3.57\pm0.82$& $3.68\pm0.99$& $6.05\pm1.47$&
        \dots\cr
100& $2.73\pm0.64$& $3.24\pm0.73$& $4.02\pm0.90$& $4.31\pm1.07$& $6.47\pm1.39$&
 $8.47\pm1.97$\cr
\noalign{\vskip 3pt\hrule\vskip 3pt}}}
\endPlancktablewide
\endgroup
\end{table*}

\subsection{Basics of CIB correlated anisotropy modelling}
%--------------------------------
\label{se:CIB_model}

The angular power spectrum of CIB correlated anisotropies is defined as:
\begin{eqnarray}
\left\langle\delta I^\nu_{\ell m}
 \delta I^{\nu^\prime}_{\ell^{\prime}m^{\prime}}\right\rangle
 = C_{{\rm d, clust}}^{\nu \times \nu^\prime}(\ell) \times
 \delta_{\ell\ell^{\prime}}\delta_{mm^{\prime}},
\label{eqn:cll0}
\end{eqnarray}
where $\nu$ and $\nu^\prime$ denote the observing frequencies and $I^{\nu, \nu^\prime}$ the
measured intensity at those frequencies. In a flat universe, the intensity
is related to the comoving emissivity $j$ via
\begin{eqnarray}
\label{eqn:intensity}
I^{\nu}&=&\int dz\frac{d\chi}{dz} a j(\nu,z)\\\nonumber
&=&\int dz\frac{d\chi}{dz} a \bar{j}(\nu,z)\left(1+\frac{\delta
    j(\nu,z)}{\bar{j}(\nu,z)}\right),
\end{eqnarray}
where $\chi(z)$ is the comoving distance to redshift $z$, and
$a=1/(1+z)$ is the scale factor. Combining Eqs.~\ref{eqn:cll0}
and \ref{eqn:intensity} and using the Limber approximation, we obtain
\begin{eqnarray}
C^{\nu \times \nu^\prime}_{\rm d, clust}(\ell)
 = \int \frac{dz}{\chi^2}\frac{d\chi}{dz} a^2
 \bar{j}(\nu,z) \bar{j}(\nu^{\prime},z)
 P^{\nu \times \nu^\prime}_{j}(k=\ell/\chi,z),
\label{eqn:cll1}
\end{eqnarray}
where $P^{\nu \times \nu^\prime}_{j}$ is the 3-D power spectrum of the emissivities
and is defined as follows:
\begin{eqnarray}
\left\langle\delta j(\vec{k},\nu)
 \delta j(\vec{k}^\prime,\nu^{\prime})\right\rangle =
 (2\pi)^3 \bar{j}(\nu)\bar{j}(\nu^{\prime})
 P^{\nu \times \nu^\prime}_{j}(\vec{k})\delta^3(\vec{k}-\vec{k}^{\prime}). 
\label{eqn:pj}
\end{eqnarray}

In the context of CIB anisotropy modelling, the simplest version of the
so-called ``halo model,'' which provides one view of the large-scale structure
of the Universe as clumps of dark matter,  consists of equating $P_j$ with the
galaxy power spectrum $P_{\rm gg}$. This is equivalent to assuming that the CIB
is sourced equally by all galaxies, so that the spatial variations in the
emissivities trace the galaxy number density,
\begin{eqnarray}
\delta j/\bar{j} = \delta n_{\rm gal}/\bar{n}_{\rm gal}.
\label{eqn:dj}
\end{eqnarray}
The dark matter halos are populated through a halo occupation distribution
(HOD) prescription. Ultimately, $P_{\rm gg}^{}(k,z)$ is written as the sum of
the contributions of  galaxies within a single dark matter halo (``1h'') and
galaxies belonging to two different halos (``2h''):
\begin{equation}
P_{\rm gg}(k,z) = P_{\rm 1h}(k,z) +P_{\rm 2h}(k,z) \, .
\end{equation}
On large scales $P_{\rm 2h}$ reduces to a constant bias (squared) times the
linear theory power spectrum, while
the 1-halo contribution encapsulates the non-linear distribution of matter.

In this paper, we developed two approaches for the modelling of CIB
anisotropies.
\begin{itemize} 
\item The first one (Sect.~\ref{mod_lin}) is very simple, and takes advantage
of the accurate measurement of CIB anisotropies with \Planck\ and IRIS at
large angular scales. As an alternative to the HOD model for $P_{\rm gg}$ we
use a constant bias model in which
$P_{\rm gg}(k,z)=b_{\rm eff}^2 P_{\rm lin}(k,z)$, where $b_{\rm eff}$ is a
redshift- and scale-independent bias and $P_{\rm lin}(k)$ is the linear
theory, dark-matter power spectrum. 
\item The second one (Sect.~\ref{mod_hod}) uses the anisotropies at all
angular scales, and takes advantage of the frequency coverage of our
measurements, to constrain a halo model with a luminosity-mass dependence.
As a matter of fact, the model described above, which assumes that emissivity
density traces galaxy number density (Eq.~\ref{eqn:dj}), implies that all
galaxies contribute equally to the emissivity density, irrespective of the
masses of their host halos. It assumes that all galaxies have the same
luminosity, which is a crude assumption, as the luminosity and the clustering
strength are closely related to the mass of the host halo.
\end{itemize}

Before embarking in the details of our models building and fitting, we want to stress that the purpose of the following sections is to build as "physical" a model as possible that reproduces our measurements. To do so, we will build on the large amount of works in the astrophysical literature that exploited the simplifying concept of halo-model. We hope that this way our work will have wider impact. Nevertheless, it has to be emphasized that these models are very phenomenological and multi-faceted. To some extent, there is no such thing as a halo model as there are many hidden assumptions standing on uneven grounds. For example, in our approach we will rely on a concentration prescription as a function of mass and redshift and use it all the way to a redshift of 6, thus pushing in a regime where it has not been validated. The same holds for our ansatz for the $L-M$ relation and more generally the concept of HOD. To explore the dependence of our conclusions on these hidden assumptions is a work that goes well beyond the scope of this paper and would certainly require a large use of simulations. To include these assumptions as Bayesian priors in our fit could be an approach but would also certainly miss serious conceptual limitations. As such, we chose, in this current modeling effort to just make our assumptions clear and justify our fixed values when possible. This approach leaves a degree of uncertainty unaccounted for in our error budget and should be keep in mind when interpreting our results.

\subsection{Fitting for a model \label{mod_fit}}
The power spectra that are computed by the models need to be colour-corrected,
from a CIB SED to our photometric convention $\nu I_{\nu}={\rm constant}$.
We use the CIB SED
from \cite{bethermin2012model} to compute the colour corrections. They are
equal to 1.076, 1.017, 1.119, 1.097, 1.068, 0.995, and 0.960 at 100, 143, 217,
353, 545, 857, and 3000\GHz, respectively. The correction to the power spectra
follows
\begin{equation}
C_{\ell,\nu,\nu^\prime}^{\rm model} \times {\rm cc}_{\nu} \times
 {\rm cc}_{\nu^\prime} = C_{\ell,\nu,\nu^\prime}^{\rm measured}.
\label{eq:cc}
\end{equation}
We use the same colour corrections (${\rm cc}$) for both the star-forming
galaxy shot noise and the CIB power spectrum. For the radio
galaxy shot noise we use a power law $S_\nu \propto \nu^\alpha$, with
$\alpha=- 0.5$. This is the average spectral index for radio sources that
mainly contribute
to the shot-noise power spectra. With such an SED, we find that the colour
corrections are all lower than 0.7\% for $100\le \nu \le 857$\GHz. We thus
neglect them.

To search for our best-fit model, we follow this scheme.
\begin{enumerate}
\item Take the residual map power spectra, as given in
Table~\ref{tab:res_map_pow_spec}.
\item Discard the first two bins at multipoles $\ell=53$ and 114
(following Sect.~\ref{dust_error} and \ref{dust_res}).
\item Correct for SZ-related residuals,
$C_{\rm SZcorr}^{\nu \times \nu^\prime}$,
and $C_{\rm CIB-SZcorr}^{\nu \times \nu^\prime}$, following Eqs.~\ref{eq:SZ2},
\ref{eq:SZ3}, \ref{eq:SZcorr}, and add the errors of these corrections
quadratically to the error bars given in Table~\ref{tab:res_map_pow_spec}.
\item Apply the colour correction to convert the theoretical model from
measured ${\rm Jy}^2\,{\rm sr}^{-1}$ to
${\rm Jy}^2\,{\rm sr}^{-1}[\nu I_\nu={\rm constant}]$ (Eq.~\ref{eq:cc}).
\item Compute the $\chi^2$ value between the theoretical model and the
observations,
further applying the correction $C_{\rm CIBcorr}^{\nu \times \nu^\prime}$
(Eq.~\ref{eq:corr_CMB}), and adding calibration errors, as described in the
next item.
\item The calibration uncertainties are treated differently than the CIB power
spectra error bars. We use an approach similar to the galaxy number counts
model of \cite{bethermin11}. 
A calibration factor $f_{\rm cal}$ is introduced. It has an initial value of
1, but can vary inside a Gaussian prior, centred on the calibration errors
given in Table~\ref{tab_conversion}. We add a term to the $\chi^2$ that takes
into account the estimated calibration uncertainties:
$\chi^2_{\rm cal} = \sum_{\rm bands} (f_{\rm cal}-1)^2/\sigma_{\rm cal}^2$,
where $\sigma_{\rm cal}^2$ are the calibration errors. The $C_\ell$ computed
for the model is thus modified according to
$C_{\ell, \nu \times \nu^\prime}^{\textrm cal} = f_{\nu}^{\rm cal}
 f_{\nu^\prime}^{\rm cal} C_{\ell,\nu \times \nu^\prime}^{\textrm{model}}$.
\end{enumerate}

\begin{figure*}
\rotatebox{0}{\resizebox{180mm}{!}{\includegraphics{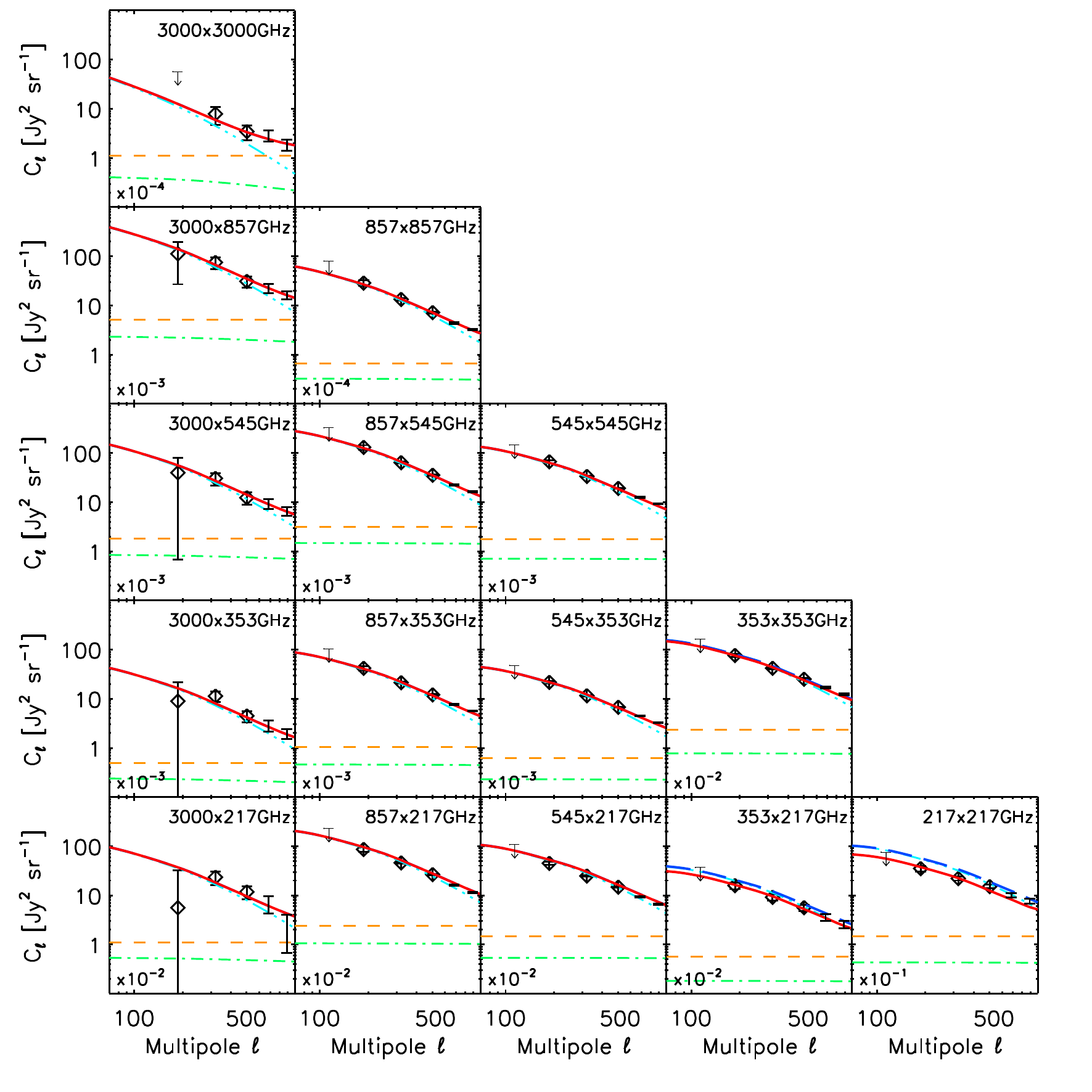}}}
\caption{(Cross-) power spectra of the CIB measured by {\it IRAS\/} and
\Planck, and the linear model. Data points are shown in black. The data used to fit
the linear model are represented by diamonds ($\ell \le 600$). High-$\ell$ points are not displayed as they are not used. The cyan dash-three-dot line (often confounded with the red continuous line) is the best-fit CIB linear model. For completeness, we also show on this figure the shot-noise level given in Table~\ref{SN_IR_jy} (orange dashed line) and
the 1-halo term predicted by \citealt{bethermin2013} (green dot-dashed line).
The red line is the sum of the linear, 1-halo and shot-noise components. It contains the spurious CIB introduced by the CMB
template (see Sect.~\ref{cmb_res}). The blue long-dashed line represents the
CIB linear best-fit model plus 1-halo and shot noise terms. It is corrected
for the CIB leakage in the CMB map, similarly to the cyan line. When the CIB leakage is negligible, the blue long-dashed line is the same as the red continuous line.}
\label{fig:clmoda}
\end{figure*}

\subsection {Constraints on SFRD and effective bias from the large-angle
linear scales}
\label{mod_lin}

\subsubsection{Fitting the linear model to the data}
\Planck\ is a unique probe of the large-scale anisotropies of the CIB.
At $\ell \lesssim 1000$, the clustering is dominated by the correlation between dark
matter halos \citep[the 2-halo term, e.g.,][]{planck2011-6.6}. \Planck\ data thus
give the opportunity to put new constraints on both star-formation history
and clustering of star-forming galaxies, using only the linear part of the
power spectra. In this modelling, we consider only the 2-halo contribution to
the cross-power spectrum between maps at frequency $\nu$ and $\nu^\prime$
(or auto-spectrum if $\nu=\nu^\prime$), which can be written as
\begin{equation}
C_{\ell,\nu,\nu^\prime}^{\rm 2h} = \int \frac{dz}{\chi^2}\frac{d\chi}{dz}
 a^2 b^2_{\rm eff}(z) \bar{j}(\nu,z) \bar{j}(\nu^{\prime},z)
 P_{\rm lin}(k=\ell/\chi,z),
\end{equation}
and we fit only for $\ell \le 600$.
Here $b_{\rm eff}$ is the effective bias of infrared galaxies at a given
redshift, i.e., the mean bias of dark matter halos hosting infrared galaxies
at a given redshift weighted by their contribution to the emissivities. This
term implicitly takes into account the fact that more massive halos are more
clustered. The link between this simple approach and the HOD approach of
Sect.~\ref{mod_hod} is discussed in Appendix~\ref{sect:effbias}. 
We compute $P_{\rm lin}(k)$ using {\tt CAMB} (\url{http://camb.info/}).

The emissivities $\bar{j}(\nu,z)$ are derived from the star formation density
$\rho_{\rm SFR}$ following (see Appendix~\ref{app_jnu})
\begin{equation}
\label{j_sfr}
\bar{j}(\nu,z) = \frac{\rho_{\rm SFR}(z) (1+z) S_{\nu,{\rm eff}}(z)
 \chi^2(z)}{K},
\end{equation}
where $K$ is the \cite{kennicutt1998} constant
(${\rm SFR}/L_{\rm IR} = 1.7\times10^{-10}$ \, ${\rm M}_\odot$\,yr$^{-1}$
for a Salpeter IMF) and $S_{\nu,{\rm eff}}(z)$ the mean effective SED of all
infrared galaxies at a given redshift. They are deduced from the
\citet{bethermin2012model} model (see Appendix~\ref{app_jnu}). These SEDs are
a mixture of secularly star-forming galaxies and starburst galaxies. The dust
temperature here increases with redshift following the measurements of
\citet{magdis2012} (see Sect.~\ref{sect:discsfh} for a discussion about the
choice of the SED library for the modelling and the impact on the results).

There are degeneracies between the evolution of the bias and of the
emissivities. In order to break them, we put some priors on the following
quantities.
\begin{itemize}
\item The local infrared luminosity density,
$\rho_{\rm SFR}(z\,{=}\,0) =
 (1.95\pm0.3)\times10^{-2}\,{\rm M}_\odot\,{\rm yr}^{-1}$ \citep{vaccari2010},
converted using the $H_0$ value measured by \textit{Planck}).
\item The local bias of infrared galaxies, $b=0.84\pm0.11$ \citep{saunders92},
converted using $\sigma_8$ measured by \Planck.
\item The mean level of the CIB deduced from galaxy number counts,
$12.6_{-1.7}^{+8.3}\,{\rm nW}\,{\rm m}^{-2}\,{\rm sr}^{-1}$ at
3000\,GHz from \cite{berta2011},
$6.5_{-1.6}^{+1.7}\,{\rm nW}\,{\rm m}^{-2}\,{\rm sr}^{-1}$ at 857\,GHz,
and $2.1_{-0.6}^{+0.7}\,{\rm nW}\,{\rm m}^{-2}\,{\rm sr}^{-1}$ at 545\,GHz
from \cite{bethermin12a}, and finally
$>0.27\,{\rm nW}\,{\rm m}^{-2}\,{\rm sr}^{-1}$ from \cite{zemcov2010} at
353\,GHz. These values are colour-corrected from PACS, SPIRE and SCUBA to
\Planck\ and {\it IRAS}, using the \cite{bethermin2012model} model.
\end{itemize}

In this simple analysis, we want to measure only two quantities: the effective
bias and its evolution with redshift, $b_{\rm eff}(z)$; and the star formation
density history, $\rho_{\rm SFR}(z)$.

Inspired by the redshift evolution of the dark matter halo bias, we chose the
following simple parametric form for the evolution of the effective bias:
\begin{equation}
b_{\rm eff}(z) = b_0 + b_1 z + b_2 z^2.
\end{equation}

For the star formation history, the values of $\rho_{\rm SFR}$ at $z=0$,
1, 2 and 4 are free parameters, and we connect these points assuming a
power law in $(1+z)$, using the two last points to extrapolate
$\rho_{\rm SFR}$ at $z>4$. We perform a Monte Carlo Markov chain analysis of
the global parameter space. We assume Gaussian uncorrelated error bars for
uncertainties, which are a linear combination of statistical and beam errors.
The calibration uncertainties are treated following the method described
Sect.~\ref{se:CIB_model}.

To be independent of the exact level of the Poisson and 1-halo power spectrum
in our linear analysis, we fit only for $\ell \leq 600$ measurements. For such $\ell$s,
contamination by the Poisson and 1-halo terms is lower than $\sim$10\%
(except for 3000$\times$3000 where it reaches $\sim$25\% at $\ell$=502, see Fig.\,\ref{fig:clmoda}). 
We nevertheless add to our model the small correction due to the 1-halo and
Poisson terms, as derived from the \cite{bethermin2013} model.

\begin{figure}
\hspace{-0.3cm}
\rotatebox{0}{\resizebox{90mm}{!}{\includegraphics{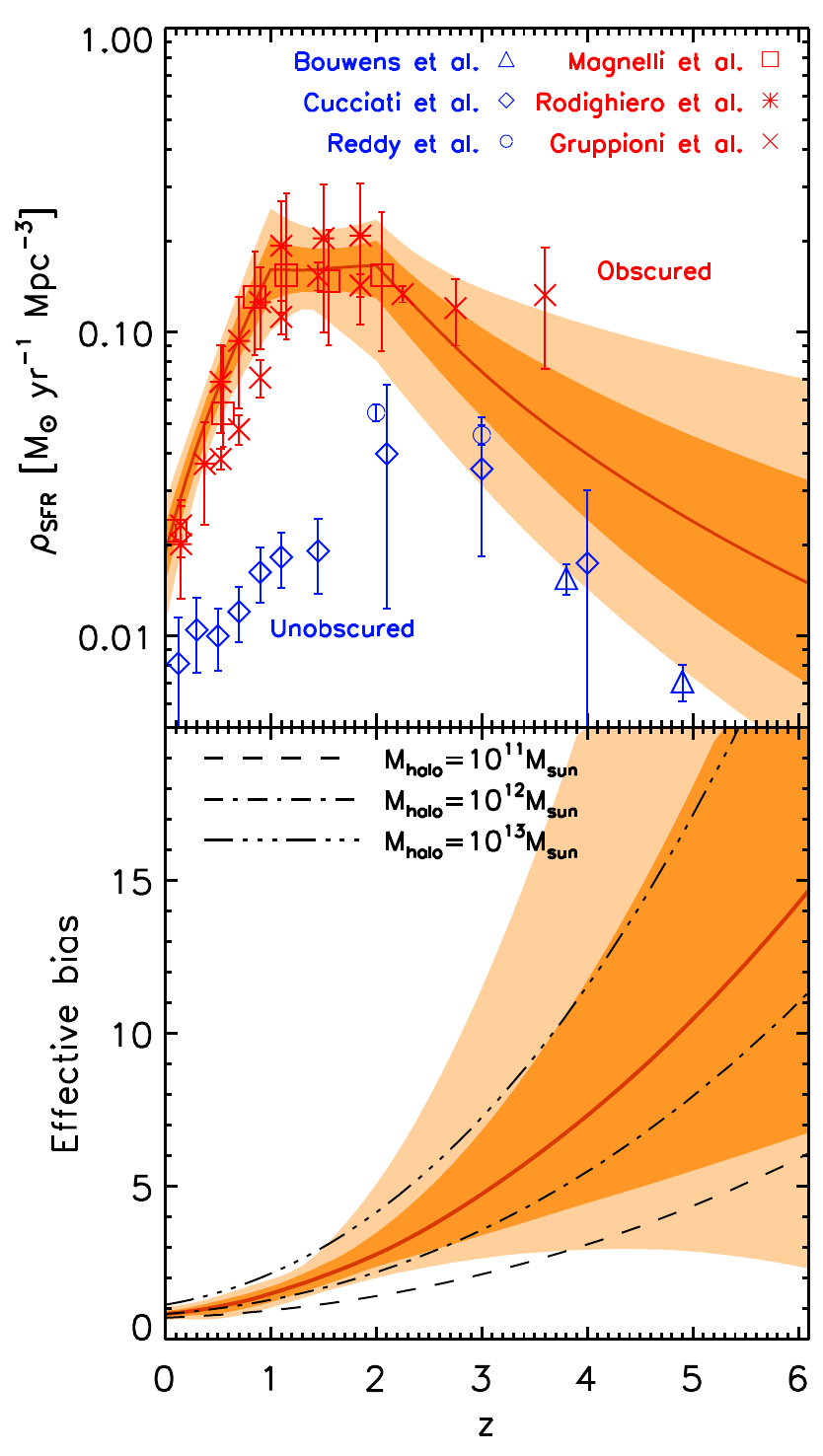}}}
\caption{Evolution of the star formation density (upper panel) and effective
bias as a function of redshift (lower panel), as constrained by the linear
part of the power spectra. In both panels, the median realization of the model
is represented with a red line, the $\pm1\,\sigma$ confidence region with a
dark orange area, and the $\pm2\,\sigma$ region with a light orange area.
In the upper panel, we added the measurements of obscured star formation from
infrared \citep[][squares]{magnelli11}, \citep[][asterisks]{rodighiero10},
\citep[][diamonds]{cucciati2012}, \citep[][crosses]{gruppioni2013}, and
unobscured star formation from uncorrected UV (\citealt{bouwens07}, triangles;
\citealt{reddy09}, circles). In the lower panel, we also plot the evolution of
the dark matter halo bias for dark matter halo mass of $10^{11}\,{\rm M}_\odot$
(dashed line), $10^{12}\,{\rm M}_\odot$ (dot-dashed line),
and $10^{13}\,{\rm M}_\odot$ (three-dots-dashed line).}
\label{fig:rhosfr_bias}
\end{figure}

\subsubsection{Results}
With a best-fit $\chi^2$ value of $35$ for $41$ degrees of freedom, we
obtain a very good fit to the data.  In Table~\ref{tab:modlin_par} we quote
median values and marginalized limits for the parameters.
The posterior value of parameters for which we imposed a Gaussian prior (local
effective bias and SFRD, plus calibration factors) are all within the
1$\,\sigma$ range of the prior values (except the 857\,GHz\ calibration factor which is at 1.2$\sigma$).

Fig.\,\ref{fig:rhosfr_bias} shows the evolution of the star formation density
with redshift (upper panel). Our derived star formation history nicely agrees
with the infrared measurements of the dust-obscured star-formation rate
density of \cite{rodighiero10} and \cite{magnelli11}, up to $z\sim2$. At
higher redshift, our determination is marginally compatible (2$\sigma$) with
\cite{gruppioni2013}, but in very good agreement with the recent work of \cite{burgarella13} at z=3.
We also compared our measurements with the UV estimate of star formation
(not corrected for dust-attenuation) from \cite{bouwens07}, \cite{cucciati2012},
and \cite{reddy09}. Below $z\sim3$, the bulk of the UV light emitted by young,
short-lived stars is reprocessed in the infrared.  Above this redshift, we
find that the star formation probed in the UV and IR regimes have roughly an
equal contribution. The infrared regime alone is thus no longer a good measure
of the total star-formation rate density.

We also studied the evolution of the effective bias (lower panel of
Fig.~\ref{fig:rhosfr_bias}). We measure an increase of the bias with redshift.
In Fig.~\ref{fig:rhosfr_bias} we compare the evolution of the galaxy dark
matter bias with that of dark matter halos of various mass (from \citealt{tinker2008}). Our results are
compatible with the track of dark matter halos with 1-3 $10^{12}\,{\rm M}_\odot$, corresponding to the halo mass of maximal efficiency of star
formation, as found in recent works
(e.g., \citealt{bethermin2012}, \citealt{wang13}, and \citealt{behroozi13},
and compatible with the related lensing magnification study of
\citealt{hildebrandt2013}).

\begin{table}[!tbh]
\begingroup
\newdimen\tblskip \tblskip=5pt
\caption{Summary of the parameters of the linear model. The values are
obtained through an MCMC analysis (median and 68\% confidence limits, CL).}
\label{tab:modlin_par}
\nointerlineskip
\vskip -3mm
\footnotesize
\setbox\tablebox=\vbox{
 \newdimen\digitwidth
 \setbox0=\hbox{\rm 0}
  \digitwidth=\wd0
  \catcode`*=\active
  \def*{\kern\digitwidth}
  \newdimen\signwidth
  \setbox0=\hbox{+}
  \signwidth=\wd0
  \catcode`!=\active
  \def!{\kern\signwidth}
\halign{\tabskip=0pt#\hfil\tabskip=1.0em&
  #\hfil\tabskip=0.1em&
  \hfil#\hfil\tabskip=0pt\cr
\noalign{\doubleline}
\noalign{\vskip -2pt}
Parameter& Definition& Median value\cr
\noalign{\vskip 3pt\hrule\vskip 3pt}
$\rho_{\rm SFR}(z\,{=}\,0)$& $z\,{=}\,0$ star formation density&
 $1.88_{-0.40}^{+0.44}\times10^{-2}\,{\rm M}_\odot\,{\rm yr}^{-1}$\cr
$\rho_{\rm SFR}(z\,{=}\,1)$& $z\,{=}\,1$ star formation density&
 $16.07_{-3.3*}^{+4.6*}\times10^{-2}\,{\rm M}_\odot\,{\rm yr}^{-1}$\cr
$\rho_{\rm SFR}(z\,{=}\,2)$& $z\,{=}\,2$ star formation density&
 $16.61_{-3.7*}^{+3.4*}\times10^{-2}\,{\rm M}_\odot\,{\rm yr}^{-1}$\cr
$\rho_{\rm SFR}(z\,{=}\,4)$& $z\,{=}\,4$ star formation density&
 $*4.0_{-1.6*}^{+2.6*}\times10^{-2}\,{\rm M}_\odot\,{\rm yr}^{-1}$\cr
\noalign{\vskip 3pt\hrule\vskip 3pt}
$b_0$& Effective bias at $z\,{=}\,0$& $0.82_{-0.10}^{+0.11}$\cr
$b_1$& First order evolution& $0.34_{-0.75}^{+0.46}$\cr
$b_2$& Second order evolution& $0.31_{-0.27}^{+0.43}$\cr
\noalign{\vskip 3pt\hrule\vskip 3pt}
$f_{3000}^{\rm cal}$& Calibration factor at 3000\,GHz&
 $1.07*_{-0.10*}^{+0.09*}$\cr
$f_{857*}^{\rm cal}$& Calibration factor at *857\,GHz&
 $1.12*_{-0.04*}^{+0.04*}$\cr
$f_{545*}^{\rm cal}$& Calibration factor at *545\,GHz&
 $1.05*_{-0.03*}^{+0.03*}$\cr
$f_{353*}^{\rm cal}$& Calibration factor at *353\,GHz&
 $1.007_{-0.014}^{+0.014}$\cr
$f_{217*}^{\rm cal}$& Calibration factor at *217\,GHz&
 $0.996_{-0.007}^{+0.008}$\cr
\noalign{\vskip 3pt\hrule\vskip 3pt}}}
\endPlancktable
\endgroup
\end{table}

\subsection{Halo model for CIB anisotropies} \label{mod_hod}
The halo model is a standard approach to describe the clustering of matter
at all scales \citep{cooray2002}. Starting from the assumption that all
galaxies live in dark matter halos, 
the clustering power spectrum can be considered as the sum of two components:
the 1-halo term (labelled $P_{\rm 1h}$), 
due to correlations of galaxies within the same halo, is responsible for the
small-scale clustering; while the 2-halo term 
($P_{\rm 2h}$), sourced by galaxy correlations in different halos, describes
the large-scale clustering.

The galaxy power spectrum is completely characterized by four main ingredients: the halo bias
between dark matter and halos;
the halo density profile, describing the spatial distribution of dark matter
inside a given halo;
the halo mass function, specifying the number density of halos with a given
mass; and a prescription for filling dark matter halos with galaxies, the
so-called Halo Occupation Distribution (HOD).

A common assumption in the simplest versions of the halo model is that all
galaxies have the same luminosity, regardless of their host dark matter
halo \citep{viero2009,amblard2011,planck2011-6.6,xia2012,viero2012}.   
However, as has already been pointed out in \cite{shang2012}, both 
galaxy clustering and galaxy luminosity are linked to host halo mass so that,
in a statistical way, 
galaxies situated in more massive halos have more stellar mass and are more luminous.
The lack of such a link between galaxy luminosity and host halo mass in the
model can lead to an interpretation of the clustering signal on small
angular scales being due to a significant overabundance 
of satellite halos (as in \citealt{amblard2011}) with respect to what is
found in numerical simulations (see discussion in \citealt{shang2012,viero2012}). However, this can instead be due to a smaller number of galaxies, but with higher luminosity.

In this paper, we assume a halo model with a galaxy luminosity-halo mass
relation similar to the one 
introduced in \cite{shang2012} and also used in \cite{viero2012}. 
We define halos as overdense regions with a mean density equal to 200 times
the mean density of the Universe and we assume an NFW profile
\citep{1997ApJ...490..493N} for the halo density profile, with a concentration 
parameter as in \cite{cooray2002}. Fitting functions of \cite{tinker2008} and the associated prescription for the
halo bias \citep[see][]{tinker2010}
will be used for the halo and sub-halo mass functions, respectively.

In the next sub-sections we will introduce the halo model that we use and we
will describe how our analysis constrains its main parameters.

\subsubsection{A halo model with luminosity dependence}
The relation between the observed flux $S_{\nu}$ and the luminosity of a
source at a comoving distance $\chi(z)$ is given by:
\begin{eqnarray}
S_{\nu}&=& \frac{(1+z)L_{\nu(1+z)}}{4\pi\chi^2(z)},
\end{eqnarray}
and the galaxy emissivity $\bar{j}_{\nu}(z)$ can be written as
\begin{eqnarray}
\label{eqn:j0}
\bar{j}_{\nu}(z)&=&\int dL \frac{dn}{dL}(L,z)
\frac{L_{(1+z)\nu}}{4\pi},
\label{eq_jnu}
\end{eqnarray}
where $dn/dL$ denotes the infrared galaxy luminosity function.

In general, in order to model the galaxy luminosity--halo mass relation,
we should introduce a scatter describing the probability density $P(L|M)$
for a halo (or a sub-halo) of mass $M$ to host a galaxy with luminosity $L$
(as in the conditional luminosity function models of, e.g.,
\citealt{yang2003}, \citealt{yang2005},  \citealt{cooray2005},
\citealt{cooray2006},  \citealt{amblard2007}, and  \citealt{db2012}).
In order to keep the analysis as simple as possible, we neglect any scatter
and introduce $L_{{\rm cen},\nu(1+z)}(M_{\rm H},z)$
(for central galaxies) and $L_{{\rm sat},\nu(1+z)}(m_{\rm SH},z)$
(for satellite galaxies), where $M_{\rm H}$ and $m_{\rm SH}$ denote the halo
and sub-halo masses, respectively, 
Eq.~\ref{eq_jnu} can be re-written as:
\begin{eqnarray}
\label{eqn:j1}
\bar{j}_{\nu}(z)&=& \int dM \frac{dN}{dM}(z)\frac{1}{4\pi}
 \Big\{\frac{}{}N_{\rm cen}L_{{\rm cen},(1+z)\nu}(M_{\rm H},z)\\
\nonumber 
& &+\int dm_{\rm SH} \frac{dn}{dm}(m_{\rm SH},z)
 L_{{\rm sat},(1+z)\nu}(m_{\rm SH},z)\Big\},
\end{eqnarray}
where $dn/dm$ denotes the sub-halo mass function and $N_{\rm cen}$ is the
number of central galaxies inside a halo.

Introducing $f^{\rm cen}_{\nu}$ and $f_{\nu}^{\rm sat}$ for central and
satellite galaxies,
\begin{eqnarray}
f_{\nu}^{\rm cen}(M,z) = N_{\rm cen}
 \frac{L_{{\rm cen},(1+z)\nu}(M_{\rm H},z)}{4\pi},
\label{eqn:fcen}
\end{eqnarray}
\begin{eqnarray}
f_{\nu}^{\rm sat}(M,z) &=& \int_{M_{\rm min}}^{M}dm
 \frac{dn_{}}{dm}(m_{\rm SH},z|M) \\\nonumber 
&& \times \frac{L_{{\rm sat},(1+z)\nu}(m_{\rm SH},z)}{4\pi},
\label{eqn:fsat}
\end{eqnarray}
then the power spectrum of CIB anisotropies at observed frequencies
$\nu,\nu^\prime$ can be written as
\begin{eqnarray}
\label{eqn:pj1h}
P^{}_{{\rm 1h},\nu\nu^{\prime}}(k,z)&=&
 \frac{1}{\bar{j}_{\nu}\bar{j}_{\nu^\prime}}\int_{M_{\rm min}}^{\infty}dM
 \frac{dN}{dM}\\\nonumber
&&\times\, \left\{f_{\nu}^{\rm cen}(M,z)f_{\nu^\prime}^{\rm sat}(M,z)u(k,M,z)
\right. \\\nonumber
&&\quad +f_{\nu^{\prime}}^{\rm cen}(M,z)f_{\nu}^{\rm sat}(M,z)
 u(k,M,z)\\\nonumber
&&\left.\quad +f_{\nu}^{\rm sat}(M,z)f_{\nu^{\prime}}^{\rm sat}(M,z)
 u^2(k,M,z)\right\},\\
\label{eqn:pj2h}
P^{}_{{\rm 2h},\nu\nu^{\prime}}(k,z)&=&
 \frac{1}{\bar{j}_{\nu}\bar{j}_{\nu^\prime}}D_{\nu}(k,z)
 D_{\nu^{\prime}}(k,z)P_{\rm lin}(k,z).
\end{eqnarray}
Here
\begin{eqnarray}
D_{\nu}(k,z)&=&\int_{M_{\rm min}}^{\infty}dM\frac{dN}{dM}b(M,z)u(k,M,z)\\
\nonumber
&&\times\, \left\{f_{\nu}^{\rm cen}(M,z)+f_{\nu}^{\rm sat}(M,z)\right\},
\label{eqn:pjf}
\end{eqnarray}
with $u(k,M,z)$ being the Fourier transform of the halo density profile.

\subsubsection{Parameterizing the $L$--$M$ relation  \label{subsec:L-M}}
In the simplest version of the halo model, where galaxies residing in halos
of different masses have the same luminosity, 
the galaxy power spectrum is fully determined by the HOD, namely the function
describing the number of central and satellite galaxies in each dark matter
halo.  In the model used here, the power spectrum depends, additionally, on
the function $L_{(1+z)\nu}(M_{\rm H},z)$, where $M$ denotes the halo mass.
The luminosity $L_{(1+z)\nu}(M_{\rm H},z)$ depends on three variables: the
redshift $z$; the mass of the host (sub)halo; and the observing frequency
$\nu$.  We will consider the following assumptions about the structure of
the luminosity-mass relation $L(M)$.
\begin{itemize}
\item We assume no difference between halos and sub-halos
with the same mass, so that 
$L(M_{\rm H},z) = L(m_{\rm SH},z)$, for $M_{\rm H} = m_{\rm SH}$. 
While recent studies \citep[e.g.,][]{rodriguez2012,rodriguez2013}
show some indication that satellite galaxies tend to have slightly more
stellar mass than central galaxies with the same halo mass, these results
depend on the subhalo mass definition used; in particular, 
the luminosity-mass relation for satellites and central galaxies has been
found to be not very different when the mass of the subhalo 
is defined at the time of accretion (as done in this paper). 

\item A very simple functional form
\citep[see][and reference therein]{blain2002} is assumed for galaxy SEDs:
\begin{eqnarray}
\Theta (\nu,z) \propto
\left\{\begin{array}{ccc}
\nu^{\beta}B_{\nu}\,(T_{\rm d}(z))&
 \nu<\nu_0\, ;\\
\nu^{-\gamma}&  \nu\ge \nu_0\, .
\end{array}\right.
\label{eqn:thetanu}
\end{eqnarray}
Here $B_{\nu}$ denotes the Planck function, while the emissivity index
$\beta$ gives information about the physical 
nature of dust and in general depends on grain composition,
temperature distribution of tunnelling states and wavelength-dependent
excitation \citep[e.g.,][]{meny2007}. The power-law function is used to
temper the exponential Wien tail 
at high frequencies and obtain a shallower SED shape, more in agreement
with observations.

The temperature is assumed to be a function of redshift according to
\begin{eqnarray}
T_{\rm d} \equiv T_0(1+z)^{\alpha}.
\label{eq:T_z}
\end{eqnarray}
This dependence of the temperature with redshift can be due to different
physical processes, such as more compact geometries for galaxies at high
redshift \citep{magdis2012}, a global evolution of the SED
\citep[e.g.,][]{addison2012,bethermin2013} or the increase of the CMB
temperature with redshift \citep{blain1999}.

The SED functions at high and low frequencies are connected smoothly at the
frequency $\nu_0$ satisfying
\begin{eqnarray}
\frac{d{\rm ln}\Theta(\nu,z)}{d\rm{ln}\nu} =-\gamma.
\end{eqnarray}
The range of variation for the parameters $\alpha$, $\gamma$, and $\nu_\ast$
is large enough to ensure that we do not exclude non-negligible regions of the
multidimensional parameter space; however we assume physically motivated
priors for both the temperature ($T_0$ in the range 20--60\,K, see measurements
in e.g., \citealt{dunne2000}, \citealt{chapman2005}, \citealt{amblard2010},
and \citealt{hwang2010}) and the emissivity index ($\beta$ in the range
1.5--2.0). The correct choice of $\beta$ is a matter of debate; measurements
of Milky Way dust, and in external galaxies \citep[e.g.,][]{boselli2012},
give values in the range 1--2, but allowing for some degree of correlation
between dust temperature and emissivity index
\citep[see, e.g.,][]{paradis2010} it is possible to obtain $\beta >2$ for
low dust temperatures ($T_{\rm d} \leq 18$K). 
On the theoretical side, while models for both insulating and conducting
materials naturally give $\beta = 2$ at long wavelengths
\citep[e.g.,][]{draine1984}, significant deviations from the value
$\beta=2$ occur when accounting for the disordered structure of the amorphous
dust grains \citep{meny2007}. Indeed, some authors \citep{shang2012,viero2012}
allow for values $\beta > 2$ when fitting CIB data. In this analysis we
prefer to be conservative and, since we assume the condition $T_{\rm d}>20\,$K,
we also impose $\beta \le 2$; this will allow us to draw solid conclusions on
the other parameters of the model, avoiding regions of the parameter space
whose physical interpretation is questionable. 
\item We assume a redshift-dependent, global normalization of the $L$--$M$
relation of the form
\begin{eqnarray}
\Phi (z)= \left(1+z\right)^{\delta}.
\label{eqn:phiz}
\end{eqnarray}
The parameter $\delta$ will be allowed to vary in the range 0--7.
Such a redshift dependence can be justified considering the evolution of the specific far infrared luminosity ($L_{\rm IR}/M_\ast$) with redshift:
if the ratio of stellar mass to halo mass evolves only mildly with redshift
\citep[see e.g.,][]{neistein2011}, then the ratio $L_{\rm IR}/M_{\rm H}$
should evolve approximately as the specific infrared luminosity. The
semi-analytic galaxy formation model of \cite{delucia2007} shows the evolution
of such a quantity with redshift as a power law with a slope of about 2.5,
while observations performed by \cite{oliver2010} indicate a much steeper
slope, around 4.4.   
\item We assume a log-normal function for the dependence of the galaxy 
luminosity on halo mass:
\begin{eqnarray}
\Sigma (M,z)= M \frac{1}{(2\pi \sigma_{L/M}^2)^{1/2}}\,
e^{-(\log_{10}(M)-\log_{10}(M_{\rm eff}))^2/
 2\sigma_{L/M}^2}.
\label{eqn:sigmam}
\end{eqnarray}
Here $M_{\rm eff}$ describes the peak of the specific IR emissivity, while the
parameter $\sigma_{L/M}$
describes the range of halo masses used for producing the IR luminosity;
we will assume that $\sigma^2_{L/M}=0.5$ throughout this paper and 
we checked that results do not significantly change when assuming
$\sigma^2_{L/M}=0.65$, as in \cite{bethermin2012}. 

The reason for choosing a log-normal functional form is that star formation
is active only over a given range of halo masses, being suppressed at both
the low- and the high-mass end by mechanisms such as photoionization,
supernovae heating, feedback from active galactic nuclei and virial shocks
\citep[see e.g.,][]{benson2003,croton2006}; it is then possible to identify a
peak in the $L$--$M$ relation, which describes the maximum in the average
infrared emissivity per unit mass.
\item At the low--mass end, we assume a minimum mass $M_{\rm min}$, which is
a free parameter in the range $10^{10}$--$10^{11}\,\msun$,
and we assume $L=0$ for $M<M_{\rm min}$. 
\end{itemize}
The equation for the luminosity-mass relation can finally be written as
\begin{eqnarray}
L_{(1+z)\nu}(M,z)=L_0 \Phi(z) \Sigma(M,z) \Theta[(1+z)\nu],
\label{eqn:lfunc}
\end{eqnarray}
where $L_0$ is a free normalization parameter (which being not physically
meaningful will not be further discussed).

\begin{figure*}
\includegraphics[width=19cm]{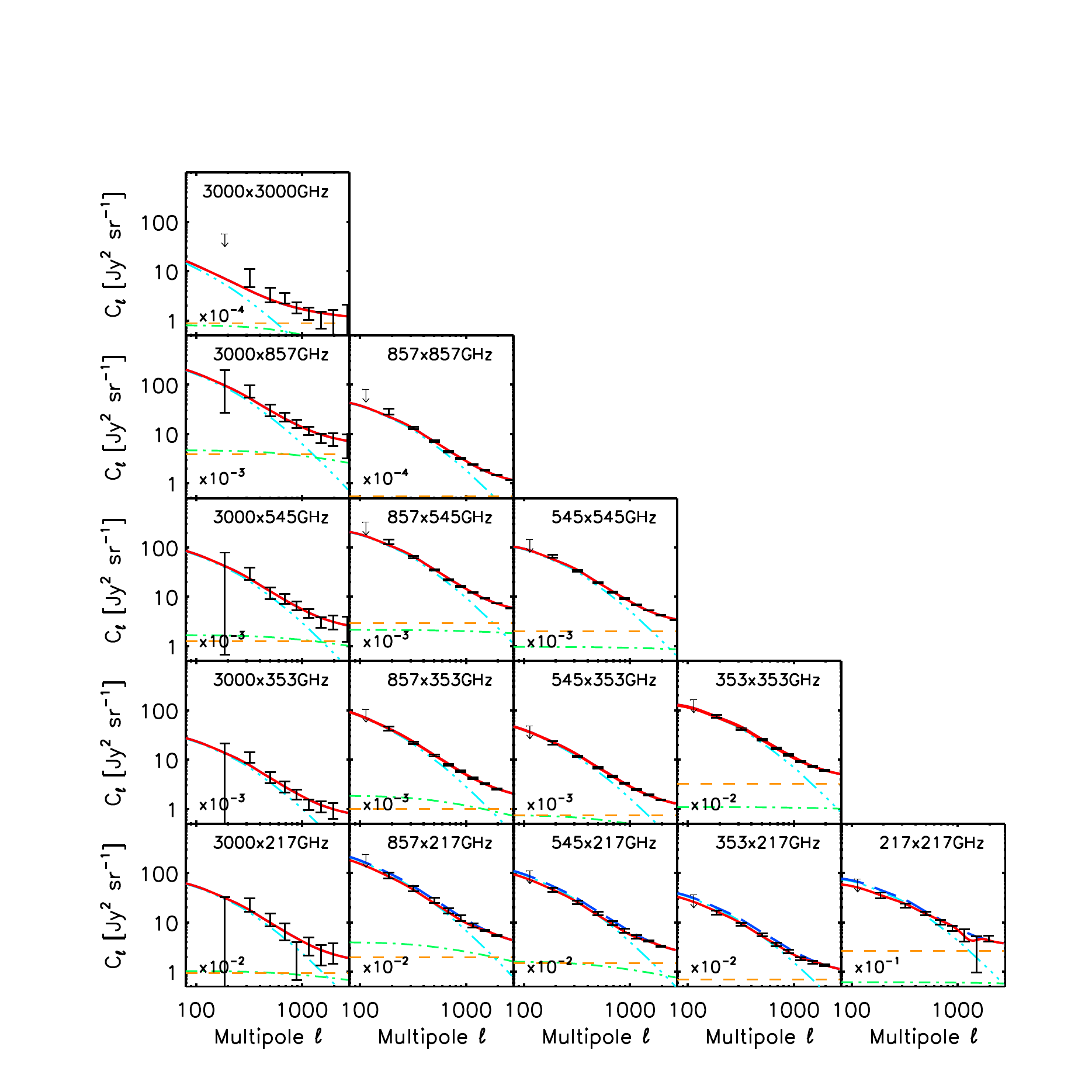}
\caption{(Cross-) power spectra of the CIB anisotropies measured by \Planck and {\it IRAS},
compared with the best-fit extended halo model. Data points are shown in black.
The red line is the sum of the linear, 1-halo and shot-noise components, which
is fitted to the data. It contains the spurious CIB introduced by the CMB
template (see Sect.~\ref{cmb_res}). The orange dashed, green dot-dashed, and cyan three-dots-dashed lines are the best-fit shot-noise level, the 1-halo and the 2-halo terms, respectively. They are corrected for the CIB leakage in the CMB. The sum of the three is the blue long-dashed line. When the CIB leakage is negligible, the blue long-dashed line is the same as the red continuous line.
\label{fig:modmes_halomod}}
\end{figure*}

\subsubsection{Method and data used}
In order to constrain the main parameters of our model, we fit for a total of $121$ data points of the $15$ possible combinations of \Planck\ auto- and
cross-power spectra at $217$, $353$, $545$, $857$ and $3000$\,\GHz, considering
the multipole range $187 \leq \ell \leq 2649$. We use the same procedure as
described in Sect.~\ref{mod_fit} in order to colour-correct our model to the
photometric convention $\nu I_{\nu}={\rm constant}$ and to include the
corrections due to CIB over-subtraction and SZ-related residuals
(see points 1--6 in Sect.~\ref{mod_fit}). There are two main differences
with respect to the linear model analysis outlined above (Sect.~\ref{mod_lin}):
\begin{itemize}
\item we keep all the calibration parameters fixed at $f^{\rm cal}=1$,
which assumption is justified by the analysis using the linear model,
allowing us not to deal with too many parameters;
\item we assume the same prior on the star formation rate density \citep{vaccari2010} as in the linear model but we do not use any constraints on the bias at redshift zero. We also assume flat priors on the mean level of the CIB at 545 and 857\,\GHz.
\end{itemize}
We perform a Monte Carlo Markov chain analysis of the global parameter space
using a modification of the publicly available code {\tt CosmoMC}
\citep{lewis2002}. 
We consider variations in the following set of eight halo model parameters:
\begin{eqnarray}
\mathscr{P} \equiv \{\alpha,\beta,\gamma,\delta,
 M_{\rm eff},M_{\rm min},T_0,L_0 \}.
\end{eqnarray} 
We assume the shot-noise levels given by the sum of the values quoted in
Tables~\ref{SN_IR_jy} and \ref{SN_rad_jy}, from \cite{bethermin2012model}
and \cite{tucci11}, respectively, and we assume flat priors around them
with width given by their 1$\sigma$ error.

\begin{figure}
\hspace{-0.3cm}
\rotatebox{0}{\resizebox{90mm}{!}{\includegraphics{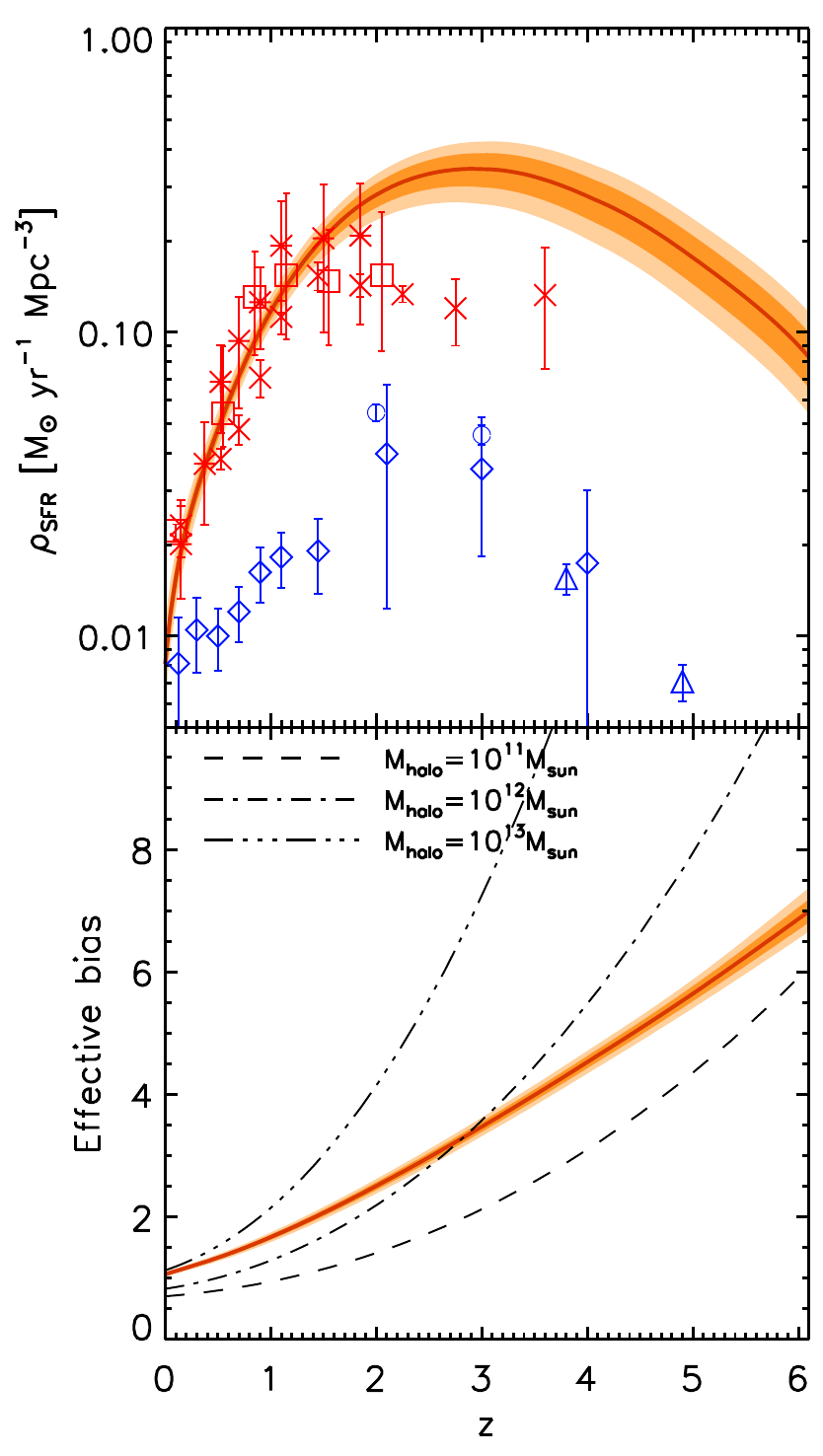}}}
\caption{Evolution of the star formation density (upper panel) and effective
bias as a function of redshift (lower panel), as constrained from our extended
halo model.  In both panels, the median realization of the model is
represented with a red line, the $\pm1\,\sigma$ confidence region in dark
orange, and the $\pm2\,\sigma$ region in light orange. In the upper panel
the reported data are the same as in Fig.~\ref{fig:rhosfr_bias}. In the lower
panel, we also plot the evolution of the dark matter-halo bias for dark matter
halo masses of $10^{11}\,{\rm M}_\odot$ (dashed line), $10^{12}\,{\rm M}_\odot$
(dot-dashed line), and $10^{13}\,{\rm M}_\odot$ (three-dots-dashed line).}
\label{fig:SFR2}
\end{figure}

The total number of free parameters in our analysis is then 23, consisting of
the sum of eight halo model parameters plus 15 shot-noise parameters. While it
is tempting to fix the shot-noise power spectra to their theoretically
modelled values (in order not to deal with too many parameters and keep the
analysis as simple as possible), we believe that, since these values are not
very tightly constrained by their underlying models, it is better to let them
vary as free parameters around their best estimates.

\begin{table*}[!tbh]
\begingroup
\newdimen\tblskip \tblskip=5pt
\caption{Mean values and marginalized $68\%$ CL for halo model
parameters and shot-noise levels (in $\mathrm{Jy^2 sr^{-1}}$). }
\label{HOD_param_values}
\nointerlineskip
\vskip -3mm
\footnotesize
\setbox\tablebox=\vbox{
 \newdimen\digitwidth
 \setbox0=\hbox{\rm 0}
  \digitwidth=\wd0
  \catcode`*=\active
  \def*{\kern\digitwidth}
  \newdimen\signwidth
  \setbox0=\hbox{+}
  \signwidth=\wd0
  \catcode`!=\active
  \def!{\kern\signwidth}
\halign{\tabskip=0pt#\hfil\tabskip=1.0em&
  #\hfil\tabskip=1.0em&
  \hfil#\hfil\tabskip=0pt\cr
\noalign{\doubleline}
\noalign{\vskip -2pt}
Parameter& Definition& Mean value\cr
\noalign{\vskip 3pt\hrule\vskip 3pt}
$\alpha$& SED: redshift evolution of the dust temperature&
 $*0.36\pm0.05$\cr
$T_0$ [K]& SED: dust temperature at $z=0$& $24.4 \pm1.9$ \cr
$\beta$& SED: emissivity index at low frequency&
 $*1.75\pm0.06$\cr
$\gamma$& SED: frequency power law index at high frequency&
 $*1.7*\pm0.2*$\cr
$\delta$& Redshift evolution of the normalization of the $L$--$M$ relation&
 $*3.6*\pm0.2*$\cr
$\log(M_{\rm eff}) [{\rm M}_{\odot}]$& Halo model most efficient mass&
 $12.6\pm0.1*$\cr
$M_{\rm min} [{\rm M}_{\odot}]$& Minimum halo mass& unconstrained\cr
\noalign{\vskip 2pt}
$S^{3000\times3000}$& Shot noise for 3000\GHz\ $\times$ 3000\GHz& $9585\pm1090$\cr
$S^{3000\times857}$& Shot noise for 3000\GHz\ $\times$ 857\GHz& $4158\pm443$\cr
$S^{3000\times545}$& Shot noise for 3000\GHz\ $\times$ 545\GHz& $1449\pm176$\cr
$S^{3000\times353}$& Shot noise for 3000\GHz\ $\times$ 353\GHz& $411\pm48$\cr
$S^{3000\times217}$& Shot noise for 3000\GHz\ $\times$ 217\GHz& $95\pm11$\cr
$S^{857\times857}$& Shot noise for 857\GHz\ $\times$ 857\GHz& $5364\pm343$\cr
$S^{857\times545}$& Shot noise for 857\GHz\ $\times$ 545\GHz& $2702\pm*124$\cr
$S^{857\times353}$& Shot noise for 857\GHz\ $\times$ 353\GHz& $*953\pm*54$\cr
$S^{857\times217}$& Shot noise for 857\GHz\ $\times$ 217\GHz& $*181\pm**6$\cr
$S^{545\times545}$& Shot noise for 545\GHz\ $\times$ 545\GHz& $1690\pm*45$\cr
$S^{545\times353}$& Shot noise for 545\GHz\ $\times$ 353\GHz& $*626\pm*19$\cr
$S^{545\times217}$& Shot noise for 545\GHz\ $\times$ 217\GHz& $*121\pm**6$\cr
$S^{353\times353}$& Shot noise for 353\GHz\ $\times$ 353\GHz& $*262\pm**8$\cr
$S^{353\times217}$& Shot noise for 353\GHz\ $\times$ 217\GHz& $**54\pm**3$\cr
$S^{217\times217}$& Shot noise for 217\GHz\ $\times$ 217\GHz& $**21\pm**2$\cr
\noalign{\vskip 3pt\hrule\vskip 3pt}}}
\endPlancktablewide
\endgroup
\end{table*}

\subsubsection{Results \label{se:results_hod}}
With a best-fit $\chi^2 $ of $100.7$ and $98$ degrees of
freedom, we obtain a remarkably good fit to the data.
In Table~\ref{HOD_param_values} we quote mean values and marginalized limits
for the model parameters. In the following, we comment on the results obtained
for some parameters of the model and for some derived quantities.

\paragraph{Peak mass $M_{\rm eff}$ --}
The mean value of the most efficient halo mass for generating the CIB,
$\log(M_{\rm eff}/\msun)=12.6 \pm 0.1$, is in good agreement with   
results obtained from a similar analysis using {\it Herschel\/}
CIB data at $250$, $350$ and $500\,\mu$m \citep{viero2012}, and with 
other analyses, using previous \Planck\ and {\it Herschel\/} data \citep[e.g.,][]{shang2012,xia2012}, 
while it is slightly higher than results from other observations and simulations
\citep[e.g.,][]{moster2010,behroozi2012,bethermin2012,wang13}. We also checked for a possible
redshift evolution of $M_{\rm eff}$ (which can be justified in the framework
of the so-called ``downsizing'' idea), performing an MCMC run with the
functional form
\begin{equation}
M_{\rm eff} = M_0 \times (1+z)^q,
\end{equation}
The large degeneracy between $M_0$ and $q$ leads to very high values of $M_0$. The bias and SFRD have the same redshift evolution as in the case $q=0$, but with much larger error bars (they are multiplied by a factor of 6 for example for the bias).

\paragraph{Constraints on the dust temperature --}
Parameterizing the average dust temperature of sources as 
\begin{equation}
T_{\rm d}(z) = T_0(1+z)^{\alpha},
\end{equation}
the data suggest a redshift evolution of the temperature, with
$T_0 = (24.4\pm1.9)\,$K and $\alpha=0.36\pm 0.05$. 
Such a trend, implying some kind of SED evolution, has been also found in
e.g., \cite{addison2012,viero2012}. Experimental results from different surveys appear to
have been quite contradictory, with systematics playing a critical role
\citep[e.g.,][]{chapman2005,coppin2008,pascale2009,amblard2010,hwang2010,
chapin2011}.
But recently, some consensus has emerged on a scenario with an increase of
dust temperature with redshift \citep{magdis2012, viero13}. The increase of
temperature may be explained by a harder interstellar radiation field at
earlier times \citep[see][for a detailed discussion]{magdis2012}.

\paragraph{Constraints on the bias --}
Galaxies are considered as a biased tracer of the dark matter field. The
galaxy overdensity  $\delta_{\rm g}(k,z)$ is assumed to trace the underlying
dark matter field $\delta_{\rm dm}(k,z)$ via
\begin{equation}
\delta_{\rm g}(k,z) = b(k,z)\delta_{\rm dm}(k,z),
\end{equation}
where $b(k,z)$ is the galaxy bias, which in general can depend not only on
scale and redshift but also on luminosity, spectral type and colour. 
On large scales, the bias is generally assumed to be scale-independent;
however, both numerical simulations \citep{kauffmann1999} and recent results
from galaxy-galaxy lensing and galaxy clustering also indicate an increase of
the bias with redshift \citep[e.g.,][]{mandelbaum2012}, 
while \cite{tegmark1998} show that the bias must be close to unity when
approaching $z=0$.     
The combination of CMB and large-scale clustering data yields a bias parameter
$b\sim 1$ \citep{verde2002} while \cite{saunders92} found
$b\sigma_8=0.84\pm 0.11$ for {\it IRAS\/} galaxies, which, assuming
$\sigma_8=0.8$, gives $b\sim0.86$. 
In Fig.~\ref{fig:rhosfr_bias} we show our estimate of the redshift dependent
bias; it is remarkable that, without assuming any prior on the value of the
bias at redshift zero, we are able to obtain a very good fit to observations,
with $b(z=0) = 1.1 \pm 0.02$. 

\begin{figure}
\hspace{-0.5cm}
\includegraphics[width=9.5cm]{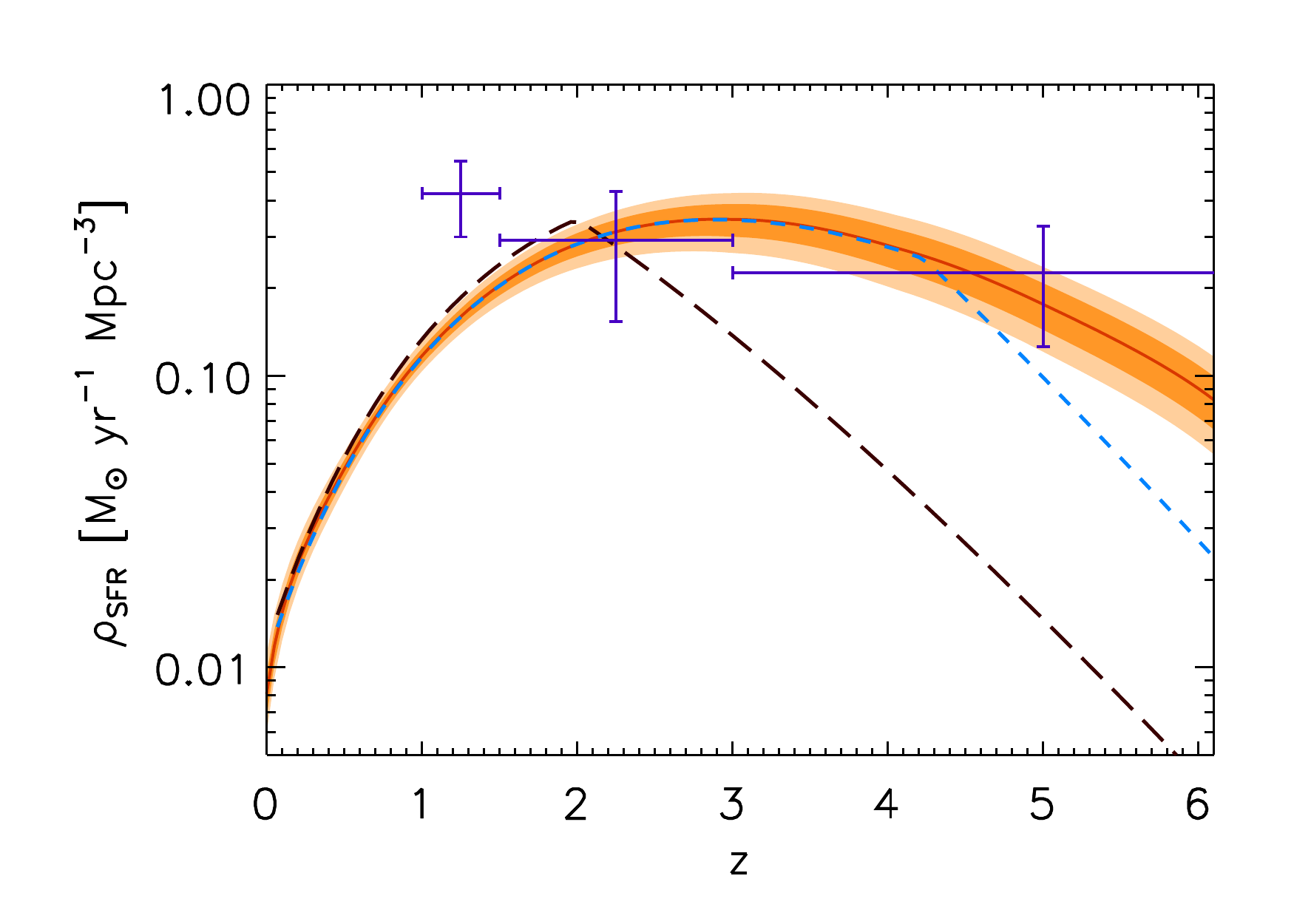}
\caption{\label{fig:SFR_EHM}
Marginalized constraints on the star formation rate density, as derived from
our extended halo model
described in Sect.~\ref{mod_hod} (red continuous line with $\pm1$ and 
$\pm2\,\sigma$ orange dashed areas). It is compared with mean values computed 
imposing the condition $\delta(z\,{\ge}\,2)=0$ (black long-dashed line), or the combined
conditions $\delta(z\,{\ge}\,z_{\rm break})=0$ and $T(z\,{=}\,z_{\rm break})
 = T(z_{\rm break})$, where $z_{\rm break}$ is found to be $4.2\pm0.5$ (blue dashed
line). The violet points with error bars are the SFR density determined from
the modelling of the CIB-CMB Lensing cross correlation by
\cite{planck2013-p13}.} 
\end{figure}

\paragraph{Constraints on star formation history --}
The mean value and $68\%$ CL bounds on the cosmic star formation rate 
density $\rho_{\rm SFR}$ are plotted in Fig.~\ref{fig:rhosfr_bias}.
The parameter $\rho_{\rm SFR}$ has been computed following Eq.~\ref{j_sfr},
replacing $s_{\nu,{\rm eff}}(z)$ by the halo model SED as given in
Eq.~\ref{eqn:thetanu}.
We use the \cite{kennicutt1998} constant to convert infrared luminosity
to star formation rate (${\rm SFR}/L_{\rm IR}
 = 1.7\times10^{-10}\,{\rm M}_\odot\,{\rm yr}^{-1}$ for a Salpeter IMF).
As can be seen from Fig.~\ref{fig:rhosfr_bias}, the star formation rate densities predicted by
both models used in this paper are in very good agreement for redshifts
$z \lesssim 2$ while there is a significant difference at higher redshifts. 
Interestingly, \cite{behroozi2012} reports a compilation of results,
showing how measurements performed before 2006 predict quite a high value
for the high redshift SFRD (then more compatible with
the halo model results, see also the compilation in \citealt{hopkins2006}),
while measurements performed after 2006 and obtained with different assumptions
about the  dust present at $z>3$, show a rapid {\it decrease\/} of the SFRD, which then becomes more compatible with results
obtained using the linear model. In an attempt to reproduce the break in the
SFRD seen in Fig.~\ref{fig:SFR_EHM} for the linear
model, we also imposed the condition $\delta=0$ at redshift $z=2$ in the
redshift normalization parameter $\Phi(z)$ of the $L$--$M$ relation.
Such a condition has already been considered in \cite{shang2012} and, while
it is motivated by some observations, it is also hard to explain from a
theoretical point of view \citep[e.g.,][]{weinmann2011}. In this case, we are
able to obtain lower values of the SFRD at high
redshifts (see Fig.~\ref{fig:SFR2}), but at the price of slightly degrading
the quality of the fit. 
We finally note that another potential reason for the discrepancy found at
high redshift can be the difference in the inferred bias evolution between
the linear model and the halo model: in fact, 
a lower value of the bias at high redshift (as found in the context of the
halo model) can be compensated by higher values for the SFRD in the same redshift range. 
We will come back to the SFRD discussion in Sect.~\ref{sect:discsfh}.\\

Finally, we are also able to determine the mean CIB intensity at the \Planck\
frequencies considered in the analysis.  The values obtained are presented
in Table~\ref{Table:CIBintensity}.
Although higher, they are compatible, within the $95\%$ CL, with results obtained from number
counts measurements \citep{bethermin12a}.

\begin{table}[!tbh]
\begingroup
\newdimen\tblskip \tblskip=5pt
\caption{Derived estimates of the CIB intensity from the extended halo model.}
\label{Table:CIBintensity}
\nointerlineskip
\vskip -3mm
\footnotesize
\setbox\tablebox=\vbox{
 \newdimen\digitwidth
 \setbox0=\hbox{\rm 0}
  \digitwidth=\wd0
  \catcode`*=\active
  \def*{\kern\digitwidth}
  \newdimen\signwidth
  \setbox0=\hbox{+}
  \signwidth=\wd0
  \catcode`!=\active
  \def!{\kern\signwidth}
\halign{\tabskip=0pt#\hfil\tabskip=1.0em&
  \hfil#\hfil\tabskip=0pt\cr
\noalign{\doubleline}
\noalign{\vskip -2pt}
Band & $\nu I_\nu$ [${\rm nW}\,{\rm m}^{-2}\,{\rm sr}^{-1}$]\cr
\noalign{\vskip 3pt\hrule\vskip 3pt}
3000\GHz& $13.1**\pm1.0**$\cr
857\GHz& $7.7**\pm0.2**$\cr
545\GHz& $2.3*\pm0.1*$\cr
353\GHz& $0.53*\pm0.02*$\cr
217\GHz& $0.077\pm0.003$\cr
\noalign{\vskip 3pt\hrule\vskip 3pt}}}
\endPlancktable
\endgroup
\end{table}

\subsection{CIB-CMB lensing cross-correlation}

We tested the validity of our approach by comparing the predictions for
our best-fit models with the measurements of the cross-correlation between the
CIB and the CMB lensing potential presented in \cite{planck2013-p13}.
For the linear model, we computed the cross-correlation following:
\begin{equation}
C_\ell^{\nu \phi} = \int b_{\rm eff}(z) \bar{j}(\nu,z) \frac{3}{\ell^2}
 \Omega_{rm m}, H_0^2 \left( \frac{\chi_\ast - \chi}{\chi_\ast \chi} \right)
 P_{\rm lin}(k=\ell/\chi,z) d\chi,
\end{equation}
where $\chi_\ast$ is the comoving distance to the CMB last-scattering surface.
We use a similar equation for the extended halo model.

\begin{figure}[!tbh]
{\includegraphics[width=8.5cm]{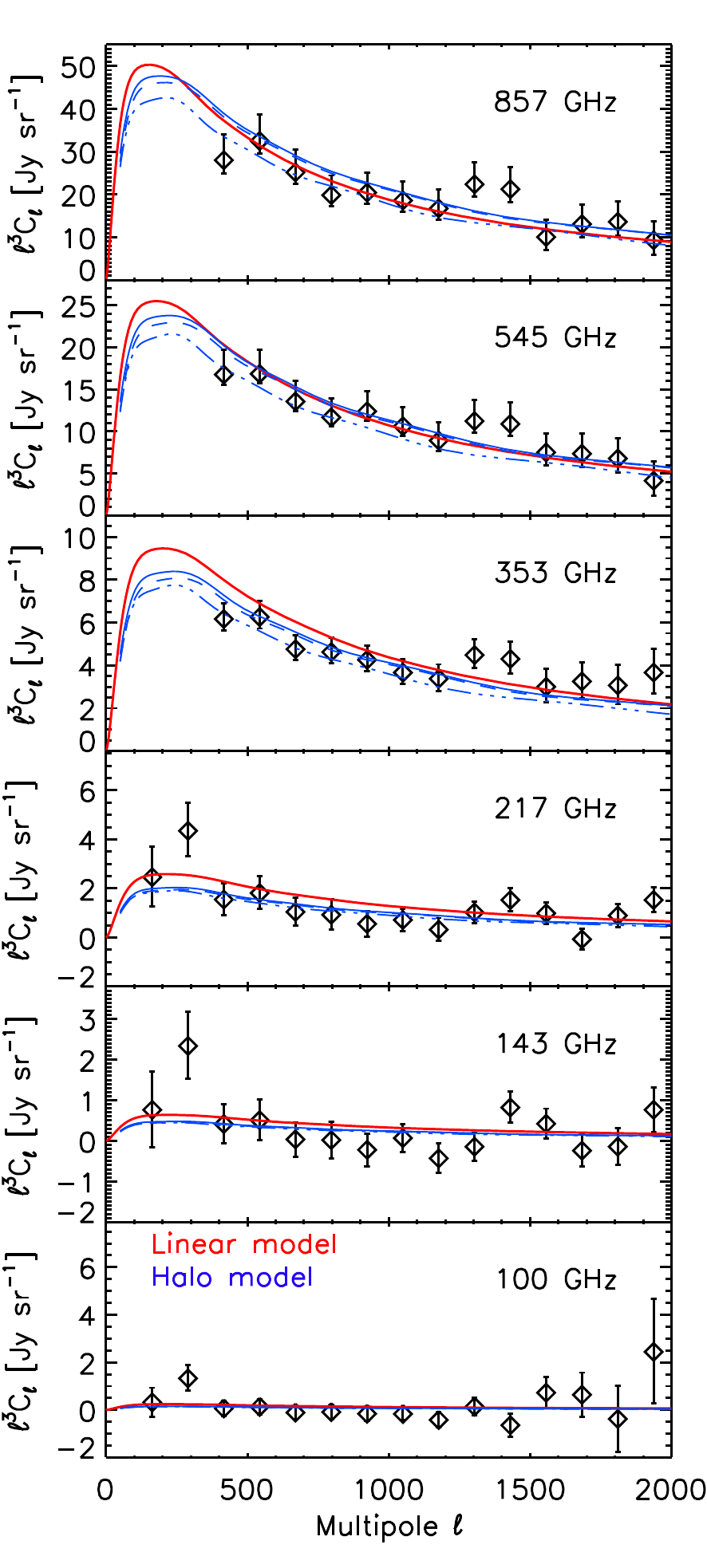}}
\caption{Comparison between the measurements of the CIB and gravitational
potential cross-correlation given in \cite{planck2013-p13} (diamonds),
with the predictions from our best-fit models of the CIB cross-power spectra
(red and blue solid lines for the linear and extended halo model,
respectively). The other curves are the two variants of the extended halo
model with: (i) a break in the global normalization of the $L$--$M$ relation
fixed at redshift $z=2$ (blue 3-dot-dashed curve); and (ii) a break in both the
temperature evolution and normalization of the $L$--$M$ relation, found at
redshift $z=4.2\pm0.5$ (blue long-dashed curve).}
\label{fig:modA_cibxlensing}
\end{figure}

Figure\,\ref{fig:modA_cibxlensing} shows a comparison between the model and
the data. The halo model (as well as its variant with a break in the
temperature and global normalization of the $L$--$M$ relation at redshift
$z\sim4$, see Sect.~\ref{sect:discsfh}) agrees remarkably well with the
measurements for all channels. The linear model gives a higher
prediction at 217 and 353\,\GHz\ (although compatible with the data points at the 1$\,\sigma$
level). 

\begin{figure}
\begin{center}
{\includegraphics[width=9.5cm]{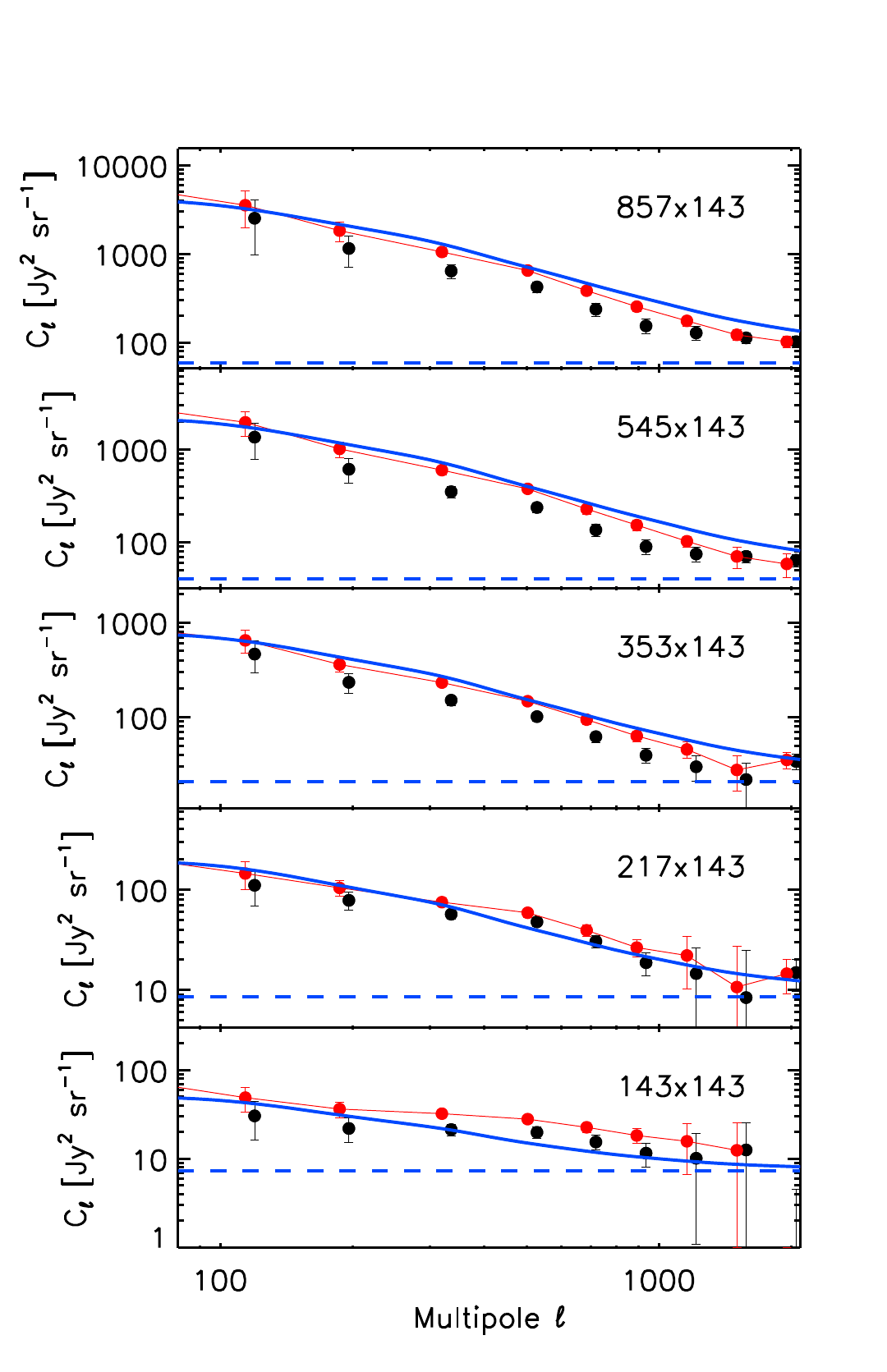}}
\caption{CIB cross- and auto- power spectra obtained at 143\GHz\ (red points).
To obtain the CIB, the cleaned CMB and Galactic dust power spectra (black
points, shifted in $\ell$ for clarity) are corrected for SZ-related residuals,
$C_{\rm SZcorr}^{\nu \times \nu^\prime}$ and
$C_{\rm CIB-SZcorr}^{\nu \times \nu^\prime}$ (following Eqs.~\ref{eq:SZ2},
\ref{eq:SZ3}, and \ref{eq:SZcorr}), and for
$C_{\rm CIBcorr}^{\nu \times \nu^\prime}$ (following Eq.~\ref{eq:corr_CMB},
and computed using the extrapolation of the best-fit halo model).
The prediction of the halo model is shown in blue (continuous for 2h+1h+shot noise;
dashed for shot noise only). \label{fig_143}
}
\end{center}
\end{figure}

\section{Discussion \label{se:discuss}}

\subsection{The 143\,\GHz\ case \label{sec:143}}
Removing the CMB anisotropy at 143\,\GHz\ is very problematic, since the CMB
power spectrum is about 5000 higher than the CIB at $\ell=100$. However,
thanks to the exceptional quality of the \Planck\ data, and the accuracy of
the 100 and 143\,\GHz\ relative photometric calibration, we can obtain
significant measurements, using the same method to clean the maps and measure
the power spectra as for the other channels. We show the measurements in
Fig.~\ref{fig_143}, together with the best-fit CIB model. This estimate has
been obtained by correcting the measurements for the SZ and spurious CIB
(induced by the use of the 100\,\GHz\ map as a CMB template). 
Those corrections are important, especially for power spectra at low
frequencies. For example, for $143\times217$, the SZ-related corrections
decrease the measurements by 10--20\%, while the correction for the spurious
CIB increases the measurements by 30--60\%. Since these corrections are
large, we have not attempted to include the 143\,\GHz\ measurements when
constraining the model.

We show in Fig.~\ref{fig_143} a comparison between extrapolation of the halo
best-fit model to the 143\,\GHz\ cross-power spectra and our CIB power-spectrum
estimates. The $143 \times \nu$ cross-power spectra agree quite well for
$\ell<1000$, at least for $\nu \ge353\,$GHz. The $143 \times 217$ CIB power spectrum lies
about 2$\,\sigma$ above the prediction at intermediate scales ($\ell=502$
and 684). This CIB overestimate increases for the  $143 \times 143$ power
spectrum, which is certainly the most difficult to obtain; this is in excess
with respect to the prediction for $300<\ell<1000$. At this frequency, however,
the CIB auto-power spectrum measurements have to be taken with caution,
as the correction for the spurious CIB can be as high as 70\%,
and is thus highly model dependent.

\subsection{Frequency decorrelation \label{sec:decohence}}
Using the power spectrum measurements, we can quantify the frequency
decoherence. We measure the correlation between bands by averaging the
quantity $C_\ell^{\nu \times \nu^\prime} / (C_\ell^{\nu \times \nu}
 \times C_\ell^{\nu^\prime \times \nu^\prime})^{1/2}$ for $150<\ell<1000$.
We restrict ourselves to this $\ell$ range to have only the clustered CIB
contribution (not the shot noise). Results are given in
Table~\ref{tab:decorrel}. We see that the CIB for the four HFI frequencies
(from 217 to 857\,\GHz) is very well correlated, the worst case being between
the 857 and 217\,\GHz\ channels, with a correlation of about 0.85. On the
other hand, the correlation of all HFI bands with IRIS is quite low, between
about 0.2 and 0.32 (with a large dispersion). This is expected, because the
redshift distribution of CIB anisotropies evolves strongly between 3000 and
$\le$857\GHz, being biased towards higher redshifts at lower frequencies
\citep[e.g.,][]{bethermin2013}.

Contrary to the range 217--857\,\GHz\ the band correlation strongly varies
with $\ell$ at 143\,\GHz, decreasing from $\ell=150$ to 1000
(see Table~\ref{tab:decorrel}). Such a decrease might be expected, based on
the high shot-noise contribution at this frequency (see Fig.~\ref{fig_143}).  Indeed,
\cite{bethermin2013} observe that the band correlation is lower for the shot
noise than for correlated anisotropies; this mimics a scale dependence.

\begin{figure}
\begin{center}
\includegraphics[width=7.3cm]{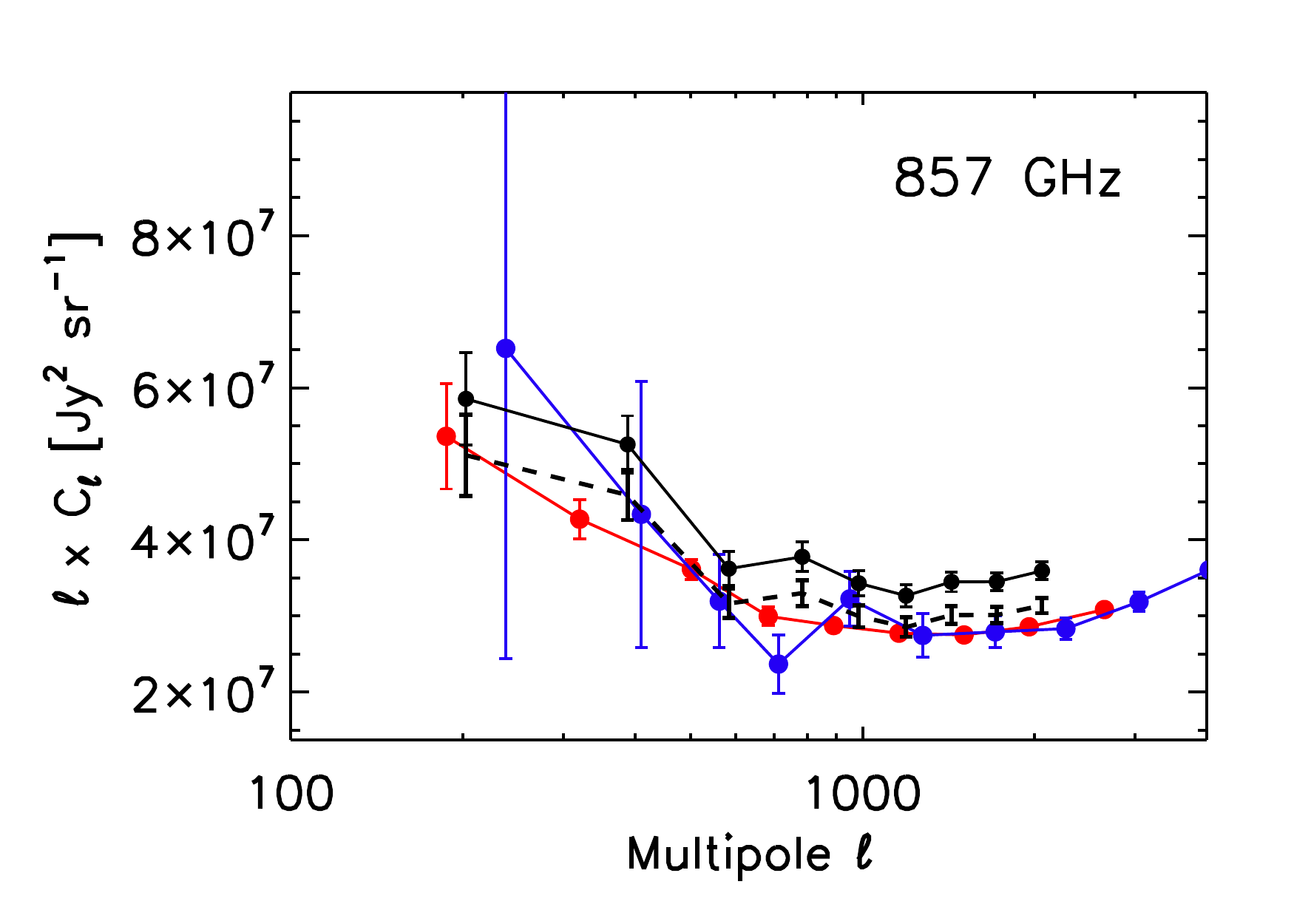}
\includegraphics[width=7.3cm]{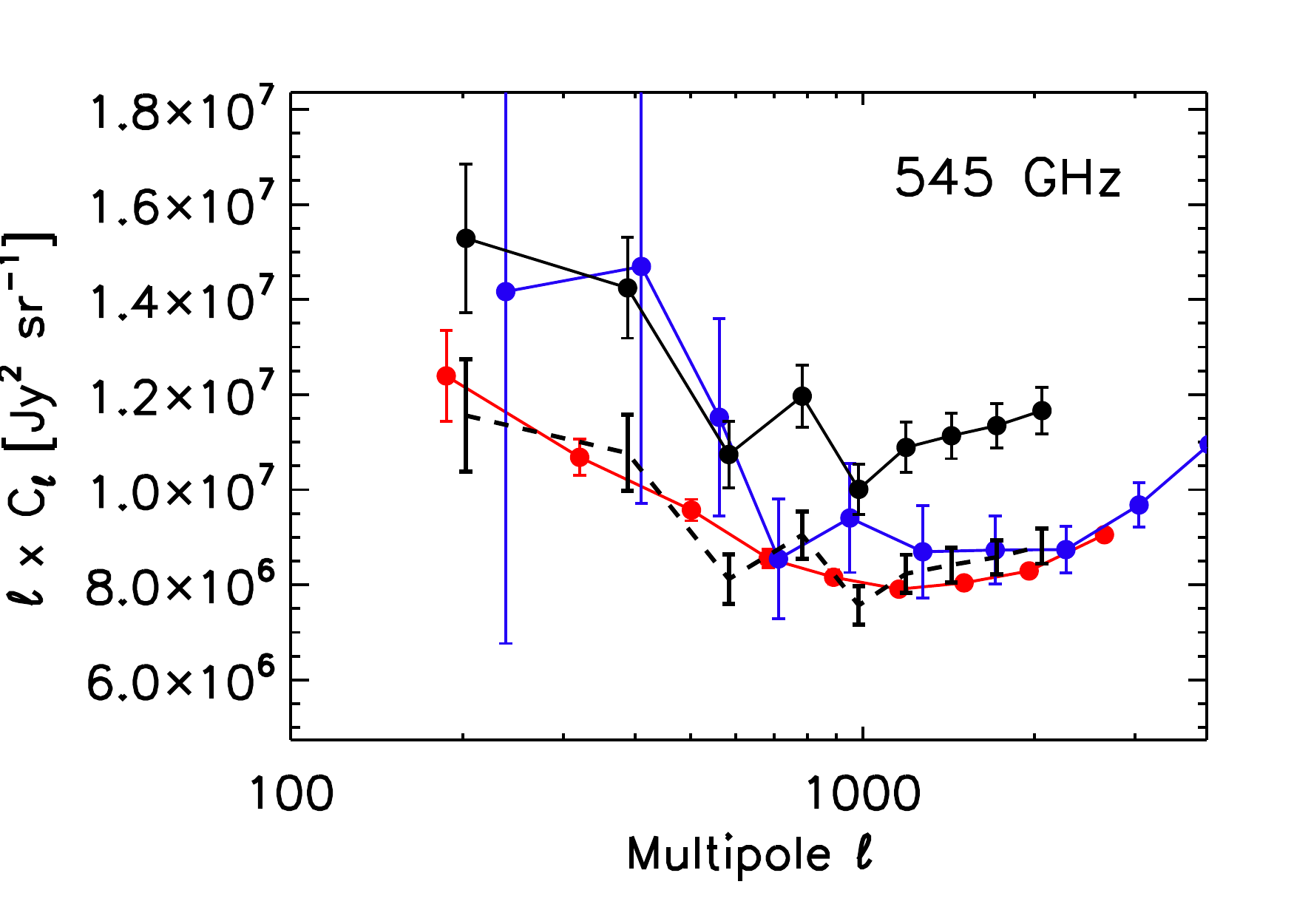}
\includegraphics[width=7.3cm]{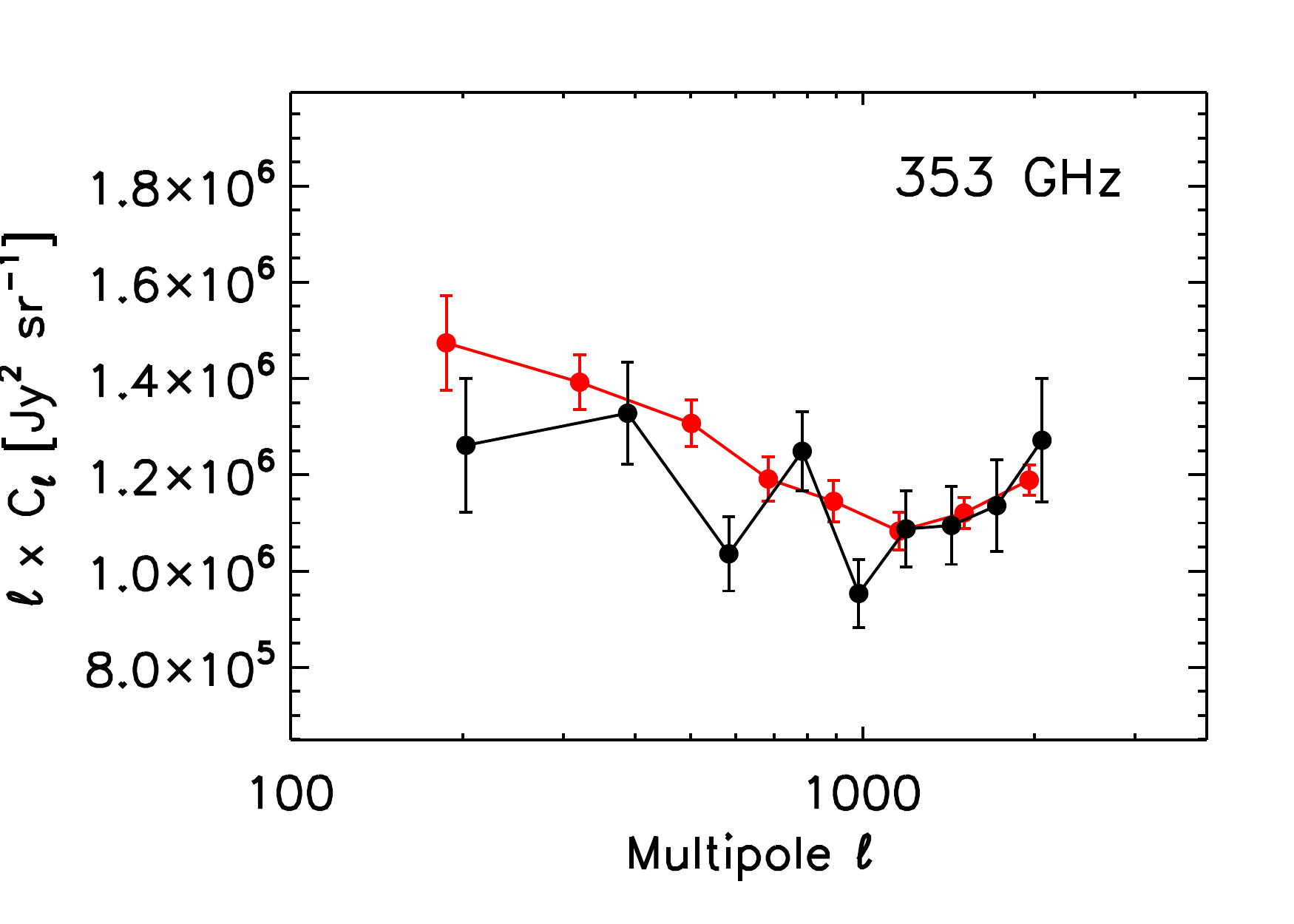}
\includegraphics[width=7.3cm]{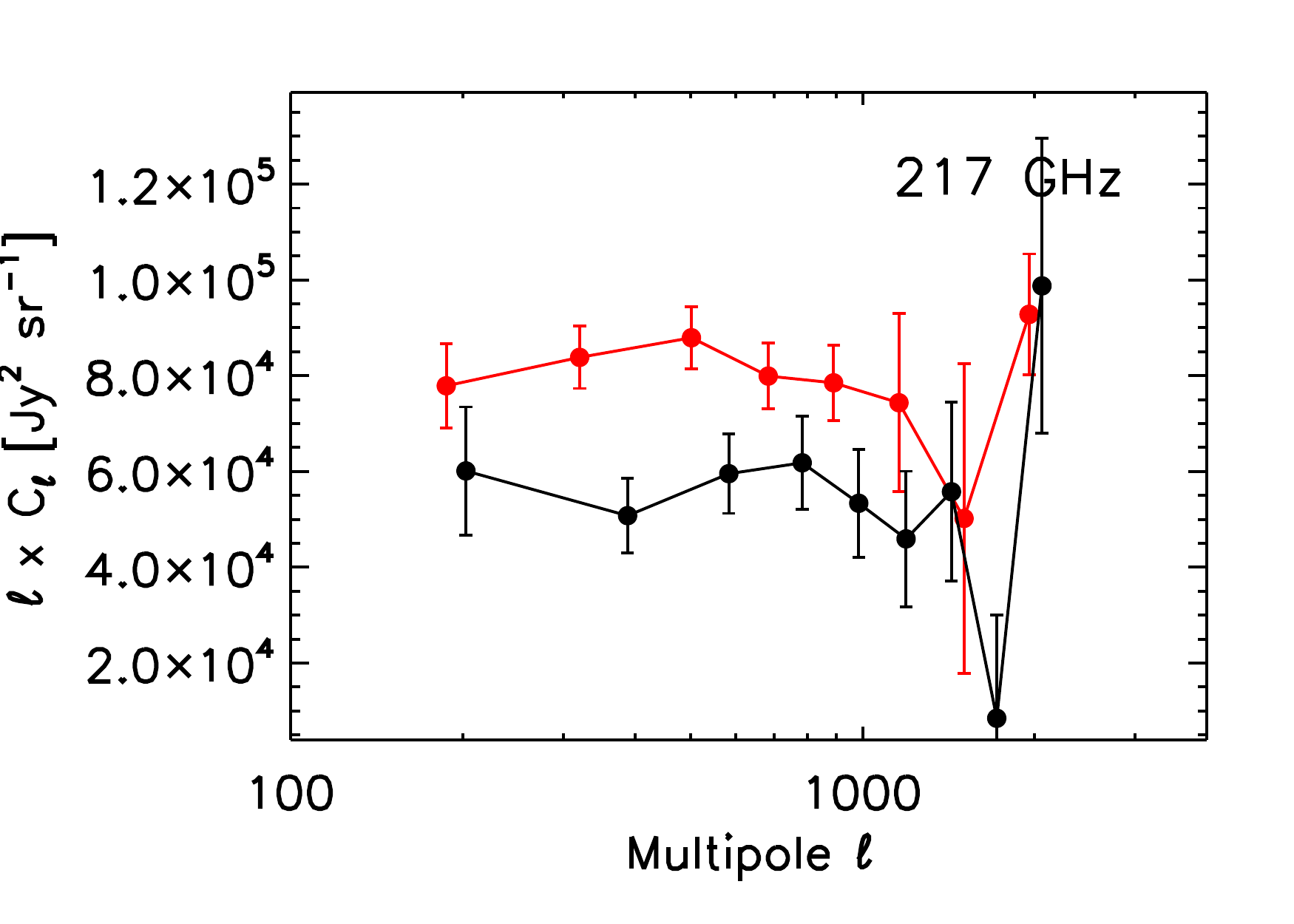}
\caption{Comparison of the CIB auto-power spectra measured using SPIRE
(blue dots, \citealt{viero2012}), earlier \Planck\ data
\citep[][black dots]{planck2011-6.6} and in this paper (red circles).
The SPIRE data have been colour corrected to be compared with HFI (see text).
The dashed lines show the \cite{planck2011-6.6} CIB measurements, rescaled at 857
and 545\GHz\ by the photometric re-calibration factors (1.07$^2$ and 1.15$^2$,
for the power spectra at 857 and 545\GHz, respectively, see
\citealt{planck2013-p03b}). At 217\GHz, the difference between the
black and red points is due to the tSZ and CIB contamination of
the CMB template that is now corrected for.} 
\label{fig:compare_CIB_others}
\end{center}
\end{figure}

\begin{table*}[!tbh]
\begingroup
\newdimen\tblskip \tblskip=5pt
\caption{Frequency decoherence of the CIB, measured by averaging
$C_\ell^{\nu \nu^\prime} /(C_\ell^{\nu \nu} \times
 C_\ell^{\nu \prime \nu \prime})^{1/2}$ for $150<\ell<1000$. The error bars
correspond to the standard deviation. The values at 143\GHz\ strongly depend
on the correction of the spurious CIB (that has been introduced by the choice
of our CMB template), which is highly model dependent. The band correlation
is strongly varying with $\ell$ at 143\GHz, increasing as one goes from
$\ell=1000$ to 150. The numbers in brackets at 143\GHz\ indicate this
variation.}
\label{tab:decorrel}
\nointerlineskip
\vskip -3mm
\footnotesize
\setbox\tablebox=\vbox{
 \newdimen\digitwidth
 \setbox0=\hbox{\rm 0}
  \digitwidth=\wd0
  \catcode`*=\active
  \def*{\kern\digitwidth}
  \newdimen\signwidth
  \setbox0=\hbox{+}
  \signwidth=\wd0
  \catcode`!=\active
  \def!{\kern\signwidth}
\halign{\tabskip=0pt\hfil#\hfil\tabskip=1.0em&
  \hfil#\hfil\tabskip=1.0em&
  \hfil#\hfil\tabskip=1.0em&
  \hfil#\hfil\tabskip=1.0em&
  \hfil#\hfil\tabskip=1.0em&
  \hfil#\hfil\tabskip=1.0em&
  \hfil#\hfil\tabskip=0pt\cr
\noalign{\doubleline}
\noalign{\vskip -2pt}
         & 3000& 857&  545& 353& 217& 143\cr
\noalign{\vskip 3pt\hrule\vskip 3pt}
3000&     1&  $0.36\pm0.06$& $0.31*\pm0.04*$& $0.29*\pm0.04*$& $0.25\pm0.07$&
 \dots\cr
*857& \dots&              1& $0.949\pm0.005$& $0.911\pm0.003$& $0.85\pm0.05$&
 $0.45\pm0.10$ [0.33--0.57]\cr
*545& \dots&          \dots&               1& $0.983\pm0.007$& $0.90\pm0.05$&
 $0.51\pm0.11$ [0.37--0.65]\cr
*353& \dots&          \dots&           \dots&               1& $0.91\pm0.05$&
 $0.54\pm0.11$ [0.41--0.68]\cr
*217& \dots&          \dots&           \dots&           \dots&             1&
 $0.78\pm0.08$ [0.66--0.84]\cr
*143& \dots&          \dots&           \dots&           \dots&         \dots&
 1\cr
\noalign{\vskip 3pt\hrule\vskip 3pt}}}
\endPlancktablewide
\endgroup
\end{table*}

\subsection{Comparison with recent measurements \label{sect_compar_previous}}
We now compare the CIB auto-spectrum measurements with the most
recent measurements from {\it Herschel}-SPIRE \citep{viero2012} and the earlier
measurements from \Planck\ \citep{planck2011-6.6}. We compute the SPIRE-HFI colour
corrections using the CIB SED from \cite{gispert2000} and the most recent
bandpasses \citep[see][]{planck2013-p03d}.
In order to compare with HFI at 857 and 545\,\GHz, the power spectra at
350 and 500~$\mu$m have to be multiplied by 1.016 and 0.805, respectively. 
We use the SPIRE power spectra with only extended sources masked 
(such that the poisson contribution is the same in both measurements).  
We see from Fig.~\ref{fig:compare_CIB_others} that the agreement between the
SPIRE and HFI measurements (red circles versus blue circles) is excellent at
857\,\GHz. At 545\,\GHz, although compatible within the error bars there is a small
difference, with the SPIRE power spectrum being higher than HFI by
about 7\% for $650<\ell<1800$ and by about 30\% for $200<\ell<600$.

Between the publication of \cite{planck2011-6.6} and this paper, the photometric calibration of the two
high-frequency HFI channels has been modified
\citep[see][]{planck2013-p03b}. Using a planet-based calibration rather than a
FIRAS-based calibration leads to a division of the calibration factors by 1.07
and 1.15 at 857 and 545\,\GHz, respectively.  After correcting for these factors,
the two \Planck\ CIB measurements agree within 1$\,\sigma$ at 545\,\GHz\, and
within 2$\,\sigma$ at 857\,\GHz. At 217\,\GHz, the discrepancy we observe between
the earlier \Planck\ CIB measurements and those presented in this paper is
explained by the SZ and CIB contamination
of the 143\,\GHz-based CMB template used in \cite{planck2011-6.6}. Note that with the
new measurements, we do not improve the error bars, since the
use of the 100\,\GHz\ channel as a CMB template adds more noise than the use of
the 143\,\GHz\ channel. The apparent difference in shape between the CIB at 217\,\GHz\ and at higher frequencies has to be attributed to the shot noise, whose contribution relative to the correlated part is higher at 217\,\GHz, making the measured CIB flatter at this frequency.
\\

\begin{figure}
\includegraphics[width=10.5cm]{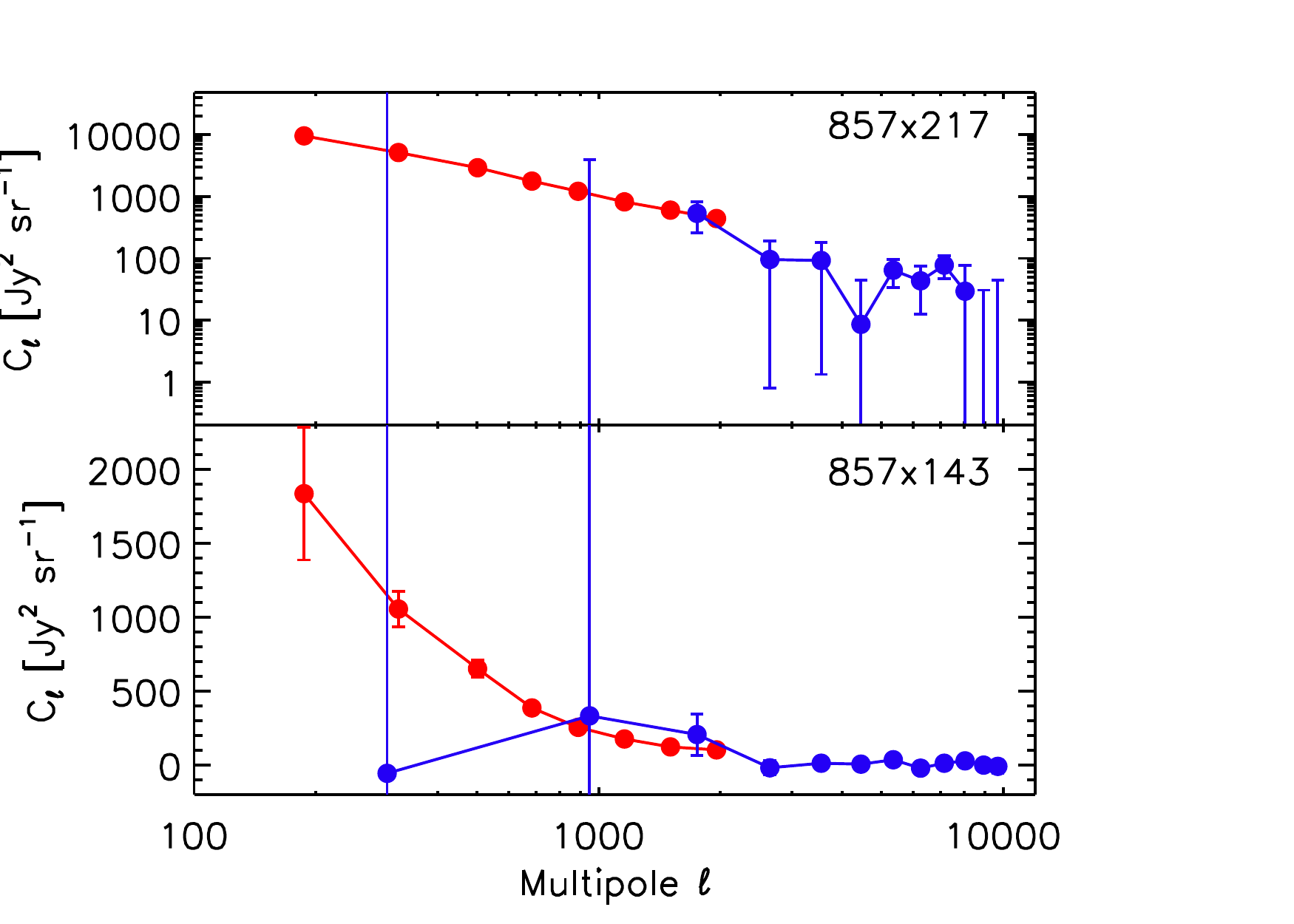}
\caption{\Planck\ (red dots) and BLAST$\times$ACT (blue dots from \citealt{hajian12}) CIB power spectra. Only the clustered CIB is shown (the shot noises have been removed as they are very different in the two measurements). No color corrections have been applied between HFI channels, and the 218\,\GHz\ (ACT) and 857\,\GHz\ (BLAST) channels. The y-axis for the $857 \times 143$ cross-correlation is in linear scale as the BLAST$\times$ACT measurement has negative values (due to the shot-noise removal).}
\label{fig:compare_hajian}
\end{figure}

For cross-power spectra, we can compare our determination with \cite{hajian12}. We show on Fig.~\ref{fig:compare_hajian} the $857 \times 217$ and $857 \times 143$ \Planck\ power spectra, and those obtained from the cross-correlation between BLAST and ACT data. For this comparison, the shot noise contributions have been removed, as they are very different for BLAST, ACT and \Planck. We do not applied any color correction. Even if three points overlap in scale, the comparison can be done only on one point, as the low-$\ell$ measurements from BLAST$\times$ACT are very noisy. For this $\ell$=1750 point, the two measurements agree within 1$\sigma$. This plot illustrates the complementarity between \Planck\ and the high-$\ell$ measurements from ACT and SPT.

 \subsection{The history of star formation density}
\label{sect:discsfh}
The star formation histories recovered from the two different modelling
approaches presented in Sects.~\ref{mod_lin} and \ref{mod_hod} are consistent
below $z = 2$, and agree with recent estimates of the obscured star-formation
density measured by {\it Spitzer\/} and {\it Herschel}. At higher redshift,
there are discrepancies between our two models and the estimate of
\citet{gruppioni2013}. The linear model is about 1 and 2$\,\sigma$ lower than
their measurements at $z=2.5$ and 3.5, respectively, while the halo model lies
about $>3 \,\sigma$ above these data points. Such estimates assume
a shape for the infrared luminosity function. They are strongly dependent on
the faint-end slope assumption, since no data are available below the break of
the luminosity function. This shows how measurement of the obscured star
formation rate density at $z>3$ is difficult.

We investigated the origin of the discrepancy between the two modelling
approaches. In particular, we modified the halo model in two ways to see how
different assumptions on the parametrization of the model can affect the
results.
\begin{enumerate}
\item We fit the data by imposing the condition $\delta=0$ for redshifts
$z\ge z_{\rm break}=2$ in the redshift normalization parameter $\Phi(z)$
of the luminosity-mass relation. This parametrization, although degrading the
quality of the fit somewhat, decreases the SFRD by a factor of about 5 for
$z=4$, which is now compatible with the linear model SFRD (see Fig.~\ref{fig:SFR2}).
\item We fit the data by imposing two conditions: $\delta=0$ for redshift
$z\ge z_{\rm break}$ and $T(z\,{\ge}\,z_{\rm break}) = {\rm constant}
 = T(z_{\rm break})$ for $z_{\rm break}$ in the range 2--5.  We find $z_{\rm break}=4.2\pm0.5$, as can be seen in
Fig.~\ref{fig:SFR2}. In this case the SFRD is only reduced at very high
redshift, by a factor $3$ at $z=6$.  Note that allowing for a redshift break in the redshift evolution of the temperature avoids reaching unphysically high values at very high
redshift. 
\end{enumerate}

We also show in Fig.~\ref{fig:SFR2} the SFRD measurements from the CIB-CMB
lensing cross-correlation \citep{planck2013-p13}. This compares favourably
with a high SFRD level at high redshift.

Part of the discrepancy between the two modelling approaches can also be attributed to the effective bias. A higher bias, as that recovered at high redshift from the linear model, favors a lower SFRD.

\begin{figure}
\includegraphics[width=8.5cm]{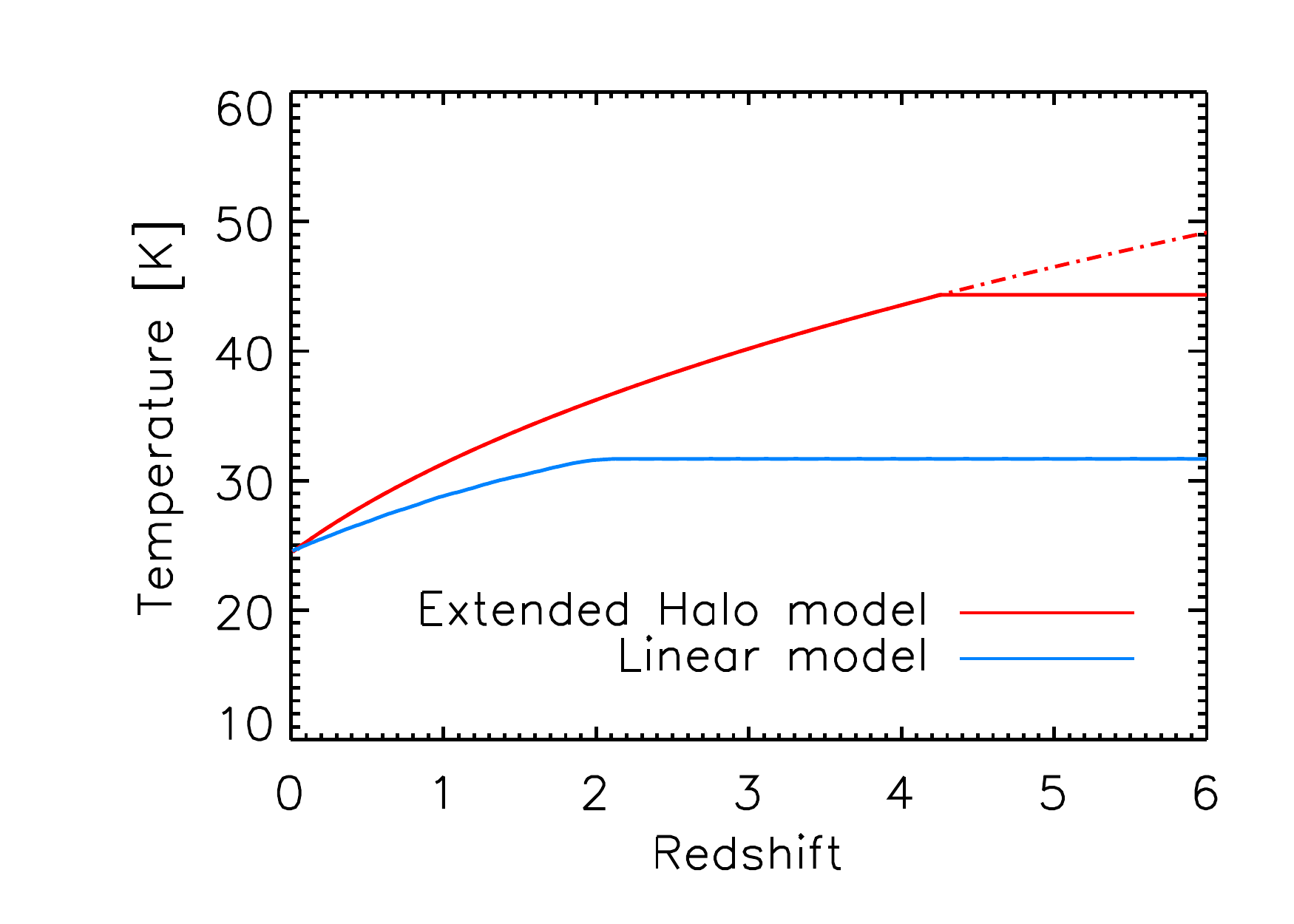}
\caption{Redshift evolution of the dust temperature of the effective SEDs used in the linear model (blue continuous line) and of the SEDs fit in the extended halo model (red continuous line with $z_{break}=4.2$, red dot-dashed line without any redshift break).} \label{fig:sed_mods}
\end{figure}

We finally compared the SEDs used in the two approaches. 
The effective SEDs present a broader peak than
the extended halo model SEDs, because
they take into account the dispersion in dust temperature and the mixing
between secularly star-forming galaxies and episodic starbursts. 
At $z>2$, there are discrepancies
in the Rayleigh-Jeans regime between the two templates, the effective SEDs being higher than the extended halo model SEDs; this discrepancy
increases with redshift. At $z=5$, for the same L$_{\rm IR}$, the parametric
SEDs of the halo model emits about 3 times less infrared light than the
effective SEDs (and hence about an order of magnitude less fluctuations).
This explains why this model requires a much higher star formation rate
density to fit CIB anisotropies than the linear model\footnote{We obtain the same SFRD if we fix in the linear model the SEDs and effective bias to those obtained from the extended halo model.}. Following the parametrisation of Eq.~\ref{eqn:thetanu}, we fit the effective SEDs with a modified black body with a $\nu^{1.75}$ emissivity law to obtain the dust temperature.
On Fig.~\ref{fig:sed_mods}, we show the redshift evolution of the temperature of the two templates. Compared to the recent average temperatures of star-forming galaxies found by \cite{viero13} up to $z\sim4$, the temperature of the effective SEDs is a bit low while the temperature of the SEDs of the extended halo model is a bit high. Knowing the SEDs of the galaxies that are responsible for the bulk of the CIB is the principal limitation in our modeling framework.
Accurate future measurement of the SEDs will be crucial to properly estimate the obscured star formation rate
density at high redshift from the CIB anisotropies. This is important
if one wants to determine whether or not the bulk of the star formation is
obscured at high redshift, and whether the UV and Lyman-break galaxy
populations are a complete tracer of the star formation in the early Universe.

\subsection{CIB non-Gaussianity \label{discussbisp}}
\cite{Lacasa2012} proposed a phenomenological prescription for the CIB
bispectrum based on its power spectrum, namely
\begin{equation}
\label{bisp_pres}
b_{\ell_1\ell_2\ell_3} = \alpha \sqrt{C_{\ell_1} \, C_{\ell_2} \, C_{\ell_3}},
\end{equation}
where $\alpha$ is a dimensionless parameter quantifying the intrinsic level of
non-Gaussianity. Using the best-fit power spectrum of the CIB model described
in Sect.~\ref{mod_hod}, we fitted this parameter $\alpha$ through a $\chi^2$
minimization using the covariance matrix described in
Sect.~\ref{Sect:bisp_pipeline}. The resulting best-fit $\alpha$, its error bar
(computed using a Fisher matrix analysis), and the $\chi^2$ value of the best
fit can be found in Table~\ref{Table:fitalphabisp}.

\begin{table}[!tbh]
\begingroup
\newdimen\tblskip \tblskip=5pt
\caption{Best-fit amplitude parameter for the bispectrum prescription
(Eq.~\ref{bisp_pres}), and the values of $\chi^2$ and number of degrees of
freedom associated with the fit.}
\label{Table:fitalphabisp}
\nointerlineskip
\vskip -3mm
\footnotesize
\setbox\tablebox=\vbox{
 \newdimen\digitwidth
 \setbox0=\hbox{\rm 0}
  \digitwidth=\wd0
  \catcode`*=\active
  \def*{\kern\digitwidth}
  \newdimen\signwidth
  \setbox0=\hbox{+}
  \signwidth=\wd0
  \catcode`!=\active
  \def!{\kern\signwidth}
\halign{\tabskip=0pt\hfil#\hfil\tabskip=1.0em&
  \hfil#\hfil\tabskip=1.0em&
  \hfil#\hfil\tabskip=1.0em&
  \hfil#\hfil\tabskip=0pt\cr
\noalign{\doubleline}
\noalign{\vskip -2pt}
Band& $\alpha$& $\chi^2$& $N_{\rm dof}$\cr
\noalign{\vskip 3pt\hrule\vskip 3pt}
217\GHz& $(1.90\pm0.5*)\times 10^{-3}$& 21.4& 37\cr
353\GHz& $(1.21\pm0.07)\times 10^{-3}$& 45.5& 39\cr
545\GHz& $(1.56\pm0.06)\times 10^{-3}$& 95.9& 35\cr
\noalign{\vskip 2pt\hrule\vskip 3pt}}}
\endPlancktable
\endgroup
\end{table}

The consistency of $\alpha$ across frequencies shows that the measured
bispectrum has a frequency dependence consistent with that of the power
spectrum. The best-fit $\alpha$ values are consistent with the values predicted using
the number counts model of \cite{bethermin11}.
The best-fit $\alpha$s are of the same order, although a little lower
than those found by \cite{Lacasa2012} on simulations by \cite{Sehgal2010},
since they found $\alpha\simeq 3\times10^{-3}$. This indicates a lower level of 
CIB non-Gaussianity than in the simulations by \cite{Sehgal2010}. 

The $\chi^2$ value of the fit shows that the prescription does not provide a
very good model of the data as frequency increases; visual inspection reveals
that this mainly comes from the fact that the measured bispectrum has a
steeper slope than the prescription. To quantify the slope of the measured
bispectrum, we fit a power law to the measurements, i.e.,
\begin{equation}
b_{\ell_1\ell_2\ell_3}
 = A \times \left(\frac{\ell_1 \, \ell_2 \, \ell_3}{\ell_0^3}\right)^{-n},
\end{equation}
where we chose as the pivot scale $\ell_0 = 320$, which is the centre of the
second multipole bin. Table~\ref{Table:fitpowerlawbisp} presents the obtained
best-fit values for the amplitude $A$ and the index $n$, as well as their
error bars and correlation (computed again with Fisher matrices) and the
$\chi^2$ value.

\begin{table}[!tbh]
\begingroup
\newdimen\tblskip \tblskip=5pt
\caption{Best-fit amplitude and index for a power-law fit to the bispectra,
as well as the associated correlation, $\chi^2$ value of the fit and number of
degrees of freedom.}
\label{Table:fitpowerlawbisp}
\nointerlineskip
\vskip -3mm
\footnotesize
\setbox\tablebox=\vbox{
 \newdimen\digitwidth
 \setbox0=\hbox{\rm 0}
  \digitwidth=\wd0
  \catcode`*=\active
  \def*{\kern\digitwidth}
  \newdimen\signwidth
  \setbox0=\hbox{+}
  \signwidth=\wd0
  \catcode`!=\active
  \def!{\kern\signwidth}
\halign{\tabskip=0pt\hfil#\hfil\tabskip=0.5em&
  \hfil#\hfil\tabskip=1.0em&
  \hfil#\hfil\tabskip=0.25em&
  \hfil#\hfil\tabskip=0.25em&
  \hfil#\hfil\tabskip=0.5em&
  \hfil#\hfil\tabskip=0pt\cr
\noalign{\doubleline}
\noalign{\vskip -2pt}
Frequency& $A\ [{\rm Jy}^3\,{\rm sr}^{-1}]$& Index& Correlation& $*\chi^2$&
 $N_{\rm dof}$\cr
\noalign{\vskip 3pt\hrule\vskip 3pt}
217\GHz& $(1.46\pm0.68)\times10^{1}$& $0.822\pm0.145$& 81.4\%& 20.6& 36\cr
353\GHz& $(5.06\pm0.49)\times10^{2}$& $0.882\pm0.070$& 82.3\%& 34.3& 38\cr
545\GHz& $(1.26\pm0.09)\times10^{4}$& $0.814\pm0.050$& 85.4\%& 82.8& 34\cr
\noalign{\vskip 2pt\hrule\vskip 3pt}}}
\endPlancktable
\endgroup
\end{table}

The power law provides a significantly better fit to the data than the
prescription of Eq.~\ref{bisp_pres}, having lower best-fit $\chi^2$ values.
The indices obtained are coherent between frequencies, and significantly
steeper than the Eq.~\ref{bisp_pres} prescription, which is $n\sim 0.6$
(since $C_\ell \propto \ell^{-1.2}$).

There is no sign of flattening of the bispectrum, showing that the shot-noise
contribution is subdominant in this multipole range. This is consistent with
shot-noise estimates based on the number counts model of \cite{bethermin11}.

A detection of CIB non-Gaussianity has recently been reported by
\cite{Crawford2013}. In this paper, they used the prescription proposed by \cite{Lacasa2012} 
to give the amplitude of the bispectrum. In comparison with this analysis, we provide a higher 
detection significance, and at several frequencies. 
Most importantly, we find an indication that the CIB bispectrum is steeper than the prescription
of \cite{Lacasa2012}, although well fitted by a power law.  However, our steeper 
CIB bispectrum at 217\,\GHz, extrapolated up to $\ell$=2000, is compatible with \cite{Crawford2013} 
bispectrum measurement at $\ell$=2000 within 1$\sigma$.
The steeper slope may be an indication that the contribution of more massive halos to the CIB
bispectrum is smaller than in the models studied by \cite{Lacasa2013} and
\cite{Penin2013}. 

\section{Conclusions \label{se:cl}}
%================
We have presented new measurements of the CIB anisotropies with \Planck. Owing
to the exceptional quality of the data, and using a complete analysis of the
different steps that lead to the CIB anisotropy power spectra, we have been
able to measure the clustering of dusty, star-forming galaxies at 143, 217,
353, 545, and 857\,GHz, with unprecedented precision. For the fist time
we also measured the bispectrum from $\ell \simeq$130 to 900 at 217, 353, and
545\GHz. The CIB power spectrum is also measured with {\it IRAS\/} at 3000\,\GHz.

We worked on 11 independent fields, chosen to have high angular-resolution
\hi\ data and low foreground contamination. The total areas used to compute
the angular power spectrum is about $2240\,{\rm deg}^2$. This improves over previous \Planck\ and
{\it Herschel\/} analyses by more than an order of magnitude.
For the bispectrum, the total area is about $4400\,{\rm deg}^2$.

To obtain the CIB, the HFI and {\it IRAS\/}  maps were cleaned using two templates: \hi\ for
Galactic cirrus; and the \Planck\ 100\GHz\ map for CMB. We used new \hi\ data
that covers very large portions of the sky. The large areas forced us to build
a dust model that takes into account the submillimetre--\hi\ emissivity
variations. However, because the \hi\ is not a perfect tracer of dust emission
(e.g., the dark gas), and clearly contains dust-deficient clouds, we had to
reduce the sky fraction to the lowest \hi\ column-density parts of the sky.
The 100\,GHz \Planck\ channel, cleaned of Galactic dust and sources, and then
filtered, provides a good template for the CMB.  This is because it has an
angular resolution close to the higher frequency channels, from which we
measure the CIB, and has the advantage of being an ``internal'' template,
meaning that its noise, data reduction processing steps, photometric
calibration, and beam are all well known. It has the drawback of contaminating
CIB measurements with tSZ signal and spurious CIB coming from the correlation
between the CIB at 100\GHz, and the CIB at higher frequency. The tSZ and spurious CIB corrections are relatively small for frequencies $\nu\ge217\,$GHz.
At 143\,\GHz, while the tSZ and tSZ--CIB corrections are still rather small
(lower than 20\%), the spurious CIB is a very large correction (between 30
and 70\% at intermediate scales) due to the high level of CIB correlation
between 100 and 143\,\GHz. Thus, the 143\,\GHz\ CIB measurements strongly rely
on the CIB model used to compute the correction. 

Due to dust contamination at high frequency, as well as radio sources, SZ and
CIB at low frequency, we conservatively restrict our bispectrum measurement to
the three frequencies 217, 353, and 545\,\GHz.  We measure the bispectrum due to
the clustering of dusty star forming galaxies from $\ell\simeq130$ to 900.
It is detected with a very high significance, $>28\,\sigma$ for all
configurations, and even $>4.5\,\sigma$ for individual configurations at 545\,\GHz.
Such measurements are completely new; they open a window for constraining
models of CIB source emission that we have not yet fully explored in this
paper.

We developed two approaches for modelling the CIB anisotropies. The first
takes advantage of the accurate measurement of CIB anisotropies performed
with \Planck\ and IRIS at large angular scales, and uses only the linear part
of the power spectra. The second approach uses the measurements at all angular
scales, and takes advantage of the frequency coverage, to constrain a halo
model with a luminosity-mass dependence. We find that both models give a very
good fit to the data. Our main findings are as follows.
\begin{itemize}
\item The models give strong constraints on the star formation history up to
redshift $\sim2.5$. At higher redshift, the accuracy of the star formation
history measurement is strongly degraded by the uncertainty on the SED of CIB
galaxies.  An accurate measurement of SEDs of galaxies that are
responsible for the bulk of the CIB will be crucial to estimate properly the
obscured star-formation rate density at high redshifts from the CIB
anisotropies.
\item As found in other recent studies, halos of mass
$M_{\rm eff}=4\times10^{12}\,\msun$ appear to be the most efficient at
actively forming stars.
\item CIB galaxies have warmer temperatures as redshift increases ($T_{\rm d}(z) = 24.4 \times (1+z)^{0.36} \, \rm K$ in the extended halo model). This is
compatible with the most recent {\it Herschel\/} observations, and can be
explained by a harder interstellar radiation field in high-$z$ galaxies.
\item The same halo occupation distribution can simultaneously fit all power
spectra.  However, the 1-halo term is significantly reduced compared to
previous studies \citep{penin2012a, planck2011-6.6}. This is due to a lower
contribution to the clustering from low-$z$ massive halos, as also
observed in \cite{bethermin2013}.
\end{itemize}
We find that the CIB bispectrum is steeper than the prescription developed by
\cite{Lacasa2012}. Just like the reduction of the 1-halo term in the power
spectrum, this may be an indication that the contribution of massive halos to
the CIB bispectrum is smaller than in the models studied by \cite{Lacasa2013}
and \cite{Penin2013}. 
The bispectrum is quite well fitted by a power law. This can be used to
provide valuable constraints on the potential contamination of measurements
of the primordial CMB bispectrum on large scales.

While our component separation process is successful in extracting the CIB
from the maps, the next step is to use a full multi-frequency fitting
procedure to separate the CIB power spectrum and bispectrum, from the tSZ
(and kSZ) effects, the CMB, and the extragalactic source contribution.
Simultaneously taking into account all the components will
improve our ability to separate them. The goal is to give unprecedented limits
on the reionization history of the Universe, as well as understanding
the history of star formation in dark matter halos.

\begin{acknowledgements}

The development of \Planck\ has been supported by: ESA; CNES and
CNRS/INSU-IN2P3-INP (France); ASI, CNR, and INAF (Italy); NASA and DoE (USA);
STFC and UKSA (UK); CSIC, MICINN and JA (Spain); Tekes, AoF and CSC (Finland);
DLR and MPG (Germany); CSA (Canada); DTU Space (Denmark); SER/SSO
(Switzerland); RCN (Norway); SFI (Ireland); FCT/MCTES (Portugal); and PRACE
(EU).  A description of the \Planck\ Collaboration and a list of its members
with the technical or scientific activities they have been involved
into, can be found at
\url{http://www.rssd.esa.int/index.php?project=PLANCK&page=PlanckCollaboration}.
The Parkes radio telescope is part of the Australia Telescope National Facility which is funded by the Commonwealth of Australia for operation as a National Facility managed by CSIRO. Some \hi\ data used in this paper are based on observations with the 100-m telescope of the MPIfR (Max-Planck-Institut f�r Radioastronomie) at Effelsberg.

\end{acknowledgements}

\bibliographystyle{aa}
\bibliography{main_bib_CIB,Planck_bib}

\appendix

\section{\hi\ data \label{hi_description}}
We describe in this Appendix the three \hi\ surveys that we use to remove
Galactic dust contamination from the frequency maps.
The 21-cm \hi\ spectra were obtained with: (1) the 100-m Green Bank Telescope
(GBT); (2) the Parkes 64-m telescope; and (3) the Effelsberg 100-m radio
telescope.

\subsection{GBT observations and data preparation}
The GBT \hi\ maps are created from \hi\ spectral observations with the Green
Bank Telescope, the details of which can be found in \cite{boothroyd2011}
and Martin et al. (in prep).  Over $800\,{\rm deg}^2$ were
mapped between 2002 and 2010, each following roughly the same
observing strategy.  The high Galactic latitude fields were mapped
using scans along lines of constant declination or constant
Galactic latitude at a scan rate equalling 3.5\arcmin\ every 4 seconds.
Each subsequent scan is offset by 3.5\arcmin\ in the corresponding
orthogonal direction.  This strategy results in a rectangular region
of the \hi\ sky sampled every 3.5\arcmin.  Some of the regions have been
scanned in this way multiple times, in order to increase the
signal-to-noise, and to investigate the stability of the system
\citep{boothroyd2011}.

The spectra are first converted from their antenna temperature scale
to a brightness temperature ($T_{\rm b}$) scale.  This involves calibration
and stray radiation corrections, as discussed in \cite{boothroyd2011}.
All $T_{\rm b}$ spectra for a corresponding region are assigned to a
3.5\arcmin\ Sanson-Flamsteed-projection grid (SFL-projection) using
convolution with an optimized tapered Bessel function, in order to minimize
noise on spatial scales smaller than the beam \citep{mangum2007}.  Note that
this observing strategy and gridding choice results in a final cube
resolution of 9.55\arcmin$\times$9.24\arcmin, slightly broader than the
inherent 9.1\arcmin$\times$9.0\arcmin\ GBT 21-cm beam. 

Each spectrum is recorded using in-band frequency-switching, resulting
in velocity coverage of $-450\,{\rm km}\,{\rm s}^{-1}
 \le V_{\mathrm LSR} \le 355\,{\rm km}\,{\rm s}^{-1}$,
with very flat baselines.  Any residual baseline is removed on a
pixel-by-pixel basis by fitting the emission-free channels in the
final calibrated cube using a third-order polynomial.

The archival LH2 field was observed using a 3\arcmin\ grid pattern with
the GBT spectral processor and was calibrated accordingly.  
As the residual baseline behaviour is different for
the archival data, only a linear polynomial was fit to the
emission-free channels.  The rest of the processing was identical to
that described above.

The individual $0.8\,{\rm km}\,{\rm s}^{-1}$ spectral channels of the cubes are
integrated to convert the $T_{\rm b}$ spectra into $N_{\mathrm{HI}}$
maps:
\begin{equation}
N_{\mathrm{HI}}(x,y) = 1.823 \times10^{18}\ \sum_{v} T_{\rm b}(x,y,v)
 \tau(T_{\rm s}) \delta v ,
\end{equation}
where the sum is over a given velocity range, $v$, and $\delta v$ is
the $0.80\,{\rm km}\,{\rm s}^{-1}$ channel spacing.
The quantity $\tau$ is the opacity correction
for spin temperature, $T_{\mathrm s}$:
\begin{equation}
\tau(T_{\mathrm s}) = - \ln( 1 - T_{\mathrm{b}} / T_{\mathrm{s}} ).
\end{equation}
For the adopted value of $T_{\mathrm{s}} = 80$~K, these corrections are
all less than 5\% for our CIB fields \citep{planck2011-6.6}.

The velocity ranges over which the integrations are performed are
selected using the observed velocity structure in each of the cubes.  The
models presented here subdivide each cube into three velocity-selected
components: a local component; intermediate-velocity clouds, IVCs; and
high-velocity clouds, HVCs.  Divisions between components are
distinguishable by reductions in structure (as measured through the
standard deviation of individual channel maps) as one progresses
through the data cube, channel by channel.  More details can be found
in \cite{planck2011-6.6}.

\subsection{GASS observations and data preparation}
\label{GASS_HI}
The GASS survey is a 21-cm line survey covering the
southern sky for all declinations $\delta \la 1\deg$.  The observations were
made with the multibeam system on the 64-m Parkes Radio Telescope. The
intrinsic angular resolution of the data is 14.4\arcmin\
full width at half maximum (FWHM). The velocity
resolution is 1.0\kms\ and the useful bandpass covers a velocity range
${|v_{lsr}|} \la 468$\kms\ for all of the observations; some data cover up to
${|v_{lsr}|} \la 500$\kms. GASS is the most sensitive, highest
angular resolution large-scale survey of Galactic \hi\ emission ever made in
the southern sky. The observations are described in \cite{2009ApJS..181..398M}. 
We used data from the final data release \citep{2010A&A...521A..17K}
that were corrected for instrumental effects and radio-frequency interference
(RFI). The data were gridded on a Cartesian grid on the Magellanic stream
(MS) coordinate system as
defined by \cite{2008ApJ...679..432N}. To minimize the noise and 
eliminate residual instrumental problems, we calculated a second 3-D
data cube with a beam of $0.5^\circ$ FWHM, smoothing at the same time in
velocity by $8\,{\rm km}\,{\rm s}^{-1}$. Emission below a 5$\,\sigma$ level
of 30\,mK in the smoothed data-cube was considered as insignificant and was
accordingly zeroed. For details in data processing and analysis see
\cite{2012A&A...547A..12V}.

When looking at the \hi\ data cube in the southern sky, one of the most
prominent structures is linked to hydrogen gas in the Magellanic stream, in
the disks of the Magellanic clouds, and in the stream's leading arm. In
particular, the Magellanic stream,  stretches over $100^\circ$ behind the
Large and Small Magellanic clouds.  We thus need to remove this contamination
to be able to use the \hi\ as a tracer of Galactic dust.

\begin{figure}
\begin{center}
\includegraphics[width=6cm]{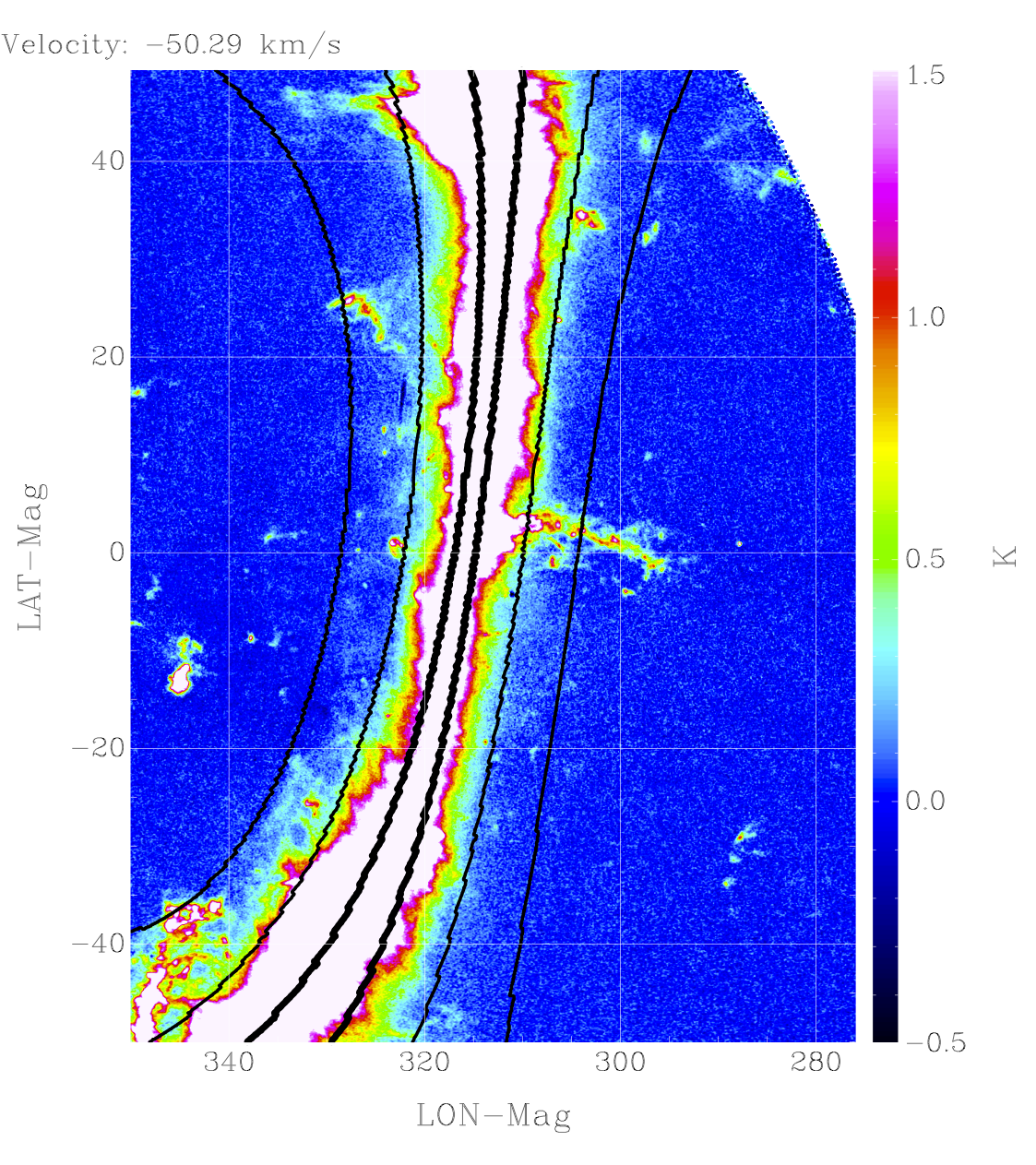}
\includegraphics[width=6cm]{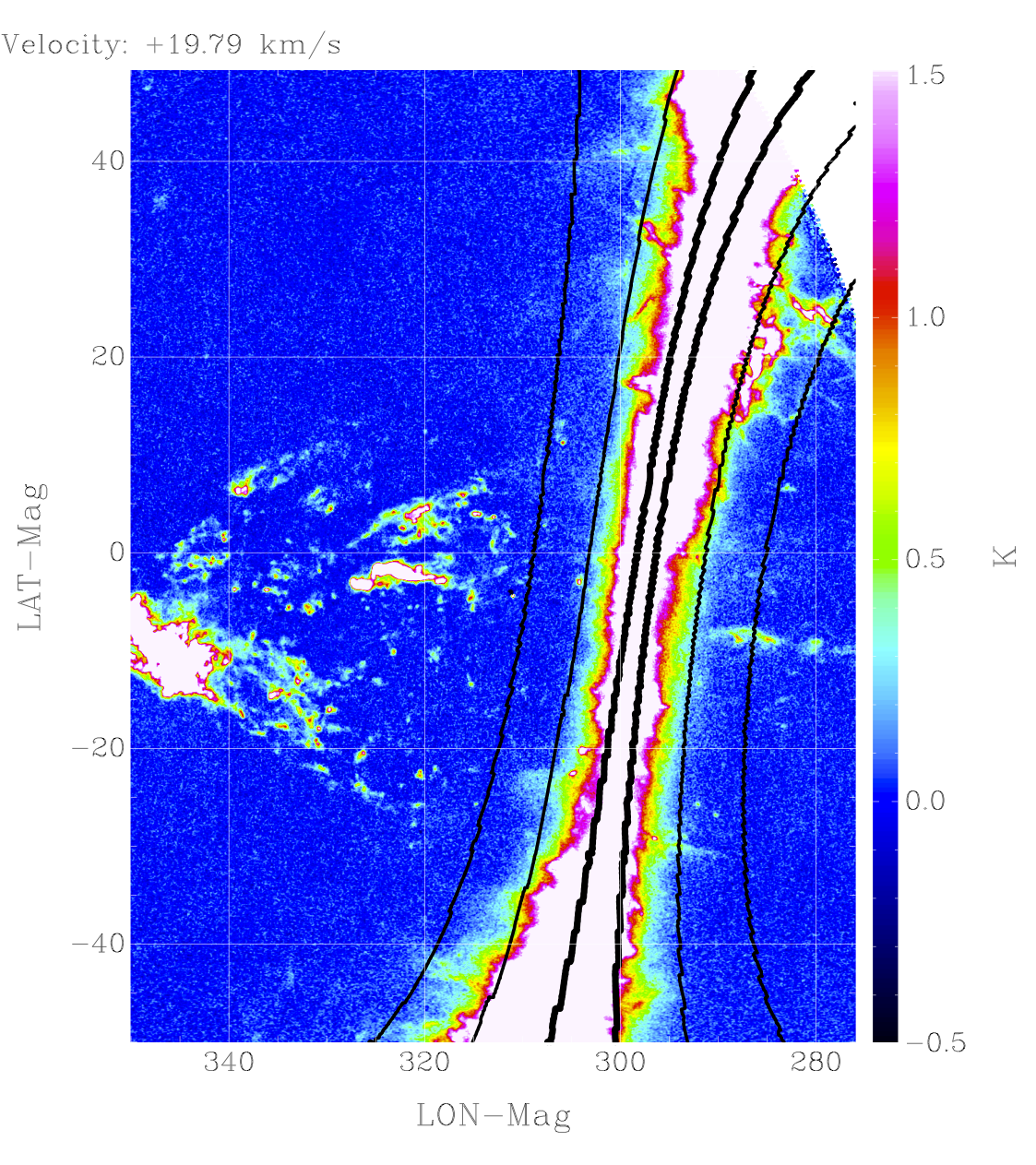}
\end{center}
\caption{\label{compa_sep_MS} Comparison of observed \hi\ emission with the Galactic model
(black isophotes are for the expected emission at levels of 0.06, 0.6 and 6.0\,K).
Magellanic coordinates (longitude and latitude)
are used. Two examples for channel maps at Galactic
standard of rest velocities of 19.8 and -50.3 km s$^{-1}$ are given.}
\end{figure}

Aiming to separate Galactic emission from the observations of the
Magellanic system we calculated the expected Milky Way
emission according to the model of
\cite{2008A&A...487..951K}. Velocities for components in direction towards
the southern Galactic pole were shifted by $-5\,{\rm km}\,{\rm s}^{-1}$
to mimic the apparent infall \citep[see][]{1974HiA.....3..423W}.
Comparisons between the emission and the model at two particular velocities
in the Galactic standard of rest frame are shown in Fig.~\ref{compa_sep_MS}.
The strong Galactic emission can be traced to weak extended line
wings that are well represented by the model. 
It is therefore feasible to use the model to
predict regions in the 3-D data cube that are most probably 
occupied by Milky Way emission. We used a clip level of 60\,mK, 
the lowest isophote in Fig.~\ref{compa_sep_MS} that delineates
the disk emission.  Such a treatment extracts most of the Galactic emission,
however we must take into account the fact that at positions with strongly
blended lines Magellanic emission also gets included. A higher clip level
would minimize this problem, although, at the same time the wings of the
Galactic emission would be affected. The chosen clip level of 60\,mK
(comparable to the instrumental noise), is a good compromise.

As the final step in the reduction of the GASS data we integrated
the \hi\ emission over the appropriate velocity range
for each individual position in the $N_{\rm side}=512$ \healpix\ database
to obtain
column densities of the Galactic gas. To avoid any interpolation
errors, we extracted profiles from the original GASS database; all
intermediate data products that have been described above served only
to discriminate Galactic emission from the Magellanic
stream. Due to the finite beam size of
the Parkes telescope and the Gaussian weighting that was used for
the gridding, the effective resolution of the \healpix\ data is
16.2\arcmin\ FWHM.

\subsection{EBHIS observations and data analyses}
\label{EBHIS_HI}
The Effelsberg-Bonn \hi\ Survey (EBHIS) comprises an \hi\ survey of the entire
northern sky for all declinations $\delta \ga -5^\circ$ with the
Effelsberg 100-m telescope. The bandwidth of 100\,MHz covers
$-1000\,{\rm km}\,{\rm s}^{-1} \leq v_{\rm LSR} \leq
 19{,}000\,{\rm km}\,{\rm s}^{-1}$. This allows us to study the detailed
Milky Way \hi\ structure as well as the local Universe up to a redshift of
$z \simeq 0.07$ \citep{kerpetal2011}, with an effective velocity resolution
of about 2.1\,${\rm km}\,{\rm s}^{-1}$.

We selected from the early survey data a clean high Galactic latitude field.
The observations of the field of interest were performed during the summer of
2011. Following the standard observing strategy \citep{kerpetal2011}
individual fields of $25\,{\rm deg}^2$ were measured. In addition to the data
reduction and calibration pipeline of \cite{winkeletal2010}, the \hi\ data
were corrected with an improved RFI mitigation detection algorithm and the
absolute offsets between the individual fields were minimized to a level of
$N_{\rm HI} \leq 3\times10^{18}\,{\rm cm}^{-2}$. As for the GBT fields, the
EBHIS data were put on a 3.5\arcmin\ SFL-projection grid.

The overall \hi\ emission in this field can be characterized by intermediate-
and low-velocity emission populating the radial velocity range
$-80\,{\rm km}\,{\rm s}^{-1} \leq v_{\rm LSR}
 \leq +20\,{\rm km}\,{\rm s}^{-1}$.

The area of interest is about $130\,{\rm deg}^2$ with \hi\ column densities
below $N_{\rm HI} = 2\times10^{20}\,{\rm cm}^{-2}$. The total \hi\ column
density range is $0.98\times 10^{20}\,{\rm cm}^{-2} \leq N_{\rm HI}
 \leq 7.5\times 10^{20}\,{\rm cm}^{-2}$. Accordingly we expect infrared excess
emission associated with molecular hydrogen towards areas of
$N_{\rm HI} \geq 3\times 10^{20}\,{\rm cm}^{-2}$, while the low-column density
regions should allow us to study the emission associated with the CIB.

In the first step we performed a linear fit of the EBHIS \hi\ data integrated
across the velocity range $-160\,{\rm km}\,{\rm s}^{-1} \leq v_{\rm LSR}
 \leq +160\,{\rm km}\,{\rm s}^{-1}$ to the HFI 857\GHz\ map:
$I_{857}=a \times N_{\rm HI} + b , $
for $N_{\rm HI} \leq 2\times 10^{20}\,{\rm cm}^{-2}$.  
The residual map shows dust excess, as expected when the total infrared
emissivity traces \hi\ and H$_{2}$ \citep{planck2011-7.0,planck2011-7.12}. In agreement
with our expectation dust excess clouds show up with \hi\ column densities
of $N_{\rm HI} \geq 3\times 10^{20}\,{\rm cm}^{-2}$.
But surprisingly, the residual map shows dust-deficient \hi\ clouds that
populate the column density range $2.0\leq N_{\rm HI} \leq
5.5\times 10^{20}\,{\rm cm}^{-2}$. The \hi\ emission of these dust-deficient
clouds can be characterized by brightness temperatures $T_{\rm B} > 15$\,K
and $\Delta \nu ({\rm FWHM}) \leq 5\,{\rm km}\,{\rm s}^{-1}$.
These \hi\ lines are associated with
individual, cold IVCs. A positional correlation between the dust-deficient
regions and the IVCs show a very strong correlation. We evaluated the dust
deficiency of the IVCs and found that they are not ``under-luminous'', but
infrared dark.

To extract the CIB in this field we produced a mask excluding the excess as
well as the dust-deficient clouds. About $90\,{\rm deg}^2$ are suitable for CIB
analysis.

\section{From infrared luminosity density to emissivities}
\label{app_jnu}
The emissivity, often written $j_\nu$, is given by \citep{penin2012a}
\begin{equation}
\label{eq:B1}
j_\nu(z) = \frac{1+z}{d\chi/dz}  \int S_\nu \frac{d^2N}{dS_\nu \, dz} dS_\nu,
\end{equation}
where $\chi$ is the comoving distance,  $S_\nu$ the flux density,
and $d^2N/dS_\nu \, dz$ the number of sources per flux density and redshift
interval. If there are several types of galaxy (labelled by ``$t$'')
with different SEDs\footnote{The result is also true if there is a scatter in
temperature.  Each range of temperature is then considered as a type.},
one can rewrite Eq.~\ref{eq:B1} as
\begin{equation}
j_\nu(z) = \frac{1+z}{d\chi/dz} \sum_{t} \int S_\nu
 \frac{d^2N_{t}}{dS_\nu \, dz} dS_\nu.
\end{equation}
We then introduce $s_{\nu, t}$ the flux density of an $L_{\rm IR}=1\,L_\odot$
source with an SED of a given type. This quantity varies only with $z$ for
a given type of SED. We can thus easily change the variable $S_\nu$ to
$L_{\rm IR}$ in the integral:
\begin{equation}
j_\nu(z) = \frac{1+z}{d\chi/dz} \sum_{t} s_\nu \int L_{\rm IR}
 \frac{d^2N_{t}}{dL_{\rm IR} \, dz} dL_{\rm IR}.
\end{equation}
We can slightly modify this expression so that it contains the bolometric
infrared luminosity function of each galaxy type
$d^2N_{t}/dL_{\rm IR} \, dV$:
\begin{equation}
\label{lab1}
j_\nu(z) = \frac{1+z}{d\chi/dz} \frac{dV}{dz}\sum_{\rm t}
 s_\nu \int L_{\rm IR} \frac{d^2N_{t}}{dL_{\rm IR} \, dV} dL_{\rm IR},
\end{equation}
where $V$ is the comoving volume.
We can simplify this expression assuming a flat $\Lambda$CDM Universe with
\begin{equation}
\frac{d \chi}{dz} = \frac{c}{H_0 \sqrt{\Omega_{\rm m}(1+z)^3 + \Omega_\Lambda}}
\end{equation}
and
\begin{equation}
\frac{dV}{dz} = \frac{c \chi^2}{H_0 \sqrt{\Omega_{\rm m}(1+z)^3 +
 \Omega_\Lambda}} = \frac{d \chi}{dz} \chi^2.
\end{equation}
We can thus simplify Eq.\,\ref{lab1}:
\begin{equation}
\label{lab2}
j_\nu(z) = (1+z) \chi^2 \sum_{t} s_\nu \int L_{\rm IR}
 \frac{d^2N_{t}}{dL_{\rm IR} \, dV} dL_{\rm IR},
\end{equation}
We introduce the contribution to the infrared luminosity density of a given
type of SED:
\begin{equation}
\rho_{{\rm IR},t} = \frac{d^2N_{t}}{dL_{\rm IR} \, dV} dL_{\rm IR},
\end{equation}
where $\rho_{\rm IR}$ is the total infrared luminosity density. We can assume
a simple conversion between $L_{\rm IR}$ and $\rho_{\rm SFR}$, the star
formation rate density, using the \cite{kennicutt1998} constant $K$.
Then Eq.\,\ref{lab2} can be rewritten as:
\begin{equation}
j_\nu(z) = (1+z) \chi^2 \frac{\rho_{\rm SFR}}{K} \times
 \left( \frac{\sum_{t} s_\nu \,
 \rho_{{\rm IR},t}}{\sum_{t}\rho_{{\rm IR},t}}\right),
\end{equation}
where $\sum_{t} s_\nu \, \rho_{{\rm IR},t}/\sum_{t}\rho_{{\rm IR},t}$
is the effective SED of infrared galaxies (noted $s_{\nu,{\rm eff}}$),
i.e., the mean SED using a weighting by the contribution to the infrared
luminosity density. Here we use the model of \cite{bethermin2012model} to
compute the contribution at each redshift. This model is based on the observed
strong correlation between stellar mass and star formation rate in these
galaxies (called main-sequence galaxies), and new SED templates from
\cite{magdis2012}. We finally obtain:
\begin{equation}
j_\nu(z) = \rho_{\rm SFR}(z) \frac{(1+z)\, \chi^2(z) \, s_{\nu,\rm eff}(z)}{K}.
\end{equation}

\section{Effective bias notion for the linear model}

\label{sect:effbias}

We discuss here the link between the approach chosen to model the 2-halo term
in Sect.~\ref{mod_lin} (linear model) and \ref{mod_hod} (HOD model). We
consider for simplicity only the case of the auto-spectrum, but this can be
easily generalized to the cross-spectrum. Eq.~\ref{eqn:pj2h} of the HOD modeli
can be re-written as
\begin{equation}
\label{eq:2hannexe}
P_{\rm 2h}^{\nu \nu}(k,z) = \int \left( \frac{a}{\chi} \right )^2
 \left( \frac{dj}{dM}(\nu,z) b(M,z) \right)^2  P_{\rm lin}(k,z) d\chi,
\end{equation}
where $dj/dM$ is the differential contribution of halos of mass $M$, at
redshift $z$, to the emissivity. It can be computed using
\begin{equation}
\frac{dj}{dM}(\nu,z) = \frac{dN}{dM} \times
 \left\{f_\nu^{\rm cen}(M,z) + f_\nu^{\rm sat}(M,z)\right\}.
\end{equation}
The total emissivity is thus
\begin{equation}
j(\nu,z) = \int \frac{dj}{dM}(\nu,z) dM.
\end{equation}
In order to simplify Eq.\,\ref{eq:2hannexe}, we introduce an effective bias,
\begin{equation}
b_{\rm eff}(z)=\frac{\int\frac{dj}{dM}(\nu,z)b(M,z) dM}
 {\int \frac{dj}{dM}(\nu,z) dM}.
\end{equation}
This effective bias is the mean bias of halos hosting the infrared galaxies,
weighted by their differential contribution to the emissivities. We finally
obtain a simpler equation, which is used in the linear model:
\begin{equation}
P_{ \rm 2h}^{\nu \nu}(k,z) = \int \left ( \frac{a}{\chi} \right )^2 j^2
 (\nu,z) b_{\rm eff}^2  P_{\rm lin}(k,z) d\chi.
\end{equation}

\section{Tables \label{se:app_tables}}

%pk_results_6
\small
\begin{table*}[!tbh]
\begingroup
\newdimen\tblskip \tblskip=5pt
\caption{
Cross-power spectra and error bars for all pairs of frequencies,
measured on Galactic dust- and CMB-free residual maps. Error bars here include
uncertainties on foreground and CMB residuals, beam and projection. To obtain the CIB power
spectra (given in Table~\ref{tab:cib_final}), the power spectra
have to be further corrected for $C_{\rm CIBcorr}^{\nu \times \nu^\prime}$
(Eq.~\ref{eq:corr_CMB}),
$C_{\rm SZcorr}^{\nu \times \nu^\prime}$ (Eq.~\ref{eq:SZ1}),
and $C_{\rm CIB-SZcorr}^{\nu \times \nu^\prime}$
(Eq.~\ref{eq:SZ3}).  Moreover, the first two points (at $\ell=53$ and 114)
have to be considered as upper limits (see Sect.~\ref{dust_res}).}
\label{tab:res_map_pow_spec}
\nointerlineskip
\vskip -3mm
\footnotesize
\setbox\tablebox=\vbox{
 \newdimen\digitwidth
 \setbox0=\hbox{\rm 0}
  \digitwidth=\wd0
  \catcode`*=\active
  \def*{\kern\digitwidth}
  \newdimen\signwidth
  \setbox0=\hbox{+}
  \signwidth=\wd0
  \catcode`!=\active
  \def!{\kern\signwidth}
\halign{\tabskip=0pt#\hfil\tabskip=0.5em&
  \hfil#\hfil\tabskip=1.0em&
  \hfil#\hfil\tabskip=1.0em&
  \hfil#\hfil\tabskip=1.0em&
  \hfil#\hfil\tabskip=1.0em&
  \hfil#\hfil\tabskip=1.0em&
  \hfil#\hfil\tabskip=1.0em&
  \hfil#\hfil\tabskip=0pt\cr
\noalign{\doubleline}
\noalign{\vskip -2pt}
&&\multispan6\hfil{Cross-power spectra, $C_\ell$, and uncertainty
 [${\rm Jy}^2\,{\rm sr}^{-1}$, $\nu \rm I_{\nu}$=cst]}\hfil\cr
\noalign{\vskip -5pt}
&&\multispan6\hrulefill\cr
& $\ell$& 3000\GHz& 857\GHz& 545\GHz& 353\GHz& 217\GHz& 143\GHz\cr
\noalign{\vskip 3pt\hrule\vskip 3pt}
3000\GHz & *187 & $(4.35 \pm 1.34)\times10^5$ & $(1.12 \pm 0.85)\times10^5$ & $(3.94 \pm 3.88)\times10^4$ & $(8.92 \pm 12.4)\times10^3$ & $(5.59 \pm 26.7)\times10^2$ & \dots \cr
& *320 & $(7.89 \pm 3.14)\times10^4$ & $(7.49 \pm 2.09)\times10^4$ & $(3.04 \pm 0.85)\times10^4$ & $(1.13 \pm 0.27)\times10^4$ & $(2.34 \pm 0.71)\times10^3$ & \dots \cr
& *502 & $(3.46 \pm 1.15)\times10^4$ & $(3.10 \pm 0.79)\times10^4$ & $(1.23 \pm 0.33)\times10^4$ & $(4.49 \pm 1.15)\times10^3$ & $(1.18 \pm 0.34)\times10^3$ & \dots \cr
& *684 & $(2.90 \pm 0.70)\times10^4$ & $(2.24 \pm 0.47)\times10^4$ & $(9.35 \pm 2.10)\times10^3$ & $(2.87 \pm 0.73)\times10^3$ & $(6.88 \pm 2.61)\times10^2$ & \dots \cr
& *890 & $(1.87 \pm 0.47)\times10^4$ & $(1.62 \pm 0.30)\times10^4$ & $(6.61 \pm 1.34)\times10^3$ & $(1.99 \pm 0.45)\times10^3$ & $(2.34 \pm 1.66)\times10^2$ & \dots \cr
& 1158 & $(1.43 \pm 0.40)\times10^4$ & $(1.17 \pm 0.22)\times10^4$ & $(4.71 \pm 0.93)\times10^3$ & $(1.25 \pm 0.31)\times10^3$ & $(3.55 \pm 1.43)\times10^2$ & \dots \cr
& 1505 & $(1.09 \pm 0.41)\times10^4$ & $(8.30 \pm 1.85)\times10^3$ & $(3.12 \pm 0.76)\times10^3$ & $(1.14 \pm 0.30)\times10^3$ & $(2.41 \pm 1.07)\times10^2$ & \dots \cr
& 1956 & $(1.05 \pm 0.58)\times10^4$ & $(8.04 \pm 2.39)\times10^3$ & $(3.16 \pm 1.00)\times10^3$ & $(9.67 \pm 3.44)\times10^2$ & $(2.58 \pm 1.14)\times10^2$ & \dots \cr
& 2649 & $(1.06 \pm 1.04)\times10^4$ & $(6.47 \pm 3.27)\times10^3$ & $(2.61 \pm 1.38)\times10^3$ & \dots & \dots & \dots \cr
\noalign{\vskip 3pt\hrule\vskip 3pt}
857\GHz & **53 & \dots & $(1.49 \pm 1.27)\times10^6$ & $(5.59 \pm 4.14)\times10^5$ & $(1.70 \pm 1.17)\times10^5$ & $(3.61 \pm 2.53)\times10^4$ & $(5.15 \pm 11.6)\times10^3$ \cr
& *114 & \dots & $(6.37 \pm 1.62)\times10^5$ & $(2.67 \pm 0.57)\times10^5$ & $(8.63 \pm 1.68)\times10^4$ & $(1.91 \pm 0.38)\times10^4$ & $(2.53 \pm 1.56)\times10^3$ \cr
& *187 & \dots & $(2.87 \pm 0.37)\times10^5$ & $(1.30 \pm 0.13)\times10^5$ & $(4.18 \pm 0.39)\times10^4$ & $(8.61 \pm 0.89)\times10^3$ & $(1.16 \pm 0.45)\times10^3$ \cr
& *320 & \dots & $(1.34 \pm 0.08)\times10^5$ & $(6.36 \pm 0.30)\times10^4$ & $(2.12 \pm 0.09)\times10^4$ & $(4.61 \pm 0.22)\times10^3$ & $(6.43 \pm 1.20)\times10^2$ \cr
& *502 & \dots & $(7.20 \pm 0.26)\times10^4$ & $(3.53 \pm 0.10)\times10^4$ & $(1.20 \pm 0.03)\times10^4$ & $(2.65 \pm 0.09)\times10^3$ & $(4.26 \pm 0.59)\times10^2$ \cr
& *684 & \dots & $(4.38 \pm 0.18)\times10^4$ & $(2.21 \pm 0.07)\times10^4$ & $(7.68 \pm 0.22)\times10^3$ & $(1.62 \pm 0.06)\times10^3$ & $(2.39 \pm 0.41)\times10^2$ \cr
& *890 & \dots & $(3.23 \pm 0.09)\times10^4$ & $(1.63 \pm 0.04)\times10^4$ & $(5.66 \pm 0.12)\times10^3$ & $(1.14 \pm 0.04)\times10^3$ & $(1.55 \pm 0.30)\times10^2$ \cr
& 1158 & \dots & $(2.40 \pm 0.05)\times10^4$ & $(1.22 \pm 0.02)\times10^4$ & $(4.12 \pm 0.07)\times10^3$ & $(8.29 \pm 0.25)\times10^2$ & $(1.29 \pm 0.23)\times10^2$ \cr
& 1505 & \dots & $(1.83 \pm 0.03)\times10^4$ & $(9.31 \pm 0.11)\times10^3$ & $(3.19 \pm 0.04)\times10^3$ & $(6.82 \pm 0.16)\times10^2$ & $(1.13 \pm 0.16)\times10^2$ \cr
& 1956 & \dots & $(1.46 \pm 0.02)\times10^4$ & $(7.38 \pm 0.07)\times10^3$ & $(2.50 \pm 0.03)\times10^3$ & $(5.36 \pm 0.11)\times10^2$ & $(1.03 \pm 0.15)\times10^2$ \cr
& 2649 & \dots & $(1.16 \pm 0.01)\times10^4$ & $(5.91 \pm 0.06)\times10^3$ & \dots & \dots & \dots \cr
\noalign{\vskip 3pt\hrule\vskip 3pt}
545\GHz & **53 & \dots & \dots & $(2.36 \pm 1.37)\times10^5$ & $(7.46 \pm 3.90)\times10^4$ & $(1.61 \pm 0.85)\times10^4$ & $(2.46 \pm 3.80)\times10^3$ \cr
& *114 & \dots & \dots & $(1.24 \pm 0.21)\times10^5$ & $(4.12 \pm 0.63)\times10^4$ & $(9.32 \pm 1.44)\times10^3$ & $(1.35 \pm 0.57)\times10^3$ \cr
& *187 & \dots & \dots & $(6.63 \pm 0.51)\times10^4$ & $(2.16 \pm 0.15)\times10^4$ & $(4.48 \pm 0.35)\times10^3$ & $(6.11 \pm 1.78)\times10^2$ \cr
& *320 & \dots & \dots & $(3.34 \pm 0.12)\times10^4$ & $(1.15 \pm 0.04)\times10^4$ & $(2.49 \pm 0.09)\times10^3$ & $(3.51 \pm 0.52)\times10^2$ \cr
& *502 & \dots & \dots & $(1.91 \pm 0.04)\times10^4$ & $(6.73 \pm 0.15)\times10^3$ & $(1.48 \pm 0.04)\times10^3$ & $(2.38 \pm 0.28)\times10^2$ \cr
& *684 & \dots & \dots & $(1.25 \pm 0.03)\times10^4$ & $(4.48 \pm 0.09)\times10^3$ & $(9.39 \pm 0.29)\times10^2$ & $(1.37 \pm 0.21)\times10^2$ \cr
& *890 & \dots & \dots & $(9.17 \pm 0.17)\times10^3$ & $(3.29 \pm 0.05)\times10^3$ & $(6.59 \pm 0.19)\times10^2$ & $(9.04 \pm 1.60)\times10^1$ \cr
& 1158 & \dots & \dots & $(6.83 \pm 0.10)\times10^3$ & $(2.45 \pm 0.03)\times10^3$ & $(4.97 \pm 0.14)\times10^2$ & $(7.53 \pm 1.33)\times10^1$ \cr
& 1505 & \dots & \dots & $(5.34 \pm 0.06)\times10^3$ & $(1.91 \pm 0.02)\times10^3$ & $(4.21 \pm 0.10)\times10^2$ & $(7.10 \pm 1.01)\times10^1$ \cr
& 1956 & \dots & \dots & $(4.24 \pm 0.04)\times10^3$ & $(1.51 \pm 0.02)\times10^3$ & $(3.29 \pm 0.07)\times10^2$ & $(6.47 \pm 1.01)\times10^1$ \cr
& 2649 & \dots & \dots & $(3.42 \pm 0.04)\times10^3$ & \dots & \dots & \dots \cr
\noalign{\vskip 3pt\hrule\vskip 3pt}
353\GHz & **53 & \dots & \dots & \dots & $(2.53 \pm 1.12)\times10^4$ & $(5.52 \pm 2.45)\times10^3$ & $(7.70 \pm 10.8)\times10^2$ \cr
& *114 & \dots & \dots & \dots & $(1.43 \pm 0.19)\times10^4$ & $(3.23 \pm 0.45)\times10^3$ & $(4.65 \pm 1.72)\times10^2$ \cr
& *187 & \dots & \dots & \dots & $(7.69 \pm 0.48)\times10^3$ & $(1.66 \pm 0.11)\times10^3$ & $(2.34 \pm 0.56)\times10^2$ \cr
& *320 & \dots & \dots & \dots & $(4.23 \pm 0.12)\times10^3$ & $(9.59 \pm 0.32)\times10^2$ & $(1.51 \pm 0.18)\times10^2$ \cr
& *502 & \dots & \dots & \dots & $(2.54 \pm 0.05)\times10^3$ & $(5.91 \pm 0.16)\times10^2$ & $(1.01 \pm 0.11)\times10^2$ \cr
& *684 & \dots & \dots & \dots & $(1.70 \pm 0.03)\times10^3$ & $(3.76 \pm 0.11)\times10^2$ & $(6.22 \pm 0.82)\times10^1$ \cr
& *890 & \dots & \dots & \dots & $(1.25 \pm 0.02)\times10^3$ & $(2.72 \pm 0.08)\times10^2$ & $(3.96 \pm 0.69)\times10^1$ \cr
& 1158 & \dots & \dots & \dots & $(9.15 \pm 0.14)\times10^2$ & $(1.87 \pm 0.09)\times10^2$ & $(2.99 \pm 0.90)\times10^1$ \cr
& 1505 & \dots & \dots & \dots & $(7.37 \pm 0.11)\times10^2$ & $(1.54 \pm 0.09)\times10^2$ & $(2.19 \pm 1.09)\times10^1$ \cr
& 1956 & \dots & \dots & \dots & $(6.05 \pm 0.09)\times10^2$ & $(1.34 \pm 0.04)\times10^2$ & $(3.41 \pm 0.64)\times10^1$ \cr
\noalign{\vskip 3pt\hrule\vskip 3pt}
217\GHz & **53 & \dots & \dots & \dots & \dots & $(1.37 \pm 0.55)\times10^3$ & $(2.16 \pm 2.41)\times10^2$ \cr
& *114 & \dots & \dots & \dots & \dots & $(7.59 \pm 1.08)\times10^2$ & $(1.10 \pm 0.42)\times10^2$ \cr
& *187 & \dots & \dots & \dots & \dots & $(4.27 \pm 0.38)\times10^2$ & $(7.80 \pm 1.60)\times10^1$ \cr
& *320 & \dots & \dots & \dots & \dots & $(2.65 \pm 0.13)\times10^2$ & $(5.68 \pm 0.60)\times10^1$ \cr
& *502 & \dots & \dots & \dots & \dots & $(1.78 \pm 0.08)\times10^2$ & $(4.77 \pm 0.47)\times10^1$ \cr
& *684 & \dots & \dots & \dots & \dots & $(1.18 \pm 0.07)\times10^2$ & $(3.05 \pm 0.43)\times10^1$ \cr
& *890 & \dots & \dots & \dots & \dots & $(8.78 \pm 0.71)\times10^1$ & $(1.87 \pm 0.48)\times10^1$ \cr
& 1158 & \dots & \dots & \dots & \dots & $(5.97 \pm 1.57)\times10^1$ & $(1.46 \pm 1.19)\times10^1$ \cr
& 1505 & \dots & \dots & \dots & \dots & $(3.08 \pm 2.15)\times10^1$ & $(8.40 \pm 16.4)\times10^0$ \cr
& 1956 & \dots & \dots & \dots & \dots & $(4.73 \pm 0.65)\times10^1$ & $(1.49 \pm 0.53)\times10^1$ \cr
\noalign{\vskip 3pt\hrule\vskip 3pt}
143\GHz & **53 & \dots & \dots & \dots & \dots & \dots & $(7.23 \pm 6.67)\times10^1$ \cr
& *114 & \dots & \dots & \dots & \dots & \dots & $(3.05 \pm 1.43)\times10^1$ \cr
& *187 & \dots & \dots & \dots & \dots & \dots & $(2.20 \pm 0.67)\times10^1$ \cr
& *320 & \dots & \dots & \dots & \dots & \dots & $(2.14 \pm 0.31)\times10^1$ \cr
& *502 & \dots & \dots & \dots & \dots & \dots & $(1.99 \pm 0.28)\times10^1$ \cr
& *684 & \dots & \dots & \dots & \dots & \dots & $(1.55 \pm 0.29)\times10^1$ \cr
& *890 & \dots & \dots & \dots & \dots & \dots & $(1.16 \pm 0.35)\times10^1$ \cr
& 1158 & \dots & \dots & \dots & \dots & \dots & $(1.02 \pm 0.91)\times10^1$ \cr
& 1505 & \dots & \dots & \dots & \dots & \dots & $(1.26 \pm 1.27)\times10^1$ \cr
\noalign{\vskip 3pt\hrule\vskip 3pt}}}
\endPlancktablewide
\endgroup
\end{table*}

%pk_results_7
\begin{table*}[!tbh]
\begingroup
\newdimen\tblskip \tblskip=5pt
\caption{Cosmic IR Background power spectra. These
estimates are obtained from the CMB- and dust-free map power spectra
(Table~\ref{tab:res_map_pow_spec}), which have been corrected for SZ
contaminations (Eqs.~\ref{eq:SZ1} and \ref{eq:SZ3}), and
for the spurious CIB contamination induced by our CMB template
(Eq.~\ref{eq:corr_CMB}, computed using our best-fit
 model described in Sect.~\ref{mod_hod}). The errors contain all the terms:
statistical uncertainty; beam uncertainty; and errors from the SZ correction.
Rows composed of entirely empty values have been omitted.}
\label{tab:cib_final}
\nointerlineskip
\vskip -3mm
\footnotesize
\setbox\tablebox=\vbox{
 \newdimen\digitwidth
 \setbox0=\hbox{\rm 0}
  \digitwidth=\wd0
  \catcode`*=\active
  \def*{\kern\digitwidth}
  \newdimen\signwidth
  \setbox0=\hbox{+}
  \signwidth=\wd0
  \catcode`!=\active
  \def!{\kern\signwidth}
\halign{\tabskip=0pt#\hfil\tabskip=0.5em&
  \hfil#\hfil\tabskip=1.0em&
  \hfil#\hfil\tabskip=1.0em&
  \hfil#\hfil\tabskip=1.0em&
  \hfil#\hfil\tabskip=1.0em&
  \hfil#\hfil\tabskip=1.0em&
  \hfil#\hfil\tabskip=1.0em&
  \hfil#\hfil\tabskip=0pt\cr
\noalign{\doubleline}
\noalign{\vskip -2pt}
&&\multispan6\hfil{Cross-power spectra, $C_\ell$, and statistical uncertainty
 [${\rm Jy}^2\,{\rm sr}^{-1}$]}\hfil\cr
\noalign{\vskip -5pt}
&&\multispan6\hrulefill\cr
& $\ell$& 3000\GHz& 857\GHz& 545\GHz& 353\GHz& 217\GHz& 143\GHz\cr
\noalign{\vskip 3pt\hrule\vskip 3pt}
3000\GHz & *187 & $<$ $5.69\; 10^{5}$ & $(1.12 \pm 0.85)\times10^5$ & $(3.94 \pm 3.88)\times10^4$ & $(8.92 \pm 12.4)\times10^3$ & $(5.59 \pm 26.7)\times10^2$ & $\cdot$ \cr
& *320 & $(7.89 \pm 3.14)\times10^4$ & $(7.49 \pm 2.09)\times10^4$ & $(3.04 \pm 0.85)\times10^4$ & $(1.13 \pm 0.27)\times10^4$ & $(2.34 \pm 0.71)\times10^3$ & $\cdot$ \cr
& *502 & $(3.46 \pm 1.15)\times10^4$ & $(3.10 \pm 0.79)\times10^4$ & $(1.23 \pm 0.33)\times10^4$ & $(4.49 \pm 1.15)\times10^3$ & $(1.18 \pm 0.34)\times10^3$ & $\cdot$ \cr
& *684 & $(2.90 \pm 0.70)\times10^4$ & $(2.24 \pm 0.47)\times10^4$ & $(9.35 \pm 2.10)\times10^3$ & $(2.87 \pm 0.73)\times10^3$ & $(6.88 \pm 2.61)\times10^2$ & $\cdot$ \cr
& *890 & $(1.87 \pm 0.47)\times10^4$ & $(1.62 \pm 0.30)\times10^4$ & $(6.61 \pm 1.34)\times10^3$ & $(1.99 \pm 0.45)\times10^3$ & $(2.34 \pm 1.66)\times10^2$ & $\cdot$ \cr
& 1158 & $(1.43 \pm 0.40)\times10^4$ & $(1.17 \pm 0.22)\times10^4$ & $(4.71 \pm 0.93)\times10^3$ & $(1.25 \pm 0.31)\times10^3$ & $(3.55 \pm 1.43)\times10^2$ & $\cdot$ \cr
& 1505 & $(1.09 \pm 0.41)\times10^4$ & $(8.30 \pm 1.85)\times10^3$ & $(3.12 \pm 0.76)\times10^3$ & $(1.14 \pm 0.30)\times10^3$ & $(2.41 \pm 1.07)\times10^2$ & $\cdot$ \cr
& 1956 & $(1.05 \pm 0.58)\times10^4$ & $(8.04 \pm 2.39)\times10^3$ & $(3.16 \pm 1.00)\times10^3$ & $(9.67 \pm 3.44)\times10^2$ & $(2.58 \pm 1.14)\times10^2$ & $\cdot$ \cr
& 2649 & $(1.06 \pm 1.04)\times10^4$ & $(6.47 \pm 3.27)\times10^3$ & $(2.61 \pm 1.38)\times10^3$ & \dots & \dots & $\cdot$ \cr
\noalign{\vskip 3pt\hrule\vskip 3pt}
857\GHz & **53 & \dots & $<$ $2.76\; 10^{6}$ & $<$ $9.73\; 10^{5}$ & $<$ $2.91\; 10^{5}$ & $<$ $6.43\; 10^{4}$ & $<$ $1.81\; 10^{4}$ \cr
& *114 & \dots & $<$ $7.99\; 10^{5}$ & $<$ $3.23\; 10^{5}$ & $<$ $1.05\; 10^{5}$ & $<$ $2.49\; 10^{4}$ & $<$ $5.12\; 10^{3}$ \cr
& *187 & \dots & $(2.87 \pm 0.37)\times10^5$ & $(1.30 \pm 0.13)\times10^5$ & $(4.30 \pm 0.41)\times10^4$ & $(9.70 \pm 1.22)\times10^3$ & $(1.84 \pm 0.45)\times10^3$ \cr
& *320 & \dots & $(1.34 \pm 0.08)\times10^5$ & $(6.36 \pm 0.30)\times10^4$ & $(2.20 \pm 0.11)\times10^4$ & $(5.26 \pm 0.53)\times10^3$ & $(1.06 \pm 0.12)\times10^3$ \cr
& *502 & \dots & $(7.20 \pm 0.26)\times10^4$ & $(3.53 \pm 0.10)\times10^4$ & $(1.25 \pm 0.06)\times10^4$ & $(3.03 \pm 0.32)\times10^3$ & $(6.52 \pm 0.59)\times10^2$ \cr
& *684 & \dots & $(4.38 \pm 0.18)\times10^4$ & $(2.21 \pm 0.07)\times10^4$ & $(7.99 \pm 0.39)\times10^3$ & $(1.88 \pm 0.22)\times10^3$ & $(3.86 \pm 0.41)\times10^2$ \cr
& *890 & \dots & $(3.23 \pm 0.09)\times10^4$ & $(1.63 \pm 0.04)\times10^4$ & $(5.88 \pm 0.27)\times10^3$ & $(1.31 \pm 0.16)\times10^3$ & $(2.55 \pm 0.30)\times10^2$ \cr
& 1158 & \dots & $(2.40 \pm 0.05)\times10^4$ & $(1.22 \pm 0.02)\times10^4$ & $(4.25 \pm 0.17)\times10^3$ & $(9.18 \pm 0.87)\times10^2$ & $(1.76 \pm 0.23)\times10^2$ \cr
& 1505 & \dots & $(1.83 \pm 0.03)\times10^4$ & $(9.31 \pm 0.11)\times10^3$ & $(3.24 \pm 0.10)\times10^3$ & $(7.00 \pm 0.23)\times10^2$ & $(1.23 \pm 0.16)\times10^2$ \cr
& 1956 & \dots & $(1.46 \pm 0.02)\times10^4$ & $(7.38 \pm 0.07)\times10^3$ & $(2.54 \pm 0.07)\times10^3$ & $(5.38 \pm 0.12)\times10^2$ & $(1.03 \pm 0.15)\times10^2$ \cr
& 2649 & \dots & $(1.16 \pm 0.01)\times10^4$ & $(5.91 \pm 0.06)\times10^3$ & \dots & \dots & \dots \cr
\noalign{\vskip 3pt\hrule\vskip 3pt}
545\GHz & **53 & \dots & \dots & $<$ $3.74\; 10^{5}$ & $<$ $1.15\; 10^{5}$ & $<$ $2.58\; 10^{4}$ & $<$ $7.04\; 10^{3}$ \cr
& *114 & \dots & \dots & $<$ $1.45\; 10^{5}$ & $<$ $4.84\; 10^{4}$ & $<$ $1.16\; 10^{4}$ & $<$ $2.53\; 10^{3}$ \cr
& *187 & \dots & \dots & $(6.63 \pm 0.51)\times10^4$ & $(2.22 \pm 0.16)\times10^4$ & $(4.97 \pm 0.48)\times10^3$ & $(1.01 \pm 0.19)\times10^3$ \cr
& *320 & \dots & \dots & $(3.34 \pm 0.12)\times10^4$ & $(1.19 \pm 0.05)\times10^4$ & $(2.79 \pm 0.21)\times10^3$ & $(5.98 \pm 0.67)\times10^2$ \cr
& *502 & \dots & \dots & $(1.91 \pm 0.04)\times10^4$ & $(6.93 \pm 0.23)\times10^3$ & $(1.65 \pm 0.12)\times10^3$ & $(3.77 \pm 0.39)\times10^2$ \cr
& *684 & \dots & \dots & $(1.25 \pm 0.03)\times10^4$ & $(4.61 \pm 0.16)\times10^3$ & $(1.06 \pm 0.09)\times10^3$ & $(2.29 \pm 0.27)\times10^2$ \cr
& *890 & \dots & \dots & $(9.17 \pm 0.17)\times10^3$ & $(3.39 \pm 0.11)\times10^3$ & $(7.41 \pm 0.63)\times10^2$ & $(1.54 \pm 0.19)\times10^2$ \cr
& 1158 & \dots & \dots & $(6.83 \pm 0.10)\times10^3$ & $(2.50 \pm 0.07)\times10^3$ & $(5.38 \pm 0.35)\times10^2$ & $(1.03 \pm 0.14)\times10^2$ \cr
& 1505 & \dots & \dots & $(5.34 \pm 0.06)\times10^3$ & $(1.93 \pm 0.04)\times10^3$ & $(4.30 \pm 0.12)\times10^2$ & $(7.09 \pm 1.80)\times10^1$ \cr
& 1956 & \dots & \dots & $(4.24 \pm 0.04)\times10^3$ & $(1.52 \pm 0.03)\times10^3$ & $(3.30 \pm 0.07)\times10^2$ & $(5.89 \pm 1.73)\times10^1$ \cr
& 2649 & \dots & \dots & $(3.42 \pm 0.04)\times10^3$ & \dots & \dots & \dots \cr
\noalign{\vskip 3pt\hrule\vskip 3pt}
353\GHz & **53 & \dots & \dots & \dots & $<$ $3.68\; 10^{4}$ & $<$ $8.01\; 10^{3}$ & $<$ $2.05\; 10^{3}$ \cr
& *114 & \dots & \dots & \dots & $<$ $1.66\; 10^{4}$ & $<$ $3.82\; 10^{3}$ & $<$ $8.26\; 10^{2}$ \cr
& *187 & \dots & \dots & \dots & $(7.88 \pm 0.53)\times10^3$ & $(1.75 \pm 0.15)\times10^3$ & $(3.61 \pm 0.62)\times10^2$ \cr
& *320 & \dots & \dots & \dots & $(4.35 \pm 0.18)\times10^3$ & $(1.02 \pm 0.06)\times10^3$ & $(2.32 \pm 0.24)\times10^2$ \cr
& *502 & \dots & \dots & \dots & $(2.60 \pm 0.10)\times10^3$ & $(6.21 \pm 0.38)\times10^2$ & $(1.48 \pm 0.14)\times10^2$ \cr
& *684 & \dots & \dots & \dots & $(1.74 \pm 0.07)\times10^3$ & $(3.97 \pm 0.27)\times10^2$ & $(9.42 \pm 1.06)\times10^1$ \cr
& *890 & \dots & \dots & \dots & $(1.29 \pm 0.05)\times10^3$ & $(2.87 \pm 0.20)\times10^2$ & $(6.33 \pm 0.83)\times10^1$ \cr
& 1158 & \dots & \dots & \dots & $(9.35 \pm 0.33)\times10^2$ & $(1.99 \pm 0.14)\times10^2$ & $(4.56 \pm 0.91)\times10^1$ \cr
& 1505 & \dots & \dots & \dots & $(7.45 \pm 0.22)\times10^2$ & $(1.59 \pm 0.10)\times10^2$ & $(2.77 \pm 1.11)\times10^1$ \cr
& 1956 & \dots & \dots & \dots & $(6.08 \pm 0.16)\times10^2$ & $(1.35 \pm 0.05)\times10^2$ & $(3.53 \pm 0.69)\times10^1$ \cr
\noalign{\vskip 3pt\hrule\vskip 3pt}
217\GHz & **53 & \dots & \dots & \dots & \dots & $<$ $1.78\; 10^{3}$ & $<$ $4.74\; 10^{2}$ \cr
& *114 & \dots & \dots & \dots & \dots & $<$ $8.47\; 10^{2}$ & $<$ $1.89\; 10^{2}$ \cr
& *187 & \dots & \dots & \dots & \dots & $(4.17 \pm 0.47)\times10^2$ & $(1.04 \pm 0.19)\times10^2$ \cr
& *320 & \dots & \dots & \dots & \dots & $(2.62 \pm 0.20)\times10^2$ & $(7.49 \pm 0.81)\times10^1$ \cr
& *502 & \dots & \dots & \dots & \dots & $(1.75 \pm 0.13)\times10^2$ & $(5.87 \pm 0.58)\times10^1$ \cr
& *684 & \dots & \dots & \dots & \dots & $(1.17 \pm 0.10)\times10^2$ & $(3.93 \pm 0.50)\times10^1$ \cr
& *890 & \dots & \dots & \dots & \dots & $(8.82 \pm 0.89)\times10^1$ & $(2.64 \pm 0.52)\times10^1$ \cr
& 1158 & \dots & \dots & \dots & \dots & $(6.42 \pm 1.61)\times10^1$ & $(2.21 \pm 1.19)\times10^1$ \cr
& 1505 & \dots & \dots & \dots & \dots & $(3.34 \pm 2.15)\times10^1$ & $(1.07 \pm 1.65)\times10^1$ \cr
& 1956 & \dots & \dots & \dots & \dots & $(4.74 \pm 0.65)\times10^1$ & $(1.45 \pm 0.54)\times10^1$ \cr
\noalign{\vskip 3pt\hrule\vskip 3pt}
143\GHz & **53 & \dots & \dots & \dots & \dots & \dots & $<$ $1.55\; 10^{2}$ \cr
& *114 & \dots & \dots & \dots & \dots & \dots & $<$ $6.41\; 10^{1}$ \cr
& *187 & \dots & \dots & \dots & \dots & \dots & $(3.64 \pm 0.73)\times10^1$ \cr
& *320 & \dots & \dots & \dots & \dots & \dots & $(3.23 \pm 0.35)\times10^1$ \cr
& *502 & \dots & \dots & \dots & \dots & \dots & $(2.81 \pm 0.30)\times10^1$ \cr
& *684 & \dots & \dots & \dots & \dots & \dots & $(2.27 \pm 0.29)\times10^1$ \cr
& *890 & \dots & \dots & \dots & \dots & \dots & $(1.84 \pm 0.35)\times10^1$ \cr
& 1158 & \dots & \dots & \dots & \dots & \dots & $(1.58 \pm 0.91)\times10^1$ \cr
& 1505 & \dots & \dots & \dots & \dots & \dots & $(1.25 \pm 1.28)\times10^1$ \cr
\noalign{\vskip 3pt\hrule\vskip 3pt}}}
\endPlancktablewide
\endgroup
\end{table*}

\begin{table*}[!tbh]
\begingroup
\newdimen\tblskip \tblskip=5pt
\caption{Measured bispectrum at 217, 353 and 545\GHz\ in
${\rm Jy}^3\,{\rm sr}^{-1}$ (see Sect.~\ref{Sect:bisp_pipeline}).
The covariance matrix is available at ESA�s
Planck Legacy Archive (PLA).}
\label{Table:217n353n545bispvaluesjy}
\nointerlineskip
\vskip -3mm
\footnotesize
\setbox\tablebox=\vbox{
 \newdimen\digitwidth
 \setbox0=\hbox{\rm 0}
  \digitwidth=\wd0
  \catcode`*=\active
  \def*{\kern\digitwidth}
  \newdimen\signwidth
  \setbox0=\hbox{+}
  \signwidth=\wd0
  \catcode`!=\active
  \def!{\kern\signwidth}
\halign{\tabskip=0pt\hfil#\hfil\tabskip=1.0em&
  \hfil#\hfil\tabskip=1.0em&
  \hfil#\hfil\tabskip=1.0em&
  \hfil#\hfil\tabskip=1.0em&
  \hfil#\hfil\tabskip=1.0em&
  \hfil#\hfil\tabskip=0pt\cr
\noalign{\doubleline}
\noalign{\vskip -2pt}
\multispan3\hfil Multipole bins\hfil& \multispan3\hfil
 Bispectrum coefficients $b_{\ell_1\ell_2\ell_3}$\hfil\cr
\noalign{\vskip 2pt}
$\ell_1$& $\ell_2$& $\ell_3$& 217\GHz& 353\GHz& 545\GHz\cr
\noalign{\vskip 3pt\hrule\vskip 3pt}
192& 192& 192& $9.098\times10^1$& $1.788\times10^3$& $3.648\times10^2$\cr
192& 192& 320& $3.351\times10^1$& $7.837\times10^2$& $2.809\times10^2$\cr
192& 320& 320& $3.540\times10^1$& $1.229\times10^3$& $3.044\times10^2$\cr
192& 320& 448& $1.706\times10^1$& $6.900\times10^2$& $1.454\times10^2$\cr
192& 448& 448& $1.537\times10^1$& $4.193\times10^2$& $1.077\times10^2$\cr
192& 448& 576& $1.128\times10^1$& $2.806\times10^2$& $8.164\times10^3$\cr
192& 576& 576& $1.797\times10^1$& $4.123\times10^2$& $6.197\times10^3$\cr
192& 576& 704&             3.924& $1.588\times10^2$& $6.147\times10^3$\cr
192& 704& 704&             1.461& $6.766\times10^1$& $5.185\times10^3$\cr
192& 704& 832&             6.888& $1.848\times10^2$& $7.090\times10^3$\cr
192& 832& 832&             3.886& $1.704\times10^2$& $6.692\times10^3$\cr
320& 320& 320& $1.167\times10^1$& $3.583\times10^2$& $1.207\times10^2$\cr
320& 320& 448& $1.058\times10^1$& $4.245\times10^2$& $9.126\times10^3$\cr
320& 320& 576& 9.259& $3.794\times10^2$& $7.184\times10^3$\cr
320& 448& 448& 9.829& $2.292\times10^2$& $6.815\times10^3$\cr
320& 448& 576& 6.725& $1.817\times10^2$& $4.186\times10^3$\cr
320& 448& 704& 6.932& $3.021\times10^2$& $6.529\times10^3$\cr
320& 576& 576& 6.978& $1.797\times10^2$& $2.683\times10^3$\cr
320& 576& 704& 5.152& $7.602\times10^2$& $2.402\times10^3$\cr
320& 576& 832& 2.121& $1.306\times10^2$& $3.140\times10^3$\cr
320& 704& 704& 2.332& $1.398\times10^2$& $5.816\times10^3$\cr
320& 704& 832& 3.241& $1.131\times10^2$& $3.826\times10^3$\cr
320& 832& 832& 3.789& $8.380\times10^2$& $5.060\times10^3$\cr
448& 448& 448& 5.096& $2.290\times10^2$& $4.523\times10^3$\cr
448& 448& 576& 7.119& $4.829\times10^2$& $3.431\times10^3$\cr
448& 448& 704& 7.120& $2.041\times10^2$& $3.519\times10^3$\cr
448& 448& 832& 2.043& $1.172\times10^2$& $2.905\times10^3$\cr
448& 576& 576& 5.575& $1.621\times10^2$& $3.765\times10^3$\cr
448& 576& 704& 7.135& $9.080\times10^2$& $1.928\times10^3$\cr
448& 576& 832& 3.883& $9.562\times10^2$& $1.939\times10^3$\cr
448& 704& 704& 7.234& $1.489\times10^2$& $3.278\times10^3$\cr
448& 704& 832& 2.235& $8.322\times10^2$& $2.006\times10^3$\cr
448& 832& 832& 2.702& $1.038\times10^2$& $2.751\times10^3$\cr
576& 576& 576& 3.026& $1.296\times10^2$& $2.842\times10^3$\cr
576& 576& 704& 4.617& $8.385\times10^2$& $1.413\times10^3$\cr
576& 576& 832& 4.747& $8.790\times10^2$& $1.669\times10^3$\cr
576& 704& 704& 2.817& $7.549\times10^2$& \dots\cr
576& 704& 832& 3.977& $4.538\times10^2$& \dots\cr
576& 832& 832& \dots& $5.934\times10^2$& \dots\cr
704& 704& 704& \dots& $1.040\times10^2$& \dots\cr
\noalign{\vskip 3pt\hrule\vskip 3pt}}}
\endPlancktablewide
\endgroup
\end{table*}

\raggedright

\end{document}